\documentclass[
 pra,
 amsmath,amssymb,
 aps,
 twocolumn,
 notitlepage,
 superscriptaddress,
]{revtex4-2}

\usepackage{float}
\makeatletter
\let\newfloat\newfloat@ltx
\makeatother

\usepackage{algorithm}
\usepackage{algpseudocode}

\usepackage[colorlinks=true]{hyperref}

\usepackage{ragged2e}
\usepackage{caption}       % ******* diff from opt paper
\usepackage{subcaption}
\DeclareCaptionJustification{justified}{\justifying}
\usepackage[capitalise]{cleveref}

\usepackage{styles}
\usepackage{fancyhdr}
\usepackage{txfonts}
\usepackage{xurl} % for urls with hyphens
\usepackage{bm} % bold math
\usepackage{physics} % kets and bras
\usepackage{tikz} %for quantum circuit diagrams
\usetikzlibrary{quantikz2} %for quantum circuit diagrams
\usepackage{tikzscale}
\usepackage[shortlabels]{enumitem}

\usepackage{bm} % bold math

\begin{document}

\title{A Comprehensive Cross-Model Framework for Benchmarking the Performance of Quantum Hamiltonian Simulations}

\author{Avimita Chatterjee}
%\orcid{0009-0001-7421-9334}
\affiliation{Department of Computer Science \& Engineering, The Pennsylvania State University, USA}
\affiliation{QED-C, managed by SRI International, USA}

\author{Sonny Rappaport}
\affiliation{QED-C, managed by SRI International, USA}

\author{Anish Giri}
\affiliation{Vanderbilt University, USA}
\affiliation{QED-C, managed by SRI International, USA}

\author{Sonika Johri}
\affiliation{Coherent Computing Inc, Cupertino, CA, USA}

\author{\\ Timothy Proctor}
\affiliation{Quantum Performance Laboratory, Sandia National Laboratories, Livermore, CA 94550, USA}

\author{David E. Bernal Neira}
%\author{David E. Bernal Neira\orcidlink{0000-0002-8308-5016}}
\affiliation{Research Institute of Advanced Computer Science, Universities Space Research Association, Mountain View, CA, USA}
\affiliation{Davidson School of Chemical Engineering, Purdue University, West Lafayette, IN, USA}

\author{Pratik Sathe} %LA-UR number 'LA-UR-24-29412'
% \author{\orcidlink{0000-0002-9978-8955}} 
\affiliation{Theoretical Division (T4), Los Alamos National Laboratory, Los Alamos, New Mexico 87545, USA}

\author{Thomas Lubinski}
%\orcid{0000-0002-3749-3430}
\affiliation{Quantum Circuits Inc, 25 Science Park, New Haven, CT 06511}
\affiliation{QED-C Technical Advisory Committee on Standards and Performance Metrics}

\collaboration{Quantum Economic Development Consortium (QED-C) collaboration} %\noaffiliation

\thanks{This work was sponsored by the Quantum Economic Development Consortium (QED-C) and was performed under the auspices of the QED-C Technical Advisory Committee on Standards and Performance Metrics. The authors acknowledge many committee members for their input to and feedback on the project and this manuscript.}

\date{\rule[15pt]{0pt}{0pt}\today}
             
% ==================================================
\begin{abstract}

\vspace{0.0cm}
Quantum Hamiltonian simulation is one of the most promising applications of quantum computing and forms the basis for many quantum algorithms. Benchmarking them is an important gauge of progress in quantum computing technology. We present a methodology and software framework to evaluate various facets of the performance of gate-based quantum computers on Trotterized quantum Hamiltonian evolution. We propose three distinct modes for benchmarking: (i) comparing simulation on a real device to that on a noiseless classical simulator, (ii) comparing simulation on a real device with exact diagonalization results, and (iii) using scalable mirror circuit techniques to assess hardware performance in scenarios beyond classical simulation methods. We demonstrate this framework on five Hamiltonian models from the HamLib library: the Fermi and Bose-Hubbard models, the transverse field Ising model, the Heisenberg model, and the Max3SAT problem. Experiments were conducted using Qiskit's Aer simulator, BlueQubit's CPU cluster and GPU simulators, and IBM's quantum hardware. Our framework, extendable to other Hamiltonians, provides comprehensive performance profiles that reveal hardware and algorithmic limitations and measure both fidelity and execution times, identifying crossover points where quantum hardware outperforms CPU/GPU simulators.

\end{abstract}

% keywords 
\keywords{Quantum Computing \and Benchmarks \and Benchmarking \and Algorithms \and Application Benchmarks \and Variational Quantum Eigensolver \and Hydrogen Lattice \and Machine Learning \and HHL }

\maketitle

\tableofcontents

% --------------------------------------------------

%%% Footer with title
\pagestyle{fancy}

\renewcommand{\headrulewidth}{0.0pt}
\lhead{}
\rhead{\thepage}

\renewcommand{\footrulewidth}{0.4pt}
\cfoot{}
\lfoot{A Framework for Benchmarking the Performance of Quantum Hamiltonian Simulations}
\rfoot{\today}
\vspace{2cm}

% ===========================================================

%****************
\begin{figure*}[t!]
\includegraphics[width=0.67\columnwidth]{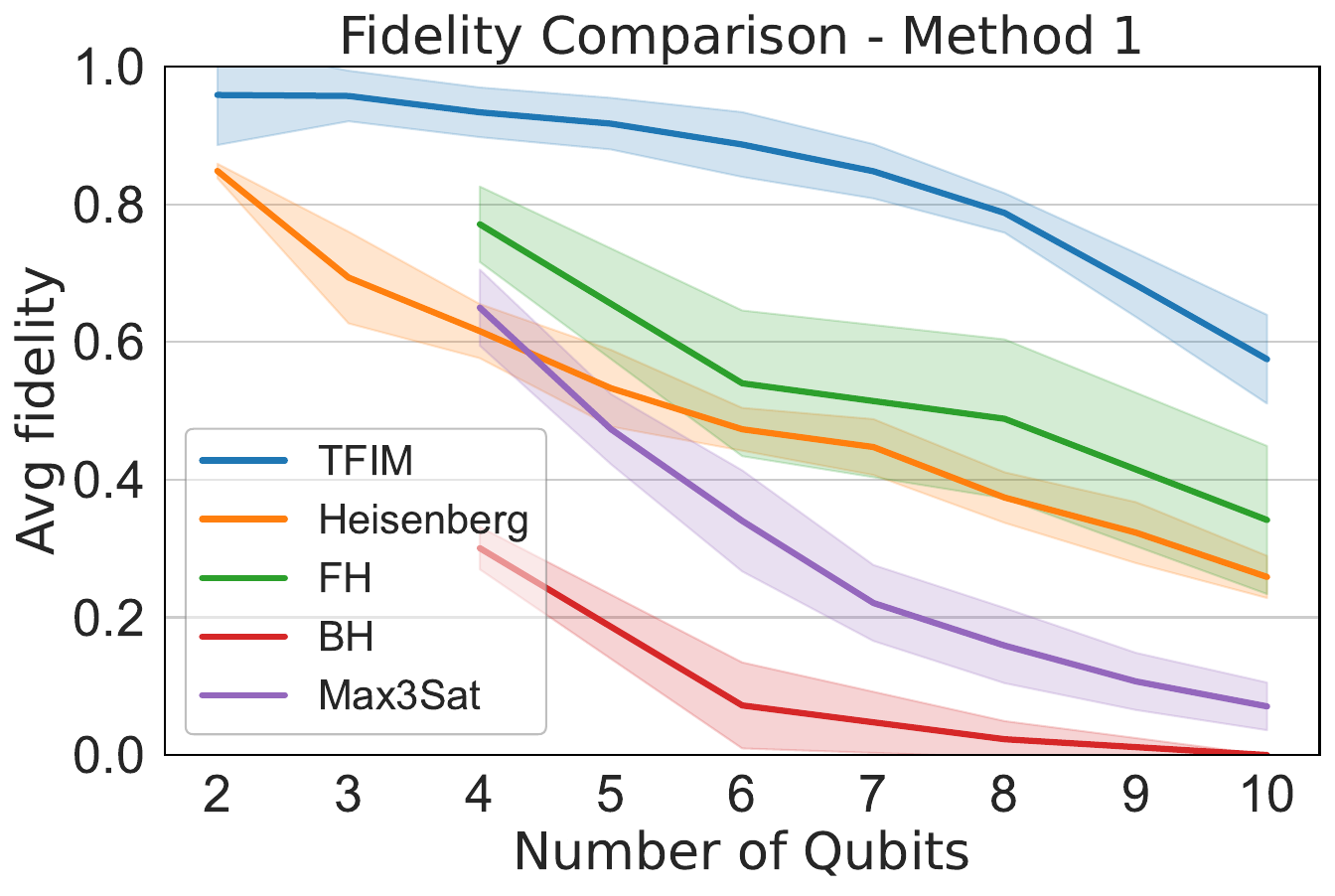}
\includegraphics[width=0.67\columnwidth]{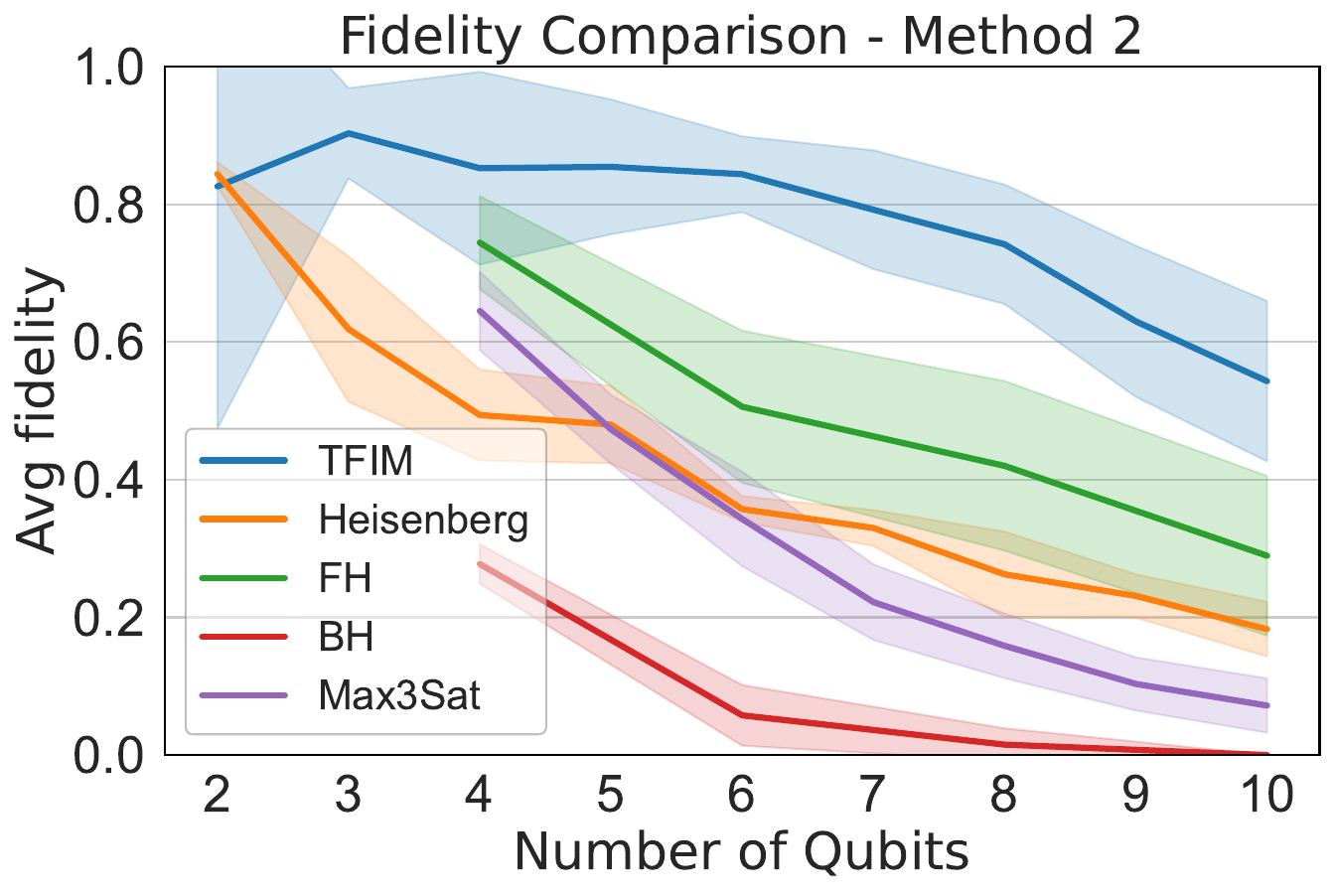}
\includegraphics[width=0.67\columnwidth]{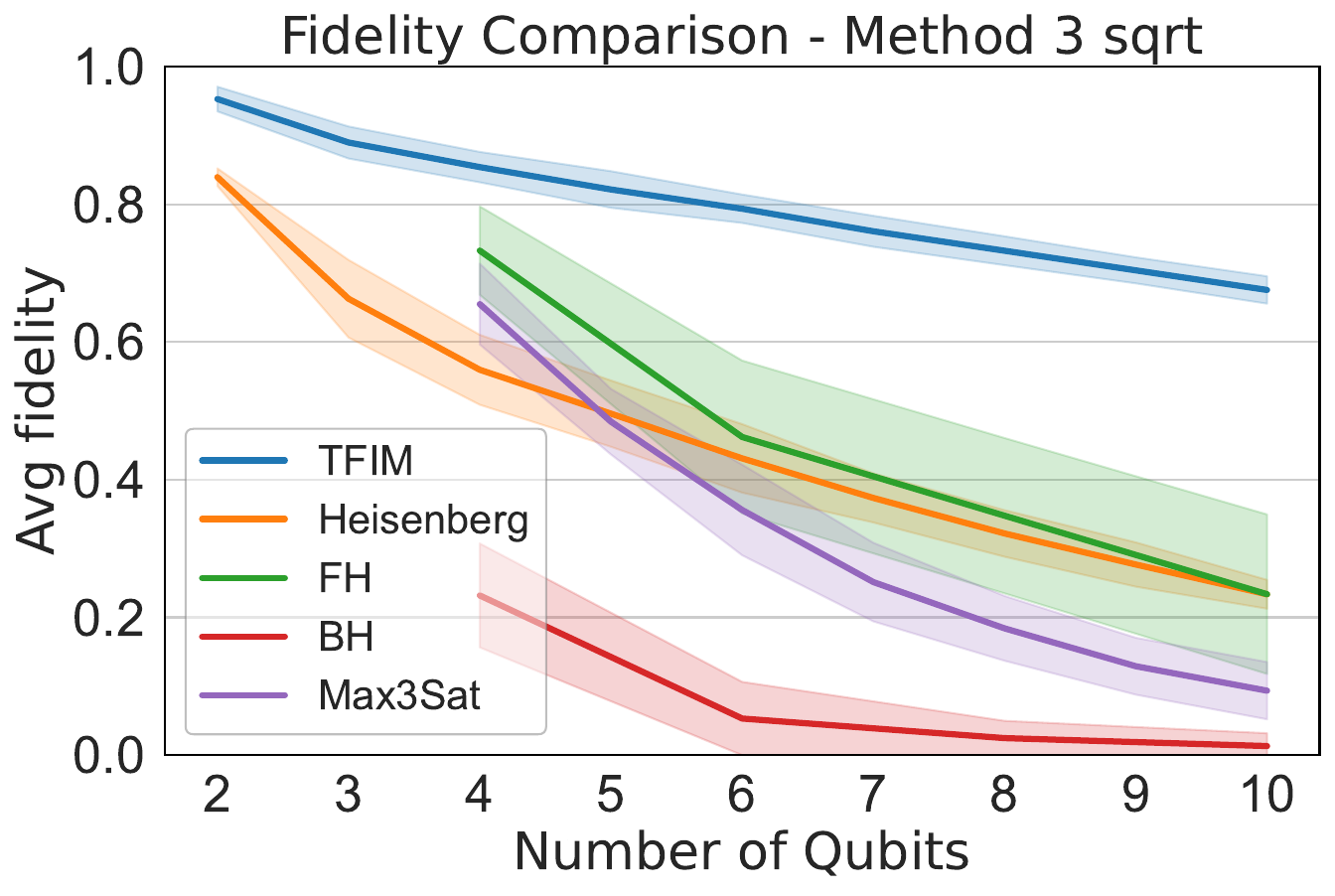}
\caption{
\textbf{Fidelity Comparison across Five Hamiltonians from HamLib using Three Benchmark Methods.}
These plots summarize results from the execution of three benchmark methods across five Hamiltonians selected from HamLib: the Transverse-Field Ising Model (TFIM), Heisenberg, Fermi-Hubbard (FH), Bose-Hubbard (BH), and Max3SAT. 
The benchmark circuits implement a Trotterized Hamiltonian simulation of 5 steps with a total time of 1.0. We execute these circuits over a range of qubit widths, from 2 to 10, and plot the fidelity of execution computed using three different methods. The execution was performed using 1000 shots on a classically implemented quantum simulator measured to mimic a quantum computer with a quantum volume of 2048. For each of the Hamiltonians, we also sweep over a range of parameter settings offered within HamLib, with results represented by the shaded region around the solid line. These benchmark methods provide insight into multiple aspects of Hamiltonian simulation, such as the impact of noise, Trotterization error, and circuit depth, as explained in the text.
}
\label{fig:hamlib_fidelity_methods}
\end{figure*}
%****************

%This figure compares fidelity and circuit depth across five Hamiltonian models (TFIM, Heisenberg, FH, BH, and Max3SAT) using four fidelity methods. The top plots show that Method 2 (noiseless) consistently has the highest fidelity, while Method 3 (mirror) has the lowest. TFIM maintains the highest fidelity overall, inversely correlating with the circuit depth shown in the bottom plots. BH, with the lowest fidelity, has the highest circuit depth, while TFIM has the least. Method 3's doubled depth underscores the inverse relationship between fidelity and circuit depth.

% ----------------------------------------------------------

\section{Introduction}
\label{sec:introduction}

Quantum computing hardware and software are advancing rapidly, enhancing the scale and sophistication of algorithms that can be tested ~\cite{brown20245}. Developments in algorithmic techniques that aim to leverage quantum advantage~\cite{preskill2018quantum} match these advances. In particular, quantum algorithms to simulate physical systems and solve combinatorial optimization problems have been proposed which require mapping the time evolution of an abstract Hamiltonian to physical qubit operations using a quantum circuit representation~\cite{Somma_2003,childs2018toward,granet2019analytical,akinci2013critical,farhiQuantumApproximateOptimization2014}. Effective benchmarking is critical for evaluating the accuracy and throughput of such real-world applications as quantum computing technology matures ~\cite{PhysRevA.77.012307, PhysRevLett.106.180504, Blume-Kohout2017-no, Cross_2019, Boixo_2018, wack_clops_2021, parekh2016benchmarking, chen2022veriqbench, li2023qasmbench, cornelissen2021scalable, tomesh2022supermarq}.

A Hamiltonian corresponds to the energy of a physical system and determines its dynamics~\cite{Eisberg_Resnick_2017a}. Hamiltonian simulation is thus fundamental to understanding and predicting the behavior of systems ranging from materials~\cite{von2021quantum},  molecules~\cite{mcardle2020quantum}, electrons, phonons~\cite{denner2023hybrid}, and atomic nuclei~\cite{savage2024quantum}, impacting fields with real-world implications such as materials science and chemistry. Simulation of specially constructed Hamiltonians also underpins key algorithms such as quantum approximate optimization~\cite{farhi2014quantum, abbas2023quantum}, HHL for linear equations~\cite{lloyd2010quantum}, and quantum approaches to solving differential equations~\cite{liu2021efficient}. 
% \textcolor{red}{@Pratik: move refs~\cite{lloyd1996universal, whitfield2011simulation, low2017optimal, childs2021theory}}. 

Given a Hamiltonian and the details of a physical quantum computer, there are several choices to be made when translating the simulation to executable operations on the hardware. 
The simplest class of algorithms involves a technique called Trotterization, which approximates a time-evolution operator as a product of short-time evolutions of simpler operators while introducing a bounded error. 
Testing and benchmarking these choices and measuring the impact of hardware and algorithmic errors allows researchers to understand the trade-offs on accuracy, computational efficiency, and scalability between different methods.

% \textbf{Motivation:} 
There is extensive theoretical work on simulating Hamiltonians with quantum computers, especially focusing on resource estimation~\cite{childs2021theory, low2023complexity, Moses_2023, dong2022quantum}. 
The impact of standard hardware noise channels and errors inherent to Trotterization is also well studied at a theoretical level~\cite{ gao2020benchmarking, saller2021benchmarking, yu2024qh9, baniasadi2018new}.
However, existing studies on benchmarking Hamiltonian simulation often focus on specific problems or hardware types, usually considering a narrow range of variables and with terminology that may present challenges for nonspecialists.
In particular, practical resources for fast and standardized benchmarking are lacking in this area (see~\autoref{sec:background}).
% zaborniak2021benchmarking, kobayashi2022parent

To address this gap, we introduce a significant practical advancement in techniques for comparing the performance of quantum circuits that implement Hamiltonian evolution based on Trotterization. 
Using several methods to compute execution fidelity, we systematically evaluate both quantum hardware and algorithmic performance. 
Specifically, we measure the impact of hardware noise on performance and trotterization errors while exploring strategies for scalable benchmarking. 

In addition to the barriers to benchmarking on the algorithmic side, current-generation quantum hardware often comes with custom interfaces that make consistent benchmarking difficult, and which may require specific training for use. 
We aim to make benchmarks user-friendly even for non-experts by leveraging QED-C's application-oriented benchmarking framework~\cite{lubinski2023_10061574,lubinski2023optimization, lubinski2024quantum}. We also utilize a published library of Hamiltonian problem instances, HamLib~\cite{sawaya2023hamlib} which enables the analysis of diverse Hamiltonian models. 
% This work extends conventional benchmarks with an extensive set of test cases.

In this paper, we focus on gate-based quantum computers, since they can simulate arbitrary Hamiltonians, thus offering greater versatility than quantum annealers that are usually restricted to a hardware-specific set of Hamiltonian classes. 
We note also that other than Trotterization, a variety of other algorithms that scale batter asymptotically have also been proposed in the literature.
(Some examples include linear combination of unitaries~\cite{Childs_2012_LCU,Berry_2015}, quantum signal processing~\cite{Low2017}, and qubitization~\cite{Low_2019}.) However, these methods are much more resource-intensive in the number of qubits and quantum gates and are generally out of reach of near-term quantum hardware for the Hamiltonians we consider.
Hence, we restrict the tested algorithmic approach to Trotterization.
Nonetheless, our platform is constructed to be general enough to extend to test other techniques when they become practical.

% \textbf{Contribution:} 
We introduce three distinct methods for calculating the fidelity of Hamiltonian simulations tailored to accurate and scalable benchmarking of quantum computing performance. We test the methods on five specific Hamiltonian models—the Fermi and Bose-Hubbard models, the transverse-field Ising model (TFIM), the Heisenberg model, and the Hamiltonian corresponding to the Max3SAT problem. By varying the number of qubits and other parameters, we evaluate performance across various computational scenarios.~\autoref{fig:hamlib_fidelity_methods} summarizes results from the execution of these benchmark methods across the five Hamiltonians.

The three methods for calculating the fidelity of the Hamiltonian simulation assess distinct aspects of the computation. Reported results from all methods involve the execution of an order-1 Trotterization circuit of five steps initialized with a checkerboard (N\'eel) state. Method 1 compares the output of quantum hardware with a noiseless simulator to assess hardware performance. The second method evaluates algorithmic accuracy by contrasting the same hardware-generated results with those from matrix diagonalization (implemented on a classical computer), revealing inherent algorithm errors. Lastly, the mirror method~\cite{proctor2022measuring, proctor2022establishing} allows for scalable benchmarking by appending the inverse of the circuit at the end of the original circuit to check if the system returns to its initial state, determining fidelity in large-scale simulations where conventional measures of fidelity are not computable.

Additionally, we compare quantum circuit simulation times on CPU and GPU platforms with execution times on quantum hardware. While simulation times increase exponentially with qubit count, quantum hardware exhibits linear or sub-linear scaling. Our quantitative comparison reveals crossover points where quantum hardware outperforms direct classical simulation of the quantum algorithm, which can inform future strategies for the effective use of quantum and classical resources. 

Our benchmarking framework enables consistent evaluation of Hamiltonians beyond the ones presented in this paper, making it relevant for a wide variety of application areas. It also permits hyperparameter tuning while maintaining problem and hardware constancy, providing developers with a means to analyze trade-offs across computing environments.

% \textbf{Paper Structure:} 
This paper is structured as follows. 
Section~\ref{sec:background} provides a background on the QED-C's application-oriented benchmarking suite, the basics of Hamiltonian simulation, and the HamLib library. In Section~\ref{sec:benchmark_ham_sim}, we discuss the three different methods for calculating fidelity and compare them. 
Following this, Section~\ref{sec:benchmarking_hamlib} details how HamLib is incorporated to benchmark the five Hamiltonian models mentioned above.
In Section~\ref{sec:results_analysis}, we shift our focus to practical performance assessments, examining execution runtime performance and exploring methods to enhance the scalability of fidelity calculations across various numbers of qubits. 
Finally, in Section~\ref{sec:future}, we discuss future research directions and conclude with a summary in Section~\ref{sec:summary-and-conclusions}.

% ==============================================

% *******************
\begin{figure}[t!]
    \includegraphics[width=0.94\columnwidth]{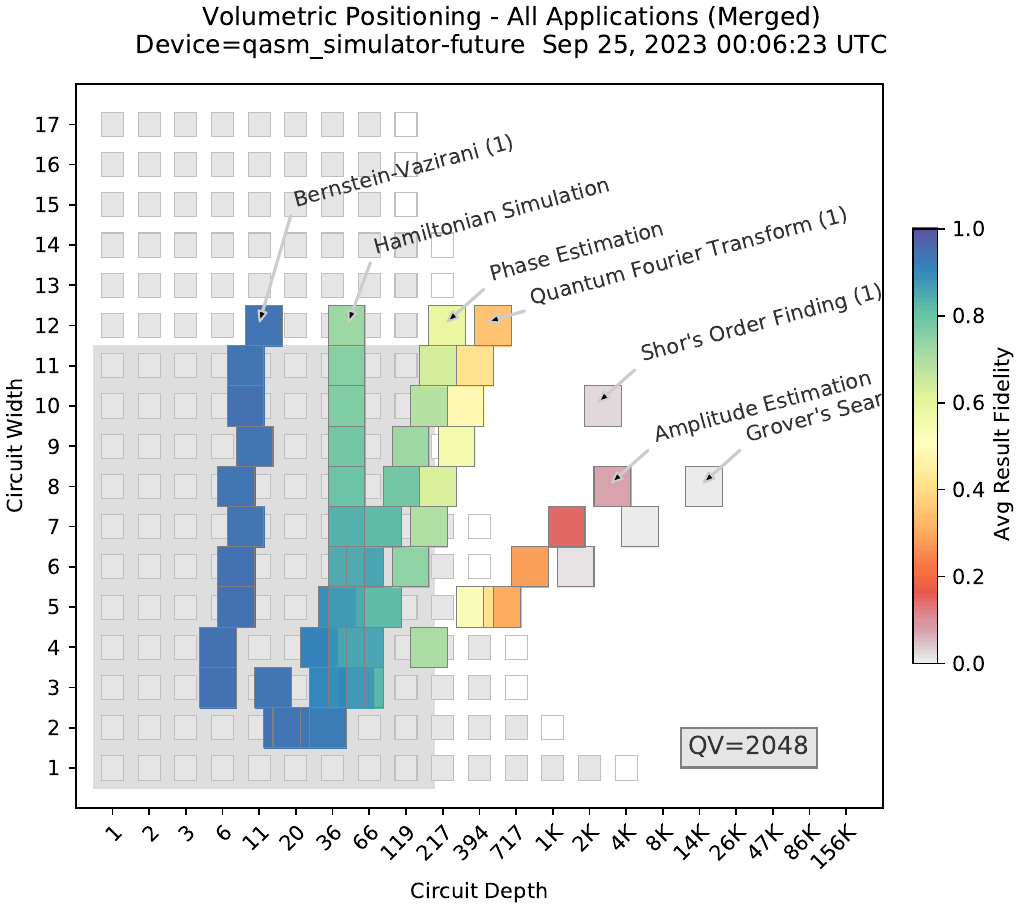}
    \caption{\textbf{QED-C Application-Oriented Benchmarks.} Shown here are results from executing several application-oriented benchmark programs from the QED-C suite on a noisy simulator of a device that reports a quantum volume of $2048$.
    The resulting quality of the application circuits (shown on a color scale) degrades as the circuits' width and depth range outside of the quantum volume region is marked by the dark gray rectangle behind the light-gray rectangles representing volumetric benchmark positions. While the success or failure of its execution can be approximately predicted from the system-level benchmarks such as QV, each application has a unique performance profile. The Hamiltonian simulation benchmarks presented in this manuscript build on the foundation established in the QED-C framework.
    }
    \label{fig:volumetric_application_profiles}
\end{figure}
% *******************

\section{Background}
\label{sec:background}

This section introduces the fundamentals of the QED-C suite of application-oriented benchmarks, followed by an explanation of the Trotterization algorithm for Hamiltonian simulation. It then introduces HamLib: a comprehensive library of Hamiltonians that has been extensively utilized in this research.
  
% -----------------------------------------------

\subsection{The QED-C Application-Oriented Benchmarks}
\label{sec:application_oriented_benchmarks}

Various methodologies for benchmarking the performance of quantum computers and characterizing improvements have become available to the community~\cite{PhysRevA.77.012307, PhysRevLett.106.180504, Blume-Kohout2017-no, Cross_2019, Boixo_2018, wack_clops_2021, parekh2016benchmarking, chen2022veriqbench, li2023qasmbench, cornelissen2021scalable, tomesh2022supermarq}. The QED-C suite of Application-Oriented Performance Benchmarks for Quantum Computing \cite{lubinski2023_10061574,lubinski2023optimization, lubinski2024quantum} adopts a methodology similar to the SPEC benchmarks used for classical computers~\cite{spec_org, hennessy_patterson_2019_all}. It utilizes various algorithms and simple applications designed as benchmarks that cover a range of problem sizes and complexities as shown in~\autoref{fig:volumetric_application_profiles}. This approach enables a comprehensive characterization of overall system performance across various application classes on different quantum computing systems. It captures metrics such as result quality, execution time, and resource usage. This suite supports both single circuit runs and iterative algorithms like QAOA~\cite{farhi2014quantum} and VQE~\cite{peruzzo2014variational}, using the normalized Hellinger fidelity to assess circuit quality, accounting for noise~\cite{lubinski2023_10061574}.

The QED-C approach can be contrasted to previous benchmarks such as Quantum Volume (QV)~\cite{cross2019validating, qiskit_measuring_quantum_volume, baldwin2022re, pelofske2022quantum} which outputs a single number based on a quantum computer passing a performance threshold, and Volumetric Benchmarking (VB)~\cite{proctor2022measuring, blume2020volumetric, proctor2022establishing} which reports performance across different circuit widths and depths. Although these other approaches provide a general assessment of a quantum system's capabilities, they may not predict performance for specific applications. The QED-C suite addresses this by providing well-defined programs that yield application-specific performance metrics and can be adapted to various quantum hardware and simulators~\cite{lubinski2023_10061574, lubinski2023optimization, lubinski2023_10061574}. 

For understanding application performance profiles, the suite employs ``volumetric positioning'' (shown in~\autoref{fig:volumetric_application_profiles}) to visualize the performance of application-specific circuits for different circuit widths and depths, which respectively correspond to a standardized definition of the number of qubits and gates in a circuit. This technique, in which the performance on a benchmark circuit is displayed against a quantum volume rectangle and volumetric background grid, helps validate application-oriented benchmark results with those predicted by system-level benchmarks~\cite{lubinski2023_10061574, lubinski2023optimization, lubinski2023_10061574, farhi2014quantum}. The QED-C suite also reports execution time, which is particularly important for iterative algorithms where run-time overhead accumulates with multiple circuit executions. Execution times are detailed as ``Elapsed'' and ``Quantum''. ``Algorithmic Depth'' and ``Normalized Depth'' provide insights into the impact of tailoring gates to different hardware backends by comparing across systems using circuits defined with a standard gate set~\cite{lubinski2023optimization}. This comprehensive approach ensures that the QED-C framework not only evaluates but also visually correlates the effectiveness of application-specific benchmarks with broader system-level benchmarks, providing a thorough evaluation of quantum computing performance.

We refer the reader to Appendix~\ref{comparing_fidelity_metrics} for an overview of the trade-offs associated with different fidelity computation techniques and the reasoning behind our choice of Hellinger fidelity for these benchmarks.

%----------------------------------------------------------

\subsection{Hamiltonian Simulation}

The Hamiltonian $\hat{H}$ is a Hermitian operator acting on the state space of a quantum system. Physical systems such as atoms and molecules evolve in time according to the Schr\"odinger equation, 
\begin{align}
    i\hbar \pdv{\psi(t)}{t} = \hat{H}(t)\psi(t). \label{eq:schro_eqn}
\end{align}
The eigenstates of the Hamiltonian correspond to the solutions of the time-independent version of the Schr\"odinger equation, $\hat{H}\psi = E\psi$, where $E$ has the physical meaning of the total energy of the system. Computing the state of the system as it evolves over time allows for understanding its dynamics while computing the lower-energy eigenstates allows for understanding its steady-state properties. On a quantum computer, algorithmic approaches for both problems typically involve implementing time evolution under the Hamiltonian.

Outside of physical systems, many other problems can also be encoded as Hamiltonians. For example, many optimization problems can be transformed into the problem of finding the ground state of an Ising model~\cite{Lucas_Karp, nielsen2010quantum, cao2019quantum, kandala2017hardware, kirby2021variational}. In this case, quantum algorithmic approaches may involve implementing approximate evolution under time-varying Hamiltonians, which are linear combinations of the target Hamiltonian and a starting Hamiltonian.

In this paper, we consider time-independent Hamiltonians.
The time-evolution under the Schr\"odinger equation [~\autoref{eq:schro_eqn}] is then given by
\begin{subequations}
\begin{align}
    \psi(t) &= \hat U(t)\ \psi_{\text{init}},\\
    \text{where } \hat U(t) &\coloneqq \exp(-i \hat H t)
\end{align}
\end{subequations}
is the `time-evolution operator', and $\psi_{\text{init}}$ is the initial state.
Hamiltonian simulation thus refers to implementing $\hat U(t)$ as accurately as possible on the target quantum hardware.
Here, to approximate $\hat U(t)$, we use the order-1 Trotterization technique, which breaks down the matrix exponentiation into a product of simpler exponentials:
%This decomposition reduces the overall complexity of the simulation by approximating the full Hamiltonian evolution as a sequence of smaller, unitary operations over short time intervals as follows:
\begin{subequations}
    \begin{align}
    \exp(-i\hat H t) &= \exp(-i \sum_j \hat H_j t) \\
    &= \bigg(\prod_j \exp(-i \hat H_j t/K) \bigg)^K  + O(t^2/K). \label{eq:trotter_op}
    \end{align}
\end{subequations}
Here, $K$ denotes the number of Trotter steps, and $\hat{H}=\sum_j \hat{H}_j$. This technique makes quantum simulations feasible if time evolution under the terms $H_j$ in the summand can be implemented by known techniques on the quantum computer. The algorithmic error is controlled by $K$, going to 0 as $K\to \infty$. However, a larger $K$ implies a longer execution time and larger circuit depth, leading to larger hardware errors. Thus, it is important to choose an optimal value of $K$, which balances algorithmic and hardware errors. Although Trotter errors have been extensively analyzed at a theoretical level~\cite{wiebe2010higher, jones2019optimising, avtandilyan2024optimal}, our benchmarking framework allows end-to-end analysis of both hardware and Trotter errors on real devices.

Next, the degrees of freedom in the Hamiltonian are mapped to qubits, and each of the exponentials in~\autoref{eq:trotter_op} is represented by a series of qubit gates. Together, these constitute a quantum circuit that is compatible with any gate-based quantum computer.

Note that for execution on real devices, there are further processing steps that convert the quantum circuit to hardware-executable quantum gates. These steps include mapping algorithmic qubits to physical qubits, scheduling gate operations, and applying error mitigation or correction techniques. Each of these steps offers opportunities for optimization, the result of which can be tested in our framework.

%----------------------------------------------------------

\subsection{HamLib: A library of Hamiltonians}

HamLib~\cite{sawaya2023hamlib} is a comprehensive dataset of quantum Hamiltonians, encompassing problem sizes ranging from 2 to 1000 qubits. The Hamiltonians are organized into several high-level categories. The first category is binary-variable optimization and related problems, including Max-K-SAT, Max-Cut, and QMaxCut. The second category covers discrete-variable optimization problems, such as Max-K-Cut and the traveling salesperson problem. Another major category includes condensed matter physics models such as the transverse-field Ising model, the Heisenberg model, the Fermi-Hubbard model, and the Bose-Hubbard model. Lastly, the dataset also includes chemistry Hamiltonians that use curated or calculated real-world parameters, with subcategories encompassing electronic structure and vibrational structure.

A useful feature of HamLib is that all problem instances have already been mapped to qubits, i.e., they are mapped to a Pauli representation of the form
$
\hat{H}_{\text{encoded}} = \sum_{i} c_i \bigotimes_k \sigma_{ik}
$,
where $\sigma_{ik}$ is a one-qubit Pauli or identity operator, i.e., $\sigma_{ik} \in \{I, X, Y, Z\}$, and $c_i$ is a real number. We select multiple Hamiltonians from this dataset, construct circuits based on them, and benchmark their performance. By systematically varying parameters within each problem, our framework evaluates how Hamiltonian simulation performance changes as the problem instance changes.

%----------------------------------------------------------

% =========================================================

\section{Benchmarking Hamiltonian Simulations}
\label{sec:benchmark_ham_sim}

% *******************
\begin{figure}[t]
    \includegraphics[width=1\columnwidth]{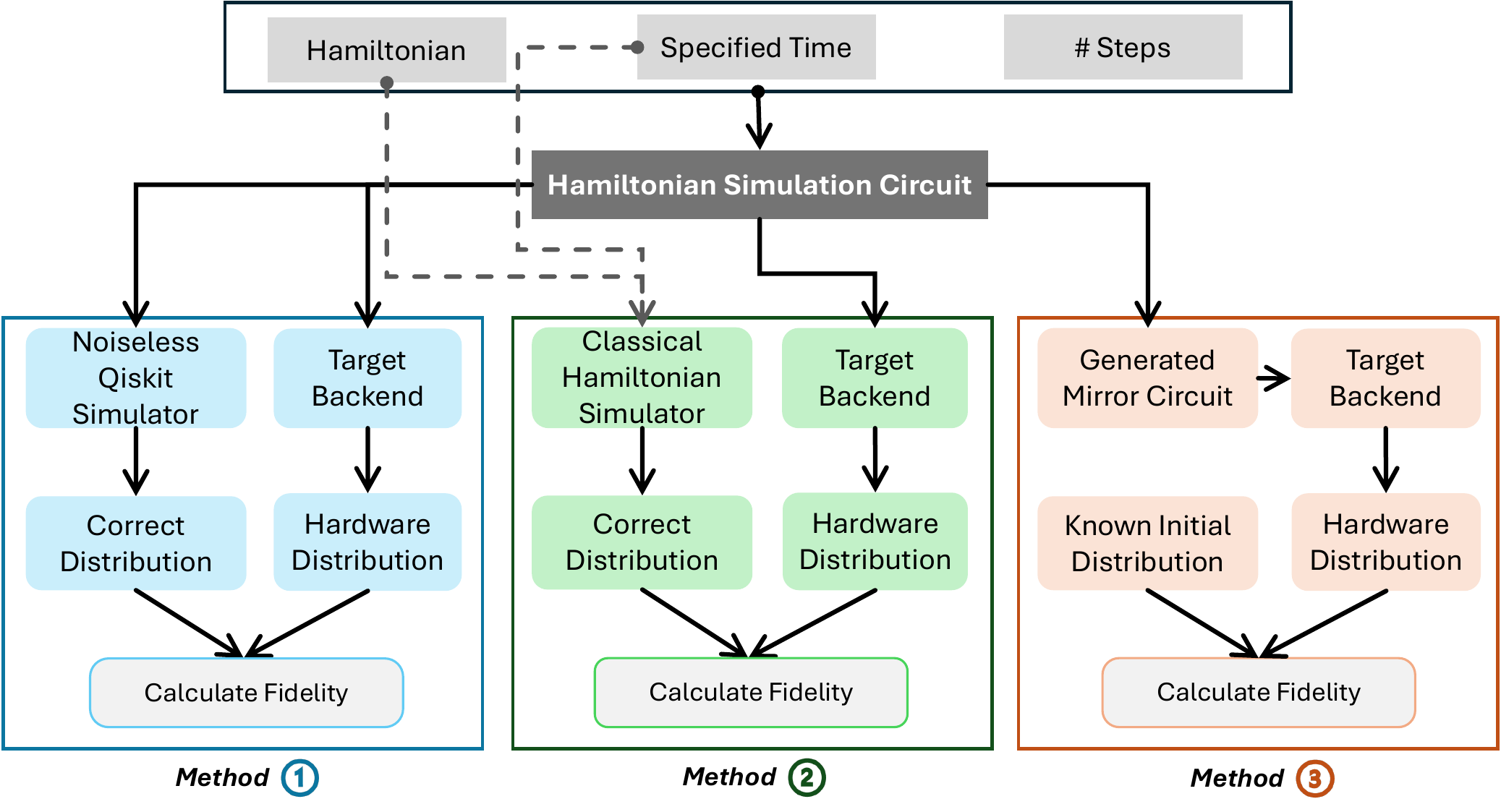}
    \caption{\textbf{Flowcharts of Quantum Circuit Evaluation Methods.} This figure presents three flowcharts corresponding to the evaluation methods detailed in the paper. Method 1 outlines the process for assessing hardware performance by comparing the shot distribution from quantum hardware against a noiseless simulator, measuring the impact of hardware noise. Method 2 depicts the evaluation of algorithmic performance, comparing hardware results with those predicted by matrix diagonalization, and emphasizes the impact of both Trotterization error and hardware noise. Method 3 illustrates scalable evaluation using the mirror method, showcasing steps to verify initial state restoration, which is vital for understanding the performance of quantum circuits in practical implementations at scale.
    }
    \label{fig:method_1_2_3_flowchart}
\end{figure}
% *******************

We employ three approaches for calculating the fidelity of Hamiltonian simulations, as shown in~\autoref{tab:method_compare}. Each method involves the construction of an order-1 Trotterization circuit tailored to a particular Hamiltonian. To constrain hardware errors arising from increased circuit depth, we choose $t=1$ and $K=5$ number of Trotterization steps as default parameter values. 
% Consequently, each circuit step approximates a time advance of $0.2$. 
For each method, we use the polarization fidelity~\cite{proctor2022measuring,lubinski2023_10061574}, a normalized version of the Hellinger fidelity, which compares the similarity between the two probability distributions. 

Although the upper limit of Trotterization error is theoretically well established~\cite{childs2021theory}, the introduction of hardware errors complicates the total error profile~\cite{Endo_2019}. Each method provides a unique measure of the impact of various error sources on the performance of the algorithm.~\autoref{fig:method_1_2_3_flowchart} presents a flowchart that outlines the procedures of these methods, which together offer a structured framework to evaluate performance.~\autoref{tab:method_compare} summarizes the four method variants, the error types addressed by each, and their scalability, emphasizing the limitations and capabilities of each approach.

% *******************
\begin{table}[]
\centering
\caption{Comparison of fidelity measurements in each method of the Hamiltonian simulation benchmarking framework.}
\begin{tabular}{c||c||c}
\textbf{Method} & \textbf{Error Represented} & \textbf{Scalability} \\ \hline \hline
1               & Hardware                   & No                   \\
2               & Hardware + Trotterization  & No                   \\
2 (noiseless)   & Trotterization             & No                   \\
3               & Hardware                   & Yes                  \\ \hline \hline
\end{tabular}
\label{tab:method_compare}
\end{table}
% *******************

%--------------------------------------------------

\subsection{Method 1: Hardware Performance}
\label{subsec:method_1}

The first method involves executing the Trotterized Hamiltonian simulation circuit which represents the operator in~\autoref{eq:trotter_op} on an ideal noiseless simulator starting from an initial state $\psi_{\text{init}}$. Measuring in the computational $Z$ basis gives a probability distribution corresponding to the ideal output of the circuit. The same Trotterized circuit is then executed on a potentially noisy target backend to derive a test distribution. The test output is then compared with the correct distribution to calculate the fidelity after execution. As both the correct and noisy test distributions originate from the same Trotterized circuit, the computed fidelity exclusively reflects the influence of hardware noise.

This method is effective for assessing hardware performance. If the hardware behaves like an ideal noiseless simulator, the fidelity deviates from 1.0 only due to sampling noise. Any additional deviations due to hardware noise further reduce the fidelity. Despite its practicality in evaluating hardware, this method relies on a noiseless circuit simulator to generate the correct distribution, which poses significant limitations due to the exponentially scaling requirements of the simulation. As the number of qubits increases, the time to measure the benchmark rises exponentially. While this approach is not feasible for systems with a large number of qubits, a study of the output for limited-size systems is still capable of providing insights into the impact of hardware choice on the simulation of different classes of Hamiltonians.

%----------------------------------------------

\subsection{Method 2: Algorithmic Performance}
\label{subsec:method_2}

The second method generates a `correct' distribution through classical matrix diagonalization~\cite{childs2018toward}, capturing state evolution up to classical numerical precision errors. The Trotterized version of the same evolution is then executed on the potentially noisy backend, producing a test distribution. Comparing these distributions yields a fidelity measure reflecting both Trotterization and hardware-induced errors. The upper bound of this method's fidelity is set by Trotterization error, which is dependent on the degree to which the Hamiltonian's constituent terms commute and the size of the Trotter step. Comparing the classical distribution with one from a noiseless simulator isolates Trotterization error and is referred to as `Method 2 (noiseless)' in subsequent sections.

This method effectively illustrates the combined impact of hardware noise and Trotterization errors, thereby providing a comprehensive measure of the implementation’s overall performance. However, similar to Method 1, this approach is constrained by the exponentially scaling simulation requirements, limiting its applicability to smaller system sizes. 

%----------------------------------------------

\subsection{Method 3: A Scalable Performance Metric}
\label{subsec:method_3}

The third method utilizes a benchmarking technique developed by Proctor et al.~\cite{proctor2022measuring, proctor2022establishing} known as ``mirror circuits''. We have implemented multiple variants of this method with different levels of robustness. The most straightforward of these (`Method 3 Simple') involves taking the Trotterized Hamiltonian simulation circuit defined above and appending its inverse, constructing a composite \emph{simple mirror circuit} \cite{proctor2022measuring} that includes both the original and its inverse. This method, also known as the inverse Hamiltonian approach, ensures that the qubits return to their initial state. Referring to the operator corresponding to the Trotterized Hamiltonian simulation circuit as $C_H$, the method effectively achieves:
$C_H^{-1} C_H \psi_{init} = \psi_{init}$.  Critical qubit gate identities used to construct the inverse circuit include $R_P(\theta) R_P(-\theta) = I$, $(CX) (CX) = I$, $H_\text{Hadamard} H_\text{Hadamard} = I$, where \( I \) denotes the identity matrix, $CX$ is the controlled X operator, and $R_P(\theta)$ is rotation around the $P$ axis.

Method 3 utilizes the initial state $\psi_{init}$ to derive the expected distribution directly, thus eliminating the need for classical simulation. To compute the fidelity measure, we execute the simple mirror circuit $C_H^{-1}C_H\psi_{init}$ on a potentially noisy backend system. The resulting distribution is then compared with the distribution computed from $\psi_{init}$. This approach, like Method 1, evaluates hardware performance but offers enhanced scalability to larger qubit systems without requiring classical simulation for the correct distribution. However, this advantage comes at the cost of approximately doubling the circuit depth.

We take the square root of the measured fidelity to address the inherent fidelity degradation caused by the doubled circuit depth. This rescaling aligns the mirror method's measurements with the standard circuit's fidelity, making them comparable to Method 1 results. To illustrate why the fidelity of a double-depth circuit (such as the mirror circuit) is the square of the fidelity of a single-depth circuit, consider the following reasoning.

Let the fidelity of a single gate be denoted by \( F \) and assume global depolarizing errors so that the fidelity of the product of $n$ gates is approximately $F^n$ \footnote{The error in this approximation decays as $O(1/2^N)$ for $N$ qubits, and it can be removed entirely by instead rescaling $F$ to ``process polarization''~\cite{proctor2022measuring}.}. For simplicity, assume that the fidelity of a gate is the same as the fidelity of its inverse, i.e., the gate's inverse introduces the same amount of error as the original gate. In a mirror circuit, the first half of the circuit consists of a series of gates, each with fidelity \( F \), followed by the inverse sequence of these gates in the second half. If the first half of the circuit has \( n \) gates, the fidelity of the entire first half is given by: $F_{\text{half}} = F^n$. The second half of the mirror circuit applies the inverse of these \( n \) gates, each with the same fidelity \( F \), resulting in:
\begin{align}
    F_{\text{total}} = F_{\text{half}} \times F_{\text{half}} = (F^n) \times (F^n) = F^{2n}
\end{align}

Under these assumptions, the fidelity of the entire double-depth mirror circuit is equal to the square of the fidelity of the single-depth base circuit. Consequently, the square root of the double-depth circuit's fidelity can be taken to represent the fidelity of the base circuit. Throughout this manuscript, we utilize the square root of the Method 3 fidelity results as our default representation of Method 3's fidelity \footnote{Note that the above analysis is simplified for our purposes. A more accurate computation rescales the fidelity to ``process polarization'' before taking the square root and rescaling it back~\cite{proctor2022measuring}.}.

The primary goal of this comparison is to use Method 3's fidelity measurements as indicators of hardware performance, similar to Method 1, but with enhanced scalability. This enables predictions of hardware performance in scenarios where Method 1 may be constrained. It's important to note that while serving similar purposes, these methods measure different, complementary performance metrics and are not expected to yield identical results.

% \vspace{0.3cm}

% *******************
\begin{table}[t!]
\centering
\caption{Variants of the third method and their formal Hamiltonian expressions.}
\begin{tabular}{c||c}
\textbf{Variants of Method 3} & \textbf{Formal Hamiltonian Expression}                                                                \\ \hline \hline
Simple                    & $C_H^{-1} C_H \psi_{init} = \psi_{init}$                                                                                       \\
Random Pauli               & $\tilde C_{H}^{-1} P_{\text{random}} C_H \psi_{init} = P_{\text{resultant}} \psi_{init}$ \\
Multiple Random Paulis         & $\underbrace{\tilde{C}_H^{-1}P_{\text{random}}C_H \psi_{init} = P_{\text{resultant}} \psi_{init}}_{\text{Repeat } N \text{ times}}$                                                                                       \\ \hline \hline
\end{tabular}
\label{tab:method3_variations}
\end{table}
% *******************

The Method 3 mirror circuits discussed above have several limitations. First, the fidelity computation does not account for the effects of quantum state preparation and measurement (SPAM) errors. Second, they may exhibit varying sensitivity to some coherent gate errors and be over-sensitive to others since coherent errors can add and cancel between the two halves of a simple mirror circuit~\cite{proctor2022measuring}. To address these issues, an alternative type of mirror circuit can be employed to more reliably measure the \emph{process fidelity} with which a quantum circuit is implemented~\cite{proctor2022measuring, proctor2022establishing}. This alternative approach mitigates the cancellation of coherent errors, enhancing its effectiveness in detecting and quantifying underlying errors and noise.

Building on this alternative approach, we developed two variants of the Method 3 mirror circuits, as outlined in~\autoref{tab:method3_variations}, to partially address the limitations of our simple Method 3 implementation. Specifically, we insert random Pauli gates in between the base circuit and its inverse. We then modify the inverse circuit to efficiently commute the Pauli gates through it, ensuring that the Pauli gates are properly propagated and accounted for throughout the entire mirror circuit. The measurement distribution expected from this modified circuit can be computed simply by applying a set of corresponding ``resultant'' Pauli operations to the initial state.

Formally, the modified Method 3 construction substitutes the inverse Hamiltonian simulation circuit (\(C_H^{-1}\)) used in a simple mirror circuit with a `quasi-inverse' (\(\tilde{C}_H^{-1}\))~\cite{proctor2022measuring}. This type of mirror circuit consists of the Hamiltonian simulation circuit, a layer of random Pauli gates (\(P_{\text{random}}\)), followed by the quasi-inverse circuit \(\tilde{C}_H^{-1}\). The overall effect is to apply a Pauli Operator (\(P_{\text{resultant}}\)), efficient to compute classically, to the initial state (\(\psi_{init}\)), expressed formally as $\tilde{C}_H^{-1} P_{\text{random}} C_H \psi_{init} = P_{\text{resultant}} \psi_{init}$. The addition of random Pauli gates disrupts the circuit symmetry, preventing systematic addition or cancellation of coherent errors and enabling the estimation of the process fidelity of $H$ from the performance of the mirror circuit.

For the Method 3 variant labeled `Method 3 Random Pauli', a fixed set of randomly generated Pauli gates is inserted into the circuit, with the same set used consistently across repeated executions. In contrast, the `Method 3 Multiple Random Paulis' variant generates a unique set of random Pauli gates for each of $N$ executions. By restructuring the circuit in this way, these variants aim to provide a more accurate depiction of errors within the Hamiltonian simulation circuit. This allows for a detailed analysis of how noise and errors influence quantum operations in practical scenarios. 

Method 1 and Method 3 evaluate complementary performance metrics. Method 3 approximates process fidelity, a standard metric of quantum operation quality. Method 1 quantifies the difference between ideal and actual probability distributions. Despite measuring different aspects, these methods often yield comparable performance metrics. Method 3 offers superior scalability, making it particularly valuable for larger quantum systems.

A comprehensive comparison between Method 1 and all variants of Method 3 is provided in~\autoref{sec:results_analysis}. For further details on the key differences between the various fidelity computation methods, refer to Appendix~\ref{comparing_fidelity_metrics}.

%----------------------------------------------

\subsection{Comparing the Methods}
\label{subsec:compare_methods}

% *******************
\begin{figure}[]
    \includegraphics[width=0.49\columnwidth]{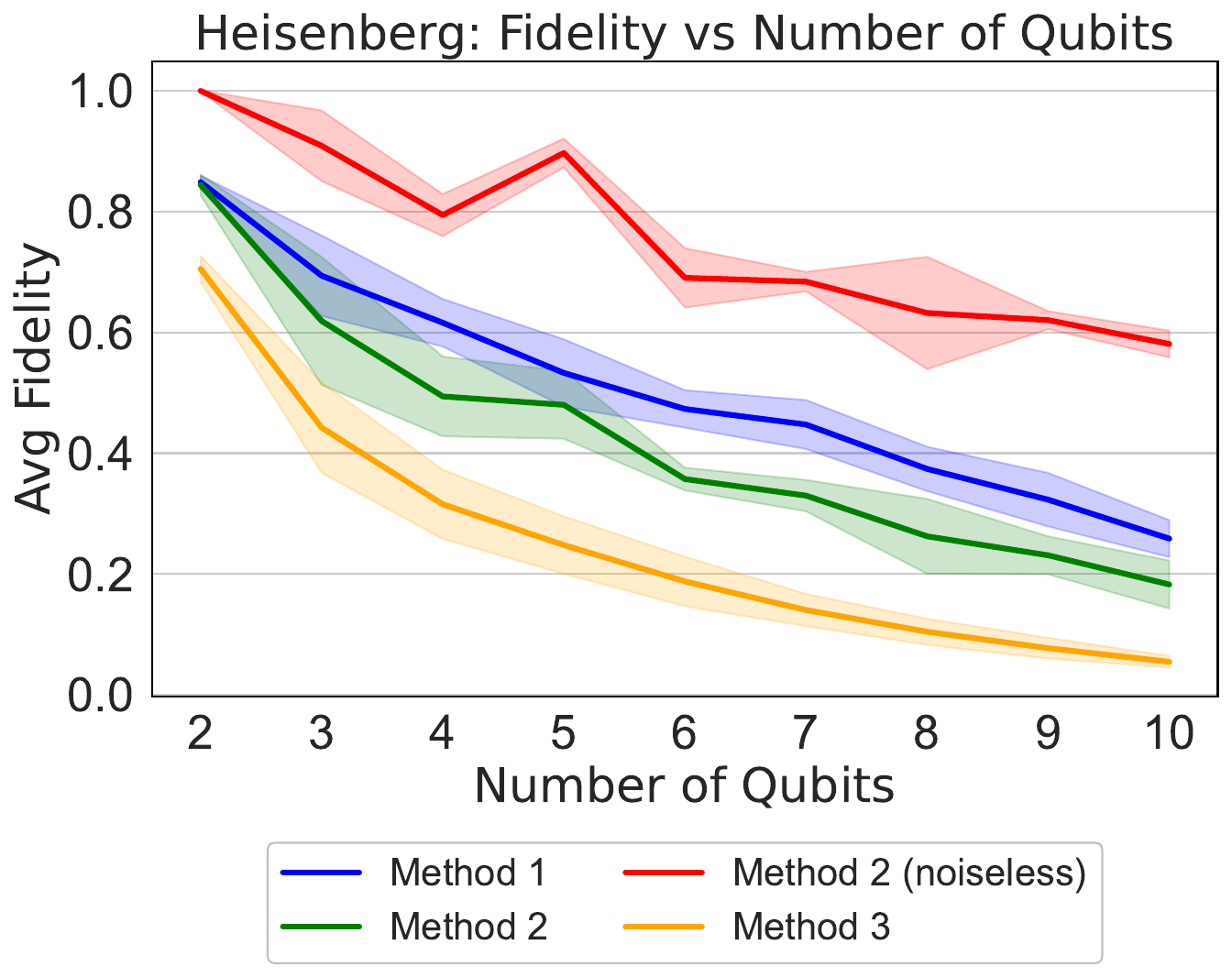}
    \includegraphics[width=0.49\columnwidth]{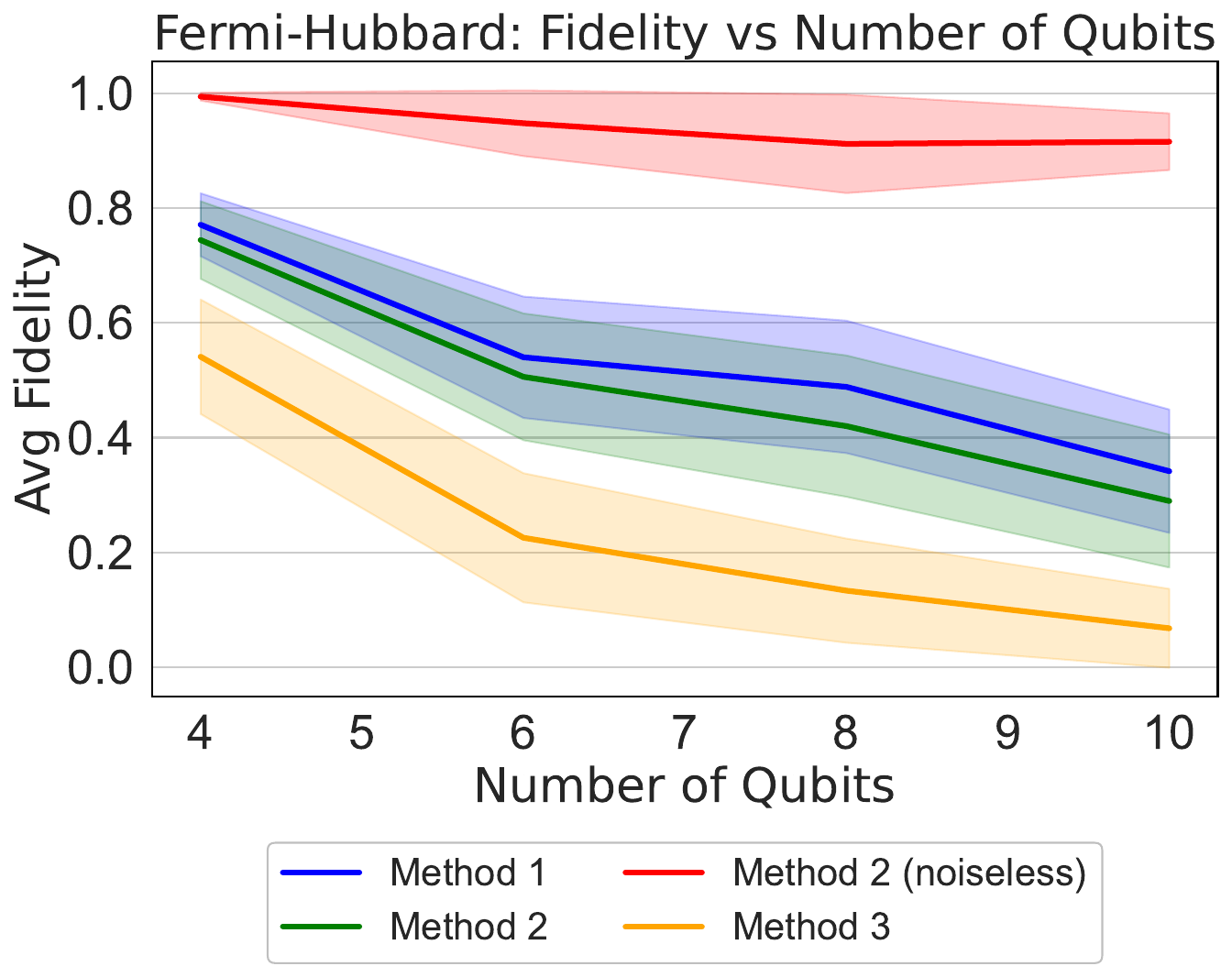}
    \caption{\textbf{Fidelity Comparisons Across Methods.} This figure shows the fidelity metrics for the Heisenberg and Fermi-Hubbard models as a function of qubit count. Each graph features four lines representing different fidelity calculation methods. Method 1 focuses on hardware noise alone. Method 2 includes both hardware noise and Trotterization errors, usually lowering fidelity compared to Method 1. Method 2 (noiseless) excludes hardware noise, focusing only on Trotterization errors, and achieves the highest fidelity. Method 3, with mirrored circuits and doubled circuit depth, records the lowest fidelity in its un-normalized form. In the remainder of this document, we take its square root to make it comparable to Methods 1 and 2.
    }
    \label{fig:method_1_2_3_fidelity}
\end{figure}
% *******************

To illustrate how these methods relate to one another and the insight each one provides, we discuss a specific example here.~\autoref{fig:method_1_2_3_fidelity} shows the fidelity metrics for the Heisenberg model and the Fermi-Hubbard model as a function of qubit count. This simulation was performed within the QED-C benchmark framework, executing $1000$ shots using the Aer simulator with our default noise model that has been measured to mimic a quantum device with a quantum volume of 2048. Unless otherwise noted, all results presented in this paper are run on this simulator with these parameters. 

Each plot includes four distinct lines representing a different fidelity calculation method. Surrounding each line is variability shading, which encapsulates the range of data points associated with that specific parameter. For instance, in the fidelity analysis of the Heisenberg model, various influencing factors, such as the calculation method, the inclusion of periodic boundary conditions, and the intensity of magnetic field interactions, are considered. The shaded region surrounding each line in these plots illustrates the variability of fidelity across all other parameters while focusing on a single parameter, in this case - the method, to isolate its direct impact on fidelity. This visualization approach is consistently applied throughout the paper to highlight parameter-specific effects.

Method 1 isolates the impact of hardware noise on the model. Conversely, Method 2 accounts for both hardware noise and Trotterization errors, typically resulting in lower fidelity compared to Method 1. Method 2 (noiseless) represents a variant of Method 2 but excludes hardware noise, focusing solely on Trotterization errors, and thus displays the highest fidelity among the methods. Theoretically, combining the results of Method 1 and Method 2 (noiseless) should mirror the results of Method 2. The `simple' variant of Method 3, similar to Method 1, assesses the impact of hardware noise but employs mirrored circuits, effectively doubling the circuit depth and generally resulting in the lowest fidelity. Given simplifying assumptions, the square root of the fidelity values from Method 3 should align with those of Method 1. However, this may not consistently hold in practice (a comprehensive comparison of Method 1 and all variants of Method 3 is detailed in~\autoref{sec:results_analysis}). As depicted in~\autoref{fig:method_1_2_3_fidelity}, these observations hold for both the Heisenberg and Fermi-Hubbard models. As we explore Hamiltonian simulations in the next section, these four methods are central to our analyses.

% ==================================================

\section{Incorporating HamLib for Benchmarking}
\label{sec:benchmarking_hamlib}

In this section, we present results from our analysis of five key models/problems taken from HamLib: the Transverse-Field Ising Model (TFIM), the quantum Heisenberg model, the Fermi-Hubbard model (FH), the Bose-Hubbard model (BH), and the Max3SAT problem. 
The parameter values associated with each problem can affect the accuracy of simulating each of these models.
Hence, we varied both the number of qubits and Hamiltonian parameters across these models. Our benchmarking approach utilized the three fidelity methods previously described. Method 3 required normalization due to its doubled circuit depth, utilizing the square root adjustment (`Method 3 sqrt').

% *******************
\begin{figure}[t!]
    \includegraphics[width=0.49\columnwidth]{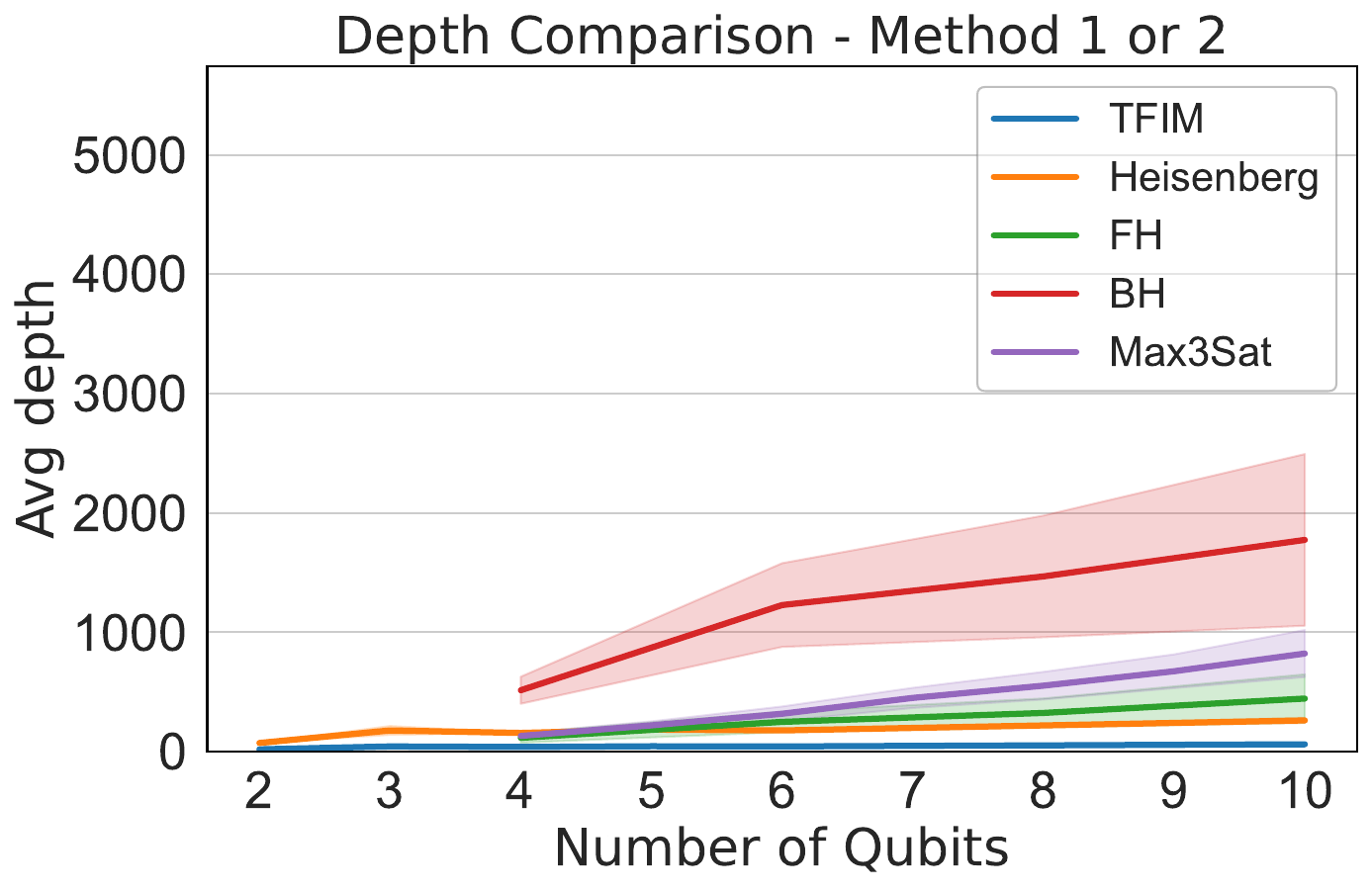}
    \includegraphics[width=0.49\columnwidth]{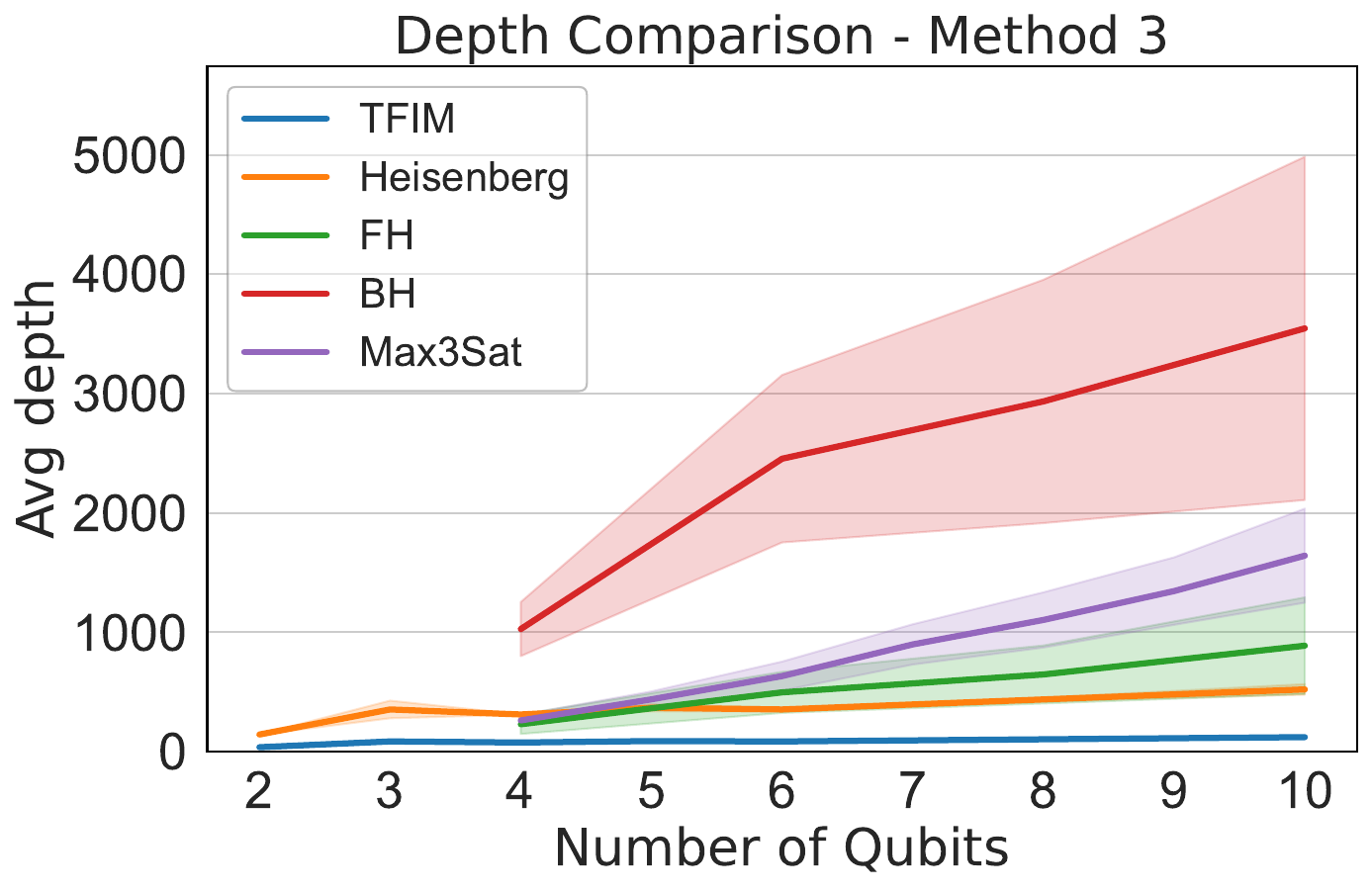}
    \caption{\textbf{Depth Comparison across Five Hamiltonians from HamLib using Three Benchmark Methods.} The left plot shows the normalized circuit depth for Methods 1 and 2, which have similar depths. The right plot shows Method 3, where the depth is doubled due to the addition of the inverse circuit. The Bose-Hubbard (BH) Hamiltonian exhibits the highest depth across all methods, followed by Max3SAT, Fermi-Hubbard (FH), Heisenberg, and TFIM, which has the lowest depth. As expected, the depth increases with the number of qubits for all Hamiltonians.
    }
    \label{fig:all_model_all_methods_depth}
\end{figure}
% *******************

Circuit depth is a critical factor in determining the quality of quantum algorithm execution on a quantum computer. To provide context for the analysis in this section,~\autoref{fig:all_model_all_methods_depth} summarizes the results of the three benchmark methods across the five Hamiltonians. We compare the average circuit depths for Methods 1 and 2 with those for Method 3 (un-normalized), which exhibits significantly increased depth due to its inverse circuit addition. The depth for all the Hamiltonians increases with qubit count, with Bose-Hubbard having the highest depth, followed by Max3SAT, Fermi-Hubbard, Heisenberg, and TFIM. For a detailed depth and complexity analysis across all Hamiltonians and their parameters, refer to~\autoref{fig:new_depth_all_models_all_params} in Appendix~\ref{apdx:subsec:detail_depth_analysis}.

% ---------------------------------------------------------------------

%\subsection{The Quantum Heisenberg and the Transverse-Field Ising Model}
\subsection{Heisenberg and Transverse-Field Ising Models}
\label{subsec:hamlib_Heis_TFIM}

% From HamLib, we implement the following Hamiltonian for the quantum Heisenberg model,
The Hamiltonian for the one-dimensional nearest-neighbor (anti-ferromagnetic) quantum Heisenberg model is given by
\[
H_{\text{Heis}} = \sum_{i=1}^{N-1} \vec{\sigma}_i \cdot \vec{\sigma}_{i+1} + \sum_{i=1}^N h Z_i,
\]
where \(\vec{\sigma}_i = (X_i, Y_i, Z_i)\).
This Hamiltonian is also known as the Heisenberg XXX model, with the additional $Z_i$ terms denoting a uniform external magnetic field applied in the $\hat z$ direction.
An exact analytical solution for this model can be obtained using the Bethe ansatz~\cite{franchini2017introduction, granet2019analytical}. 
We simulated the model with periodic as well as open (or non-periodic) boundary conditions while varying \( h \) across the set \{0, 0.1, 0.5, 1, 2, 3, 5\} to explore different magnetic field strengths.

% *******************
\begin{figure}[]
    \includegraphics[width=0.49\columnwidth]{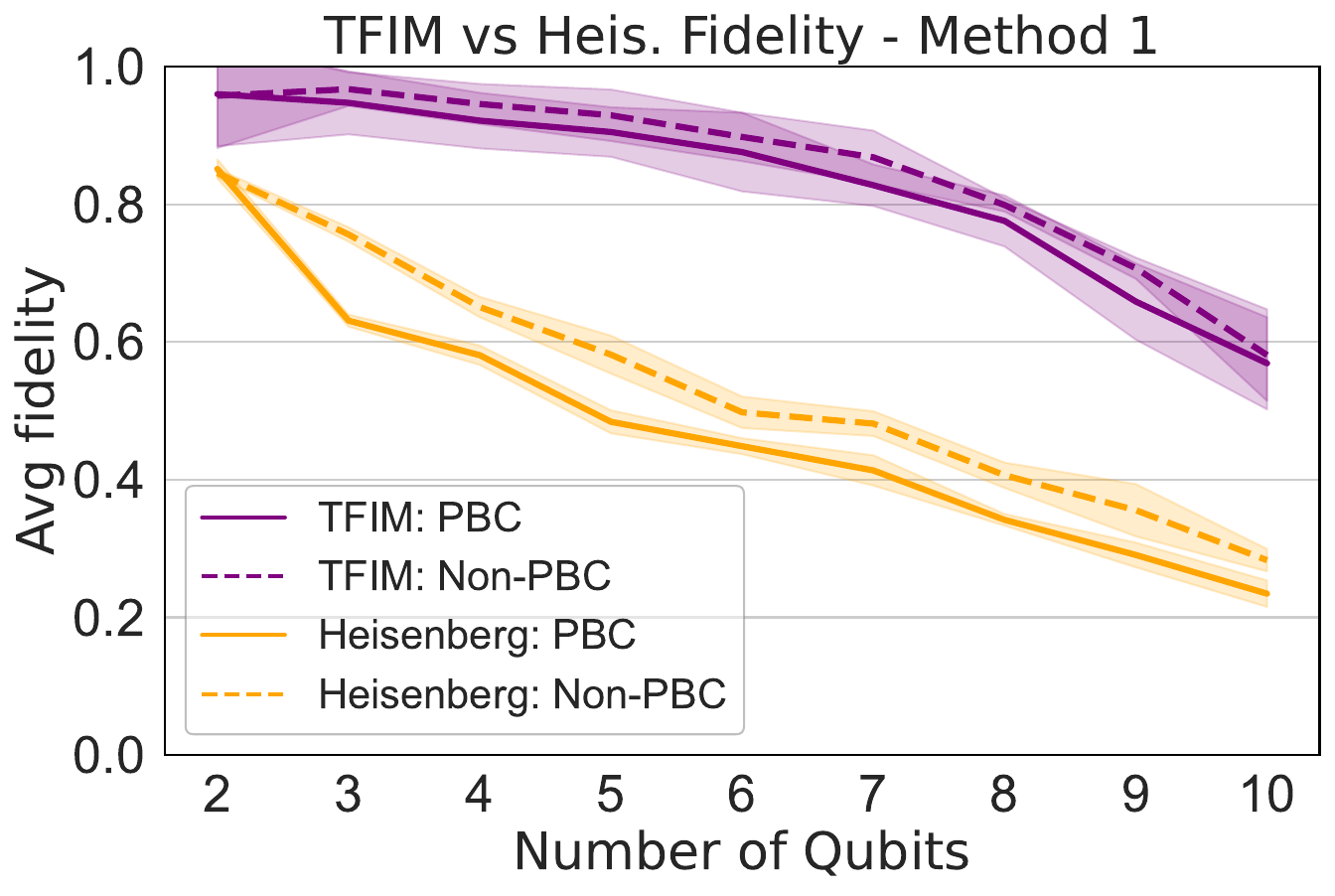}
    \includegraphics[width=0.49\columnwidth]{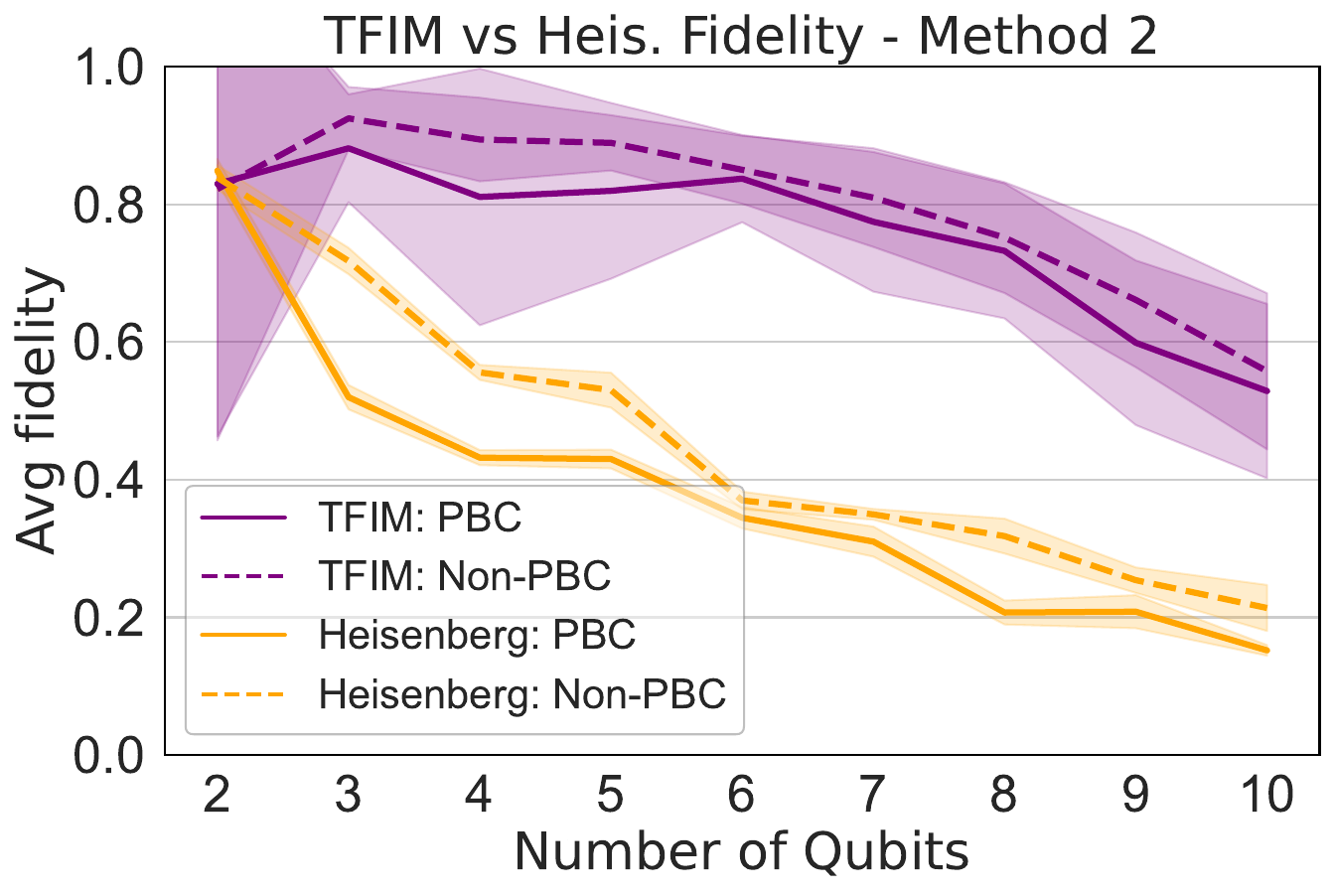}
    \includegraphics[width=0.49\columnwidth]{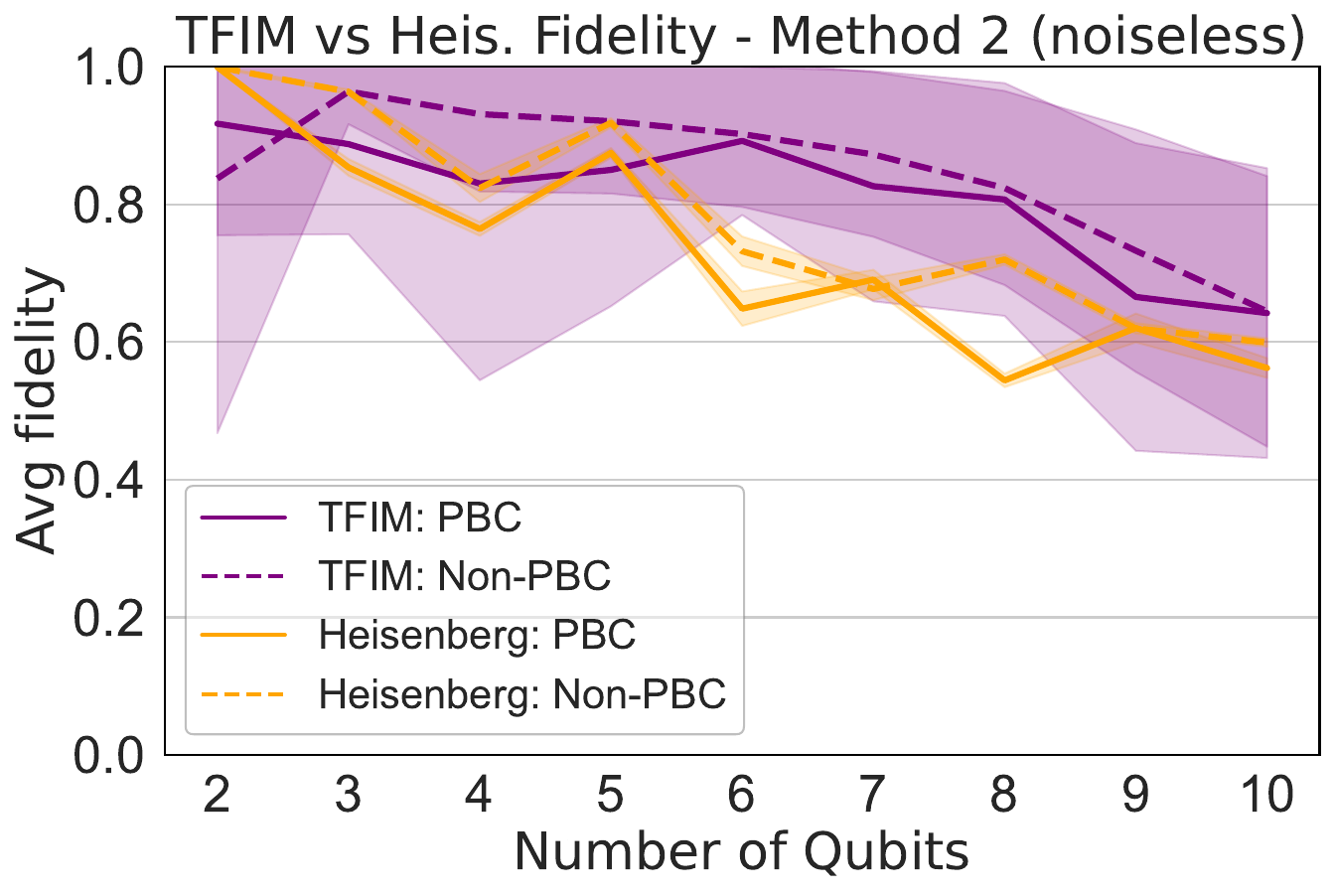}
    \includegraphics[width=0.49\columnwidth]{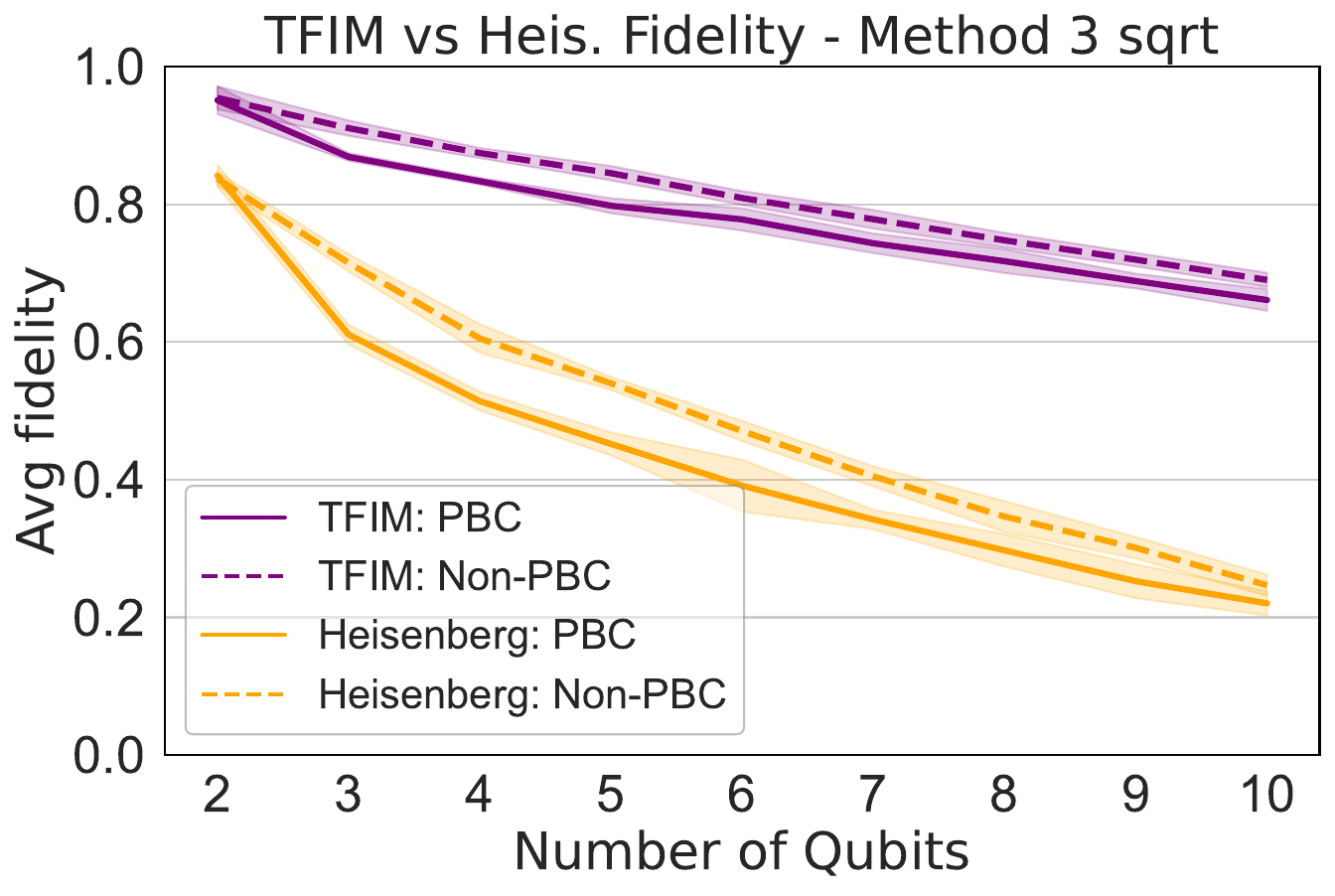}
    \caption{\textbf{Fidelity Comparisons in TFIM and Heisenberg Models Under Different Boundary Conditions.} This figure illustrates the fidelity of the Transverse Field Ising Model (TFIM) and the Heisenberg model as affected by varying periodic boundary conditions (PBC and Non-PBC) across four distinct fidelity calculation methods. Generally, the TFIM model demonstrates a higher fidelity than the Heisenberg model under both boundary conditions. Consistent across all methods, the normalized Method 3 yields a comparable fidelity, while Method 2 (noiseless) achieves the highest fidelity, illustrating the differential impact of calculation methodologies on model performance.
    }    \label{fig:heis_tfim_all_methods_all_pbc_fidelity}
    \vspace{-10pt}
\end{figure}
% *******************

% *******************
% \begin{figure}[t!]
%     \includegraphics[width=1\columnwidth]{figures/hamlib_heis_tfim_basic_simulation/heis_tfim_all_methods_depth.pdf}
%     \caption{\textbf{Circuit Depth Analysis in TFIM and Heisenberg Models.} This figure examines the circuit depth for the Transverse Field Ising Model (TFIM) and the Heisenberg model using three methods. Methods 1 and 2 have identical circuit depths, which are notably lower for the TFIM compared to the Heisenberg model. In addition, periodic boundary conditions increase the circuit depth for both models. Method 3, applied to both models, exhibits a circuit depth that is twice that of Methods 1 or 2, reflecting the increased computational complexity introduced by mirrored circuit designs. This layout allows for a clear comparison of how each method and boundary condition impacts the physical requirements of quantum simulations.
%     }
%     \label{fig:heis_tfim_all_methods_depth}
% \end{figure}
% *******************

The transverse-field Ising model (TFIM) represents the simplest form among the condensed matter models we consider. 
Notably, in one dimension, it is classically tractable, even with disorder. The Hamiltonian for this model is expressed as
\[
H = \sum_i h_i X_i + \sum_{\langle i, j \rangle} Z_i Z_j,
\]
where the summation extends over each edge \(\langle i, j \rangle\) within the lattice. Quantum critical points for the TFIM are observed at \(h \approx 3\) in two-dimensional models \cite{kallin2013entanglement} and at \(h \approx 5.16\) in three-dimensional implementations \cite{tepaske2021three}. For our analysis, we explore the effects at and around these critical points by employing one-dimensional Hamiltonians with field strengths \(h\) spanning from 0 to 6, inclusive, at fine increments \(\{0, 0.1, 0.5, 1, 2, 3, 4, 5, 6\}\) while investigating both periodic and non-periodic boundary conditions \cite{blote2002cluster}.

To gain deeper insights into these two models, we examine~\autoref{fig:heis_tfim_all_methods_all_pbc_fidelity} that displays the fidelity of both the Transverse Field Ising Model (TFIM) and the Heisenberg model under varying periodic boundary conditions (PBC and Non-PBC) using the four different fidelity calculation methods previously described.~\autoref{fig:new_depth_all_models_all_params} in Appendix~\ref{apdx:subsec:detail_depth_analysis} explores the circuit depth for Methods 1 and 2, which are identical. We note that the magnetic field strength, or \(h\), only affects single qubit gate angles and, therefore, has no impact on the circuit depth or fidelity.

The TFIM implementation shows higher fidelity than the Heisenberg model, primarily due to its lower circuit depth. 
TFIM's simpler interactions, focusing on single-axis spin interactions and a transverse magnetic field, require fewer quantum gates. 
In contrast, the Heisenberg model's interactions along all three axes (X, Y, Z) necessitate a more complex gate array, increasing circuit depth \cite{dutta2010quantum, biamonte2008realizable, white2004real}. Periodic boundary conditions further increase the depth and reduce fidelity in both models compared to their Non-PBC variants.
Fidelity trends across methods remain consistent: Method 3 `sqrt' shows comparable but slightly lower fidelity than Method 1, while Method 2 (noiseless) consistently demonstrates the highest fidelity, representing Trotterization error in Hamiltonian evolution.

%-----------------------------------------------------

\subsection{Fermi and Bose-Hubbard Models}
\label{sec:hamlib_FH_BH}

% *******************
\begin{figure}[t!]
    \includegraphics[width=0.49\columnwidth]{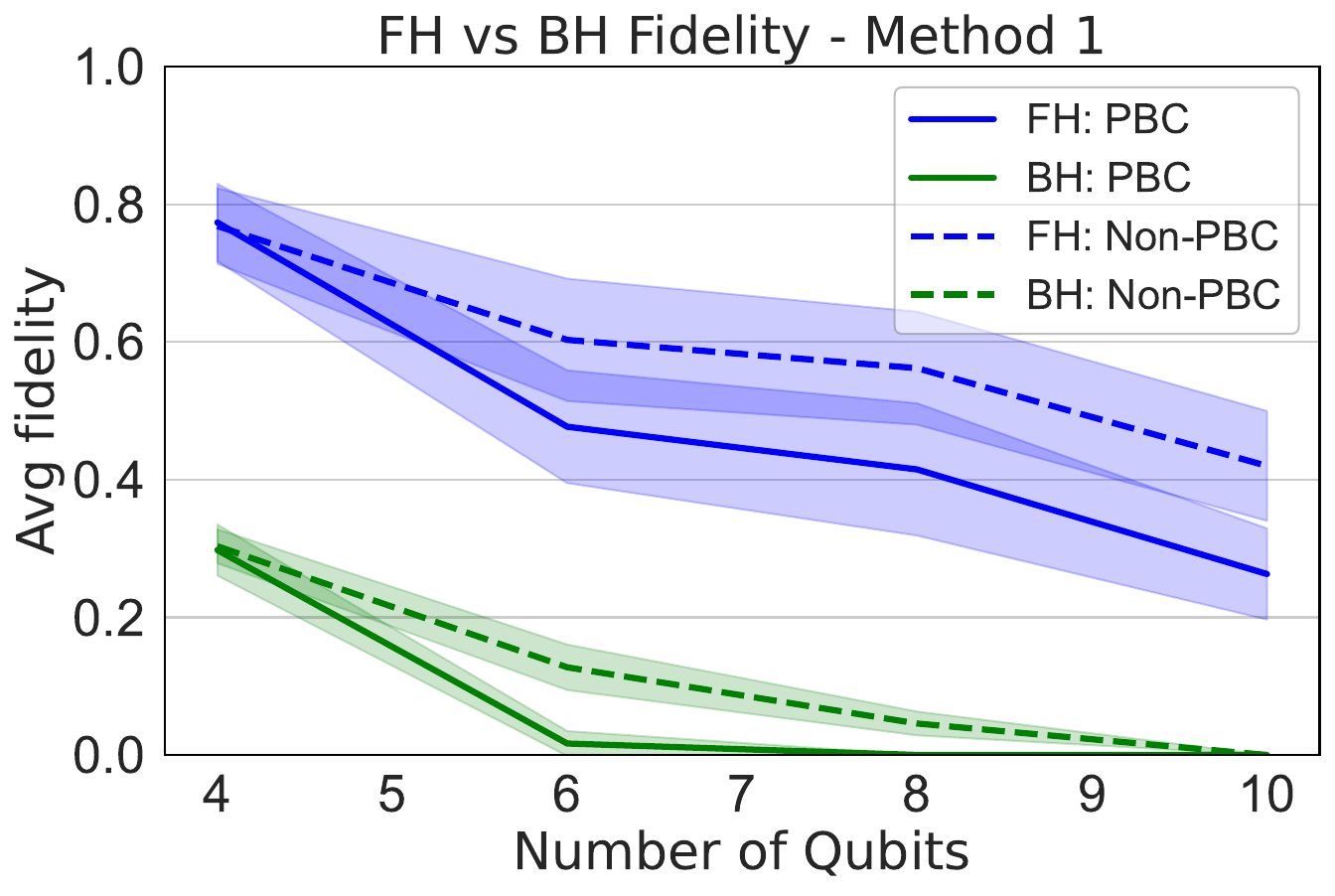}
    \includegraphics[width=0.49\columnwidth]{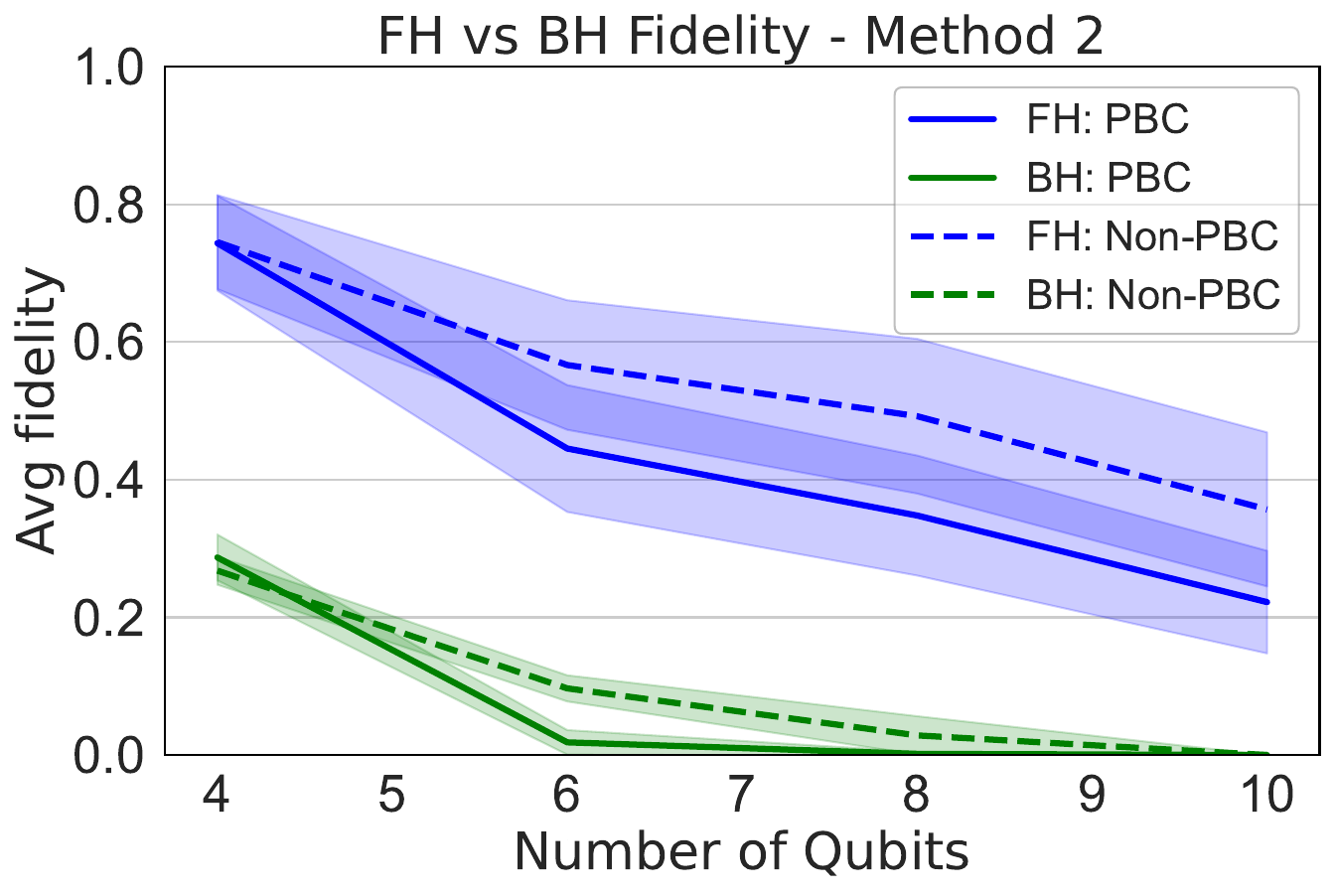}
    \includegraphics[width=0.49\columnwidth]{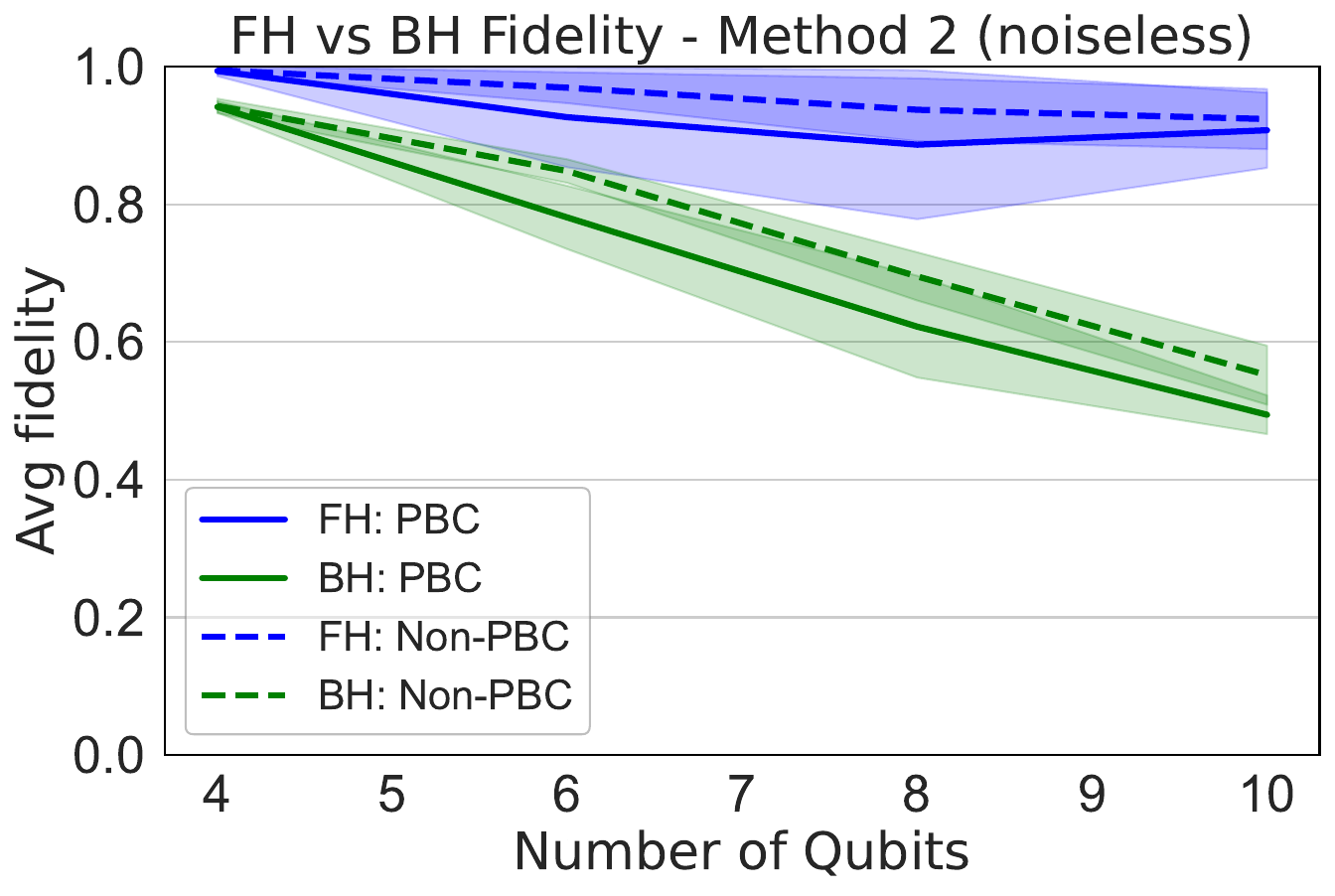}
    \includegraphics[width=0.49\columnwidth]{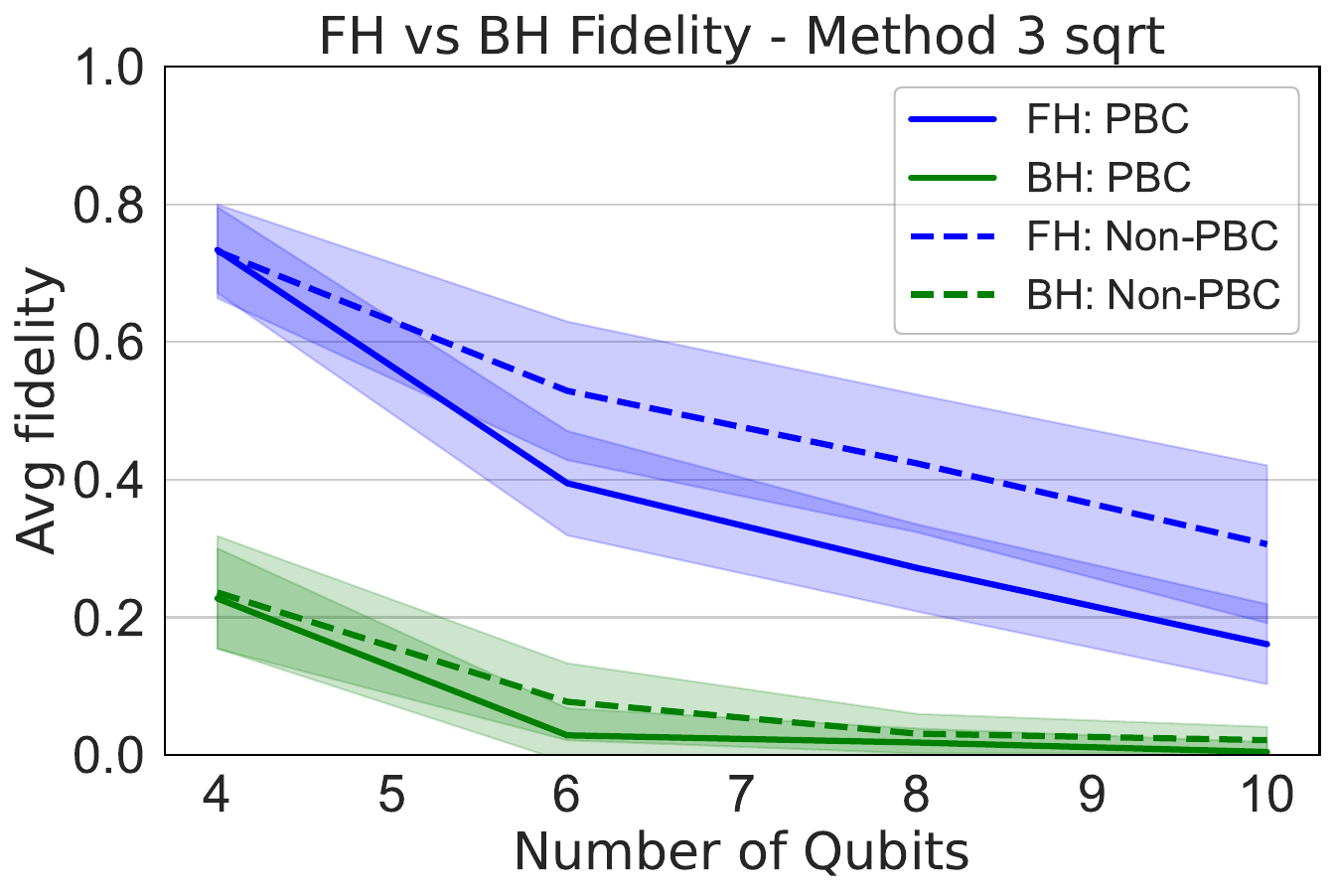}
    \caption{\textbf{Impact of Periodic Boundary Conditions on Fidelity in Hubbard Models.} This figure compares the fidelity effects of periodic boundary conditions on the Fermi-Hubbard and Bose-Hubbard models across four calculation methods. Models with periodic boundary conditions consistently show lower fidelity. The Fermi-Hubbard model generally has higher fidelity than the Bose-Hubbard model, highlighting their differing responses to boundary conditions.
    }
    \label{fig:fh_bh_all_methods_all_pbc_fidelity}
\end{figure}
% *******************

The Fermi-Hubbard Hamiltonian~\cite{hubbard1964electron} can be expressed as
\[
H_{FH} = -t \sum_{\langle i, j \rangle, \sigma} (c_{i,\sigma}^\dagger c_{j,\sigma} + c_{j,\sigma}^\dagger c_{i,\sigma}) + U \sum_i n_{i,\uparrow} n_{i,\downarrow},
\]
where \(\langle i, j \rangle\) denotes adjacent lattice sites \(i\) and \(j\), \(\sigma\) represents the fermion spin, \(c\) and \(c^\dagger\) are the fermionic annihilation and creation operators, respectively, and \(n_{j,\sigma} = c_{j,\sigma}^\dagger c_{j,\sigma}\) is the number operator. The first term of the Hamiltonian describes the tunneling of fermions between adjacent sites with amplitude \(t\), representing the non-interacting dynamics, while the second term captures the on-site fermion interaction with strength \(U\). Although the Fermi-Hubbard model is solvable analytically when \(U = 0\) or \(t = 0\), a general analytical solution exists only in 1D~\cite{lieb2003one}. In higher dimensions, the model requires extensive numerical simulation, particularly challenging in the intermediate coupling regime (\(U/t = 4, 6, 8\)) near half-filling, due to its complexity~\cite{leblanc2015solutions}.

We utilize HamLib's Fermi-Hubbard Hamiltonian implementations, focusing on one-dimensional configurations with three fermion-to-qubit mappings: Jordan-Wigner, parity, and Bravyi-Kitaev~\cite{cao2019quantum}. Our analysis covers both periodic and non-periodic boundary conditions, with interaction parameter $U$ varying across 0, 2, 4, 6, 8, and 12. While $U$ doesn't affect circuit depth or fidelity, it may influence the algorithmic Trotter error. We also explore Bravyi-Kitaev ($bk$), Jordan-Wigner ($jw$), and parity ($parity$) encoding strategies.

On the other hand, the Bose-Hubbard model is expressed as
\[
H_{BH} = -t \sum_i (b_i^\dagger b_{i+1} + b_{i+1}^\dagger b_i) + \frac{U}{2} \sum_i n_i(n_i - 1),
\]
where \(b_i^\dagger\) and \(b_i\) denote bosonic creation and annihilation operators, respectively; \(n_i = b_i^\dagger b_i\) is the particle number operator at site \(i\), \(t\) is the tunneling strength (assumed to be \(t = 1\) in this dataset), and \(U\) is the interaction energy per site. Typically, the model also includes a term proportional to the chemical potential \(\mu\) to regulate particle numbers, which we exclude here, assuming that users will initialize the particle count as needed. The model exhibits two distinct phases: a Mott insulator and a superfluid~\cite{fisher1989boson, fisher1989boson}. Extensions of this model introduce phases such as density waves~\cite{pai2005superfluid} and supersolids~\cite{batrouni1995supersolids}, reflecting its complexity. Complementing extensive theoretical studies on bosonic systems~\cite{somma2005quantum, macridin2018digital, macridin2018electron, klco2019digitization, sawaya2020resource, sawaya2020connectivity, tong2022provably, liu2022towards}, HamLib provides a dataset crafted for experimental validations using qubits.

For our investigations, we employ values for \(U\) that are dimension-dependent, derived from established phase diagrams \cite{freericks1994phase}. Our study focuses on the one-dimensional Bose-Hubbard model, examining a series of \(U/t\) ratios: \{2, 10, 20, 30, 40\}. The interaction energy per site \(U\) does not influence the circuit depth, and consequently, it does not impact the fidelity. We considered both periodic and non-periodic boundary conditions and investigated different encoding strategies, specifically $gray$ and standard binary encodings ($stdbinary$).

% *******************
\begin{figure}[]
    \includegraphics[width=0.49\columnwidth]{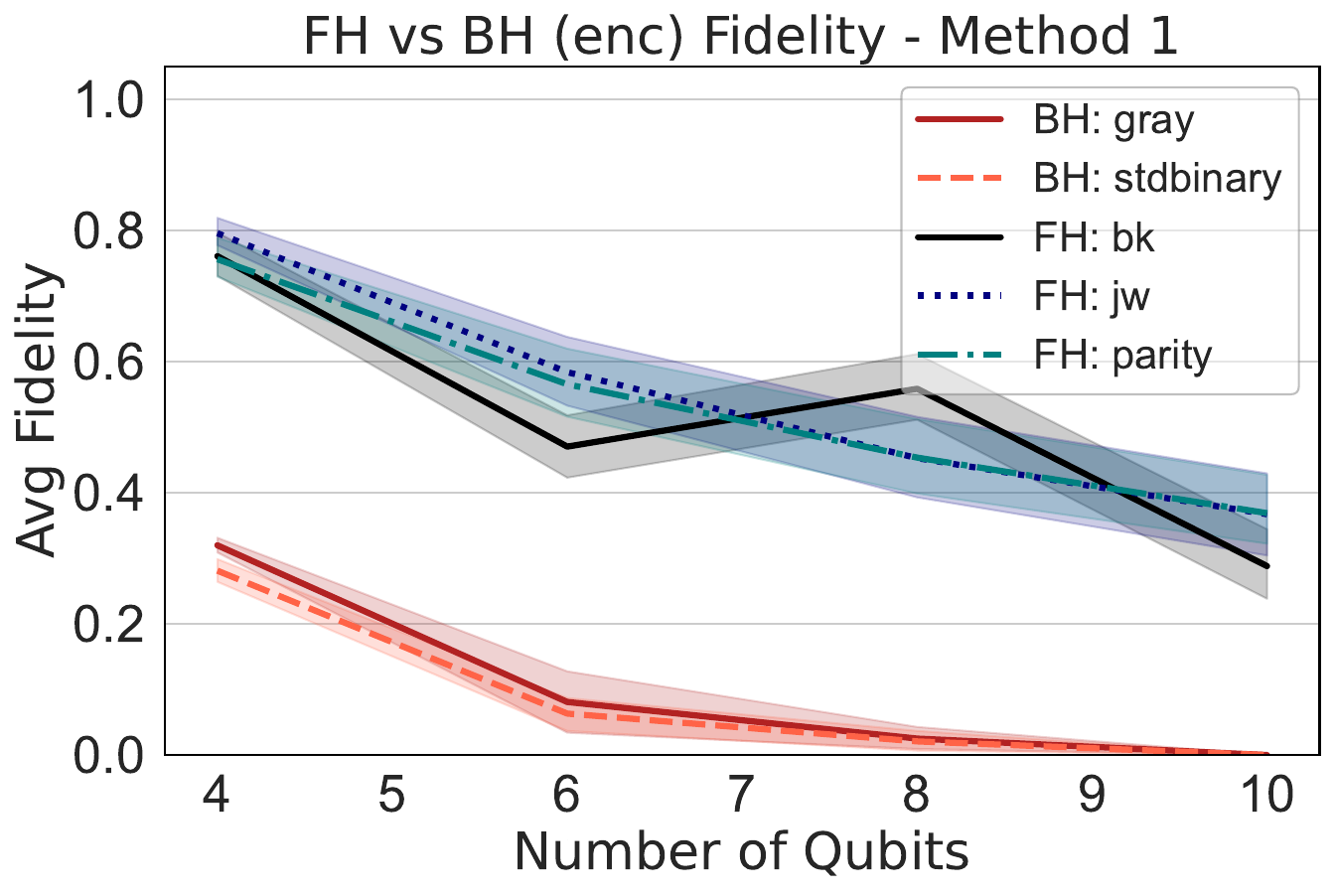}
    \includegraphics[width=0.49\columnwidth]{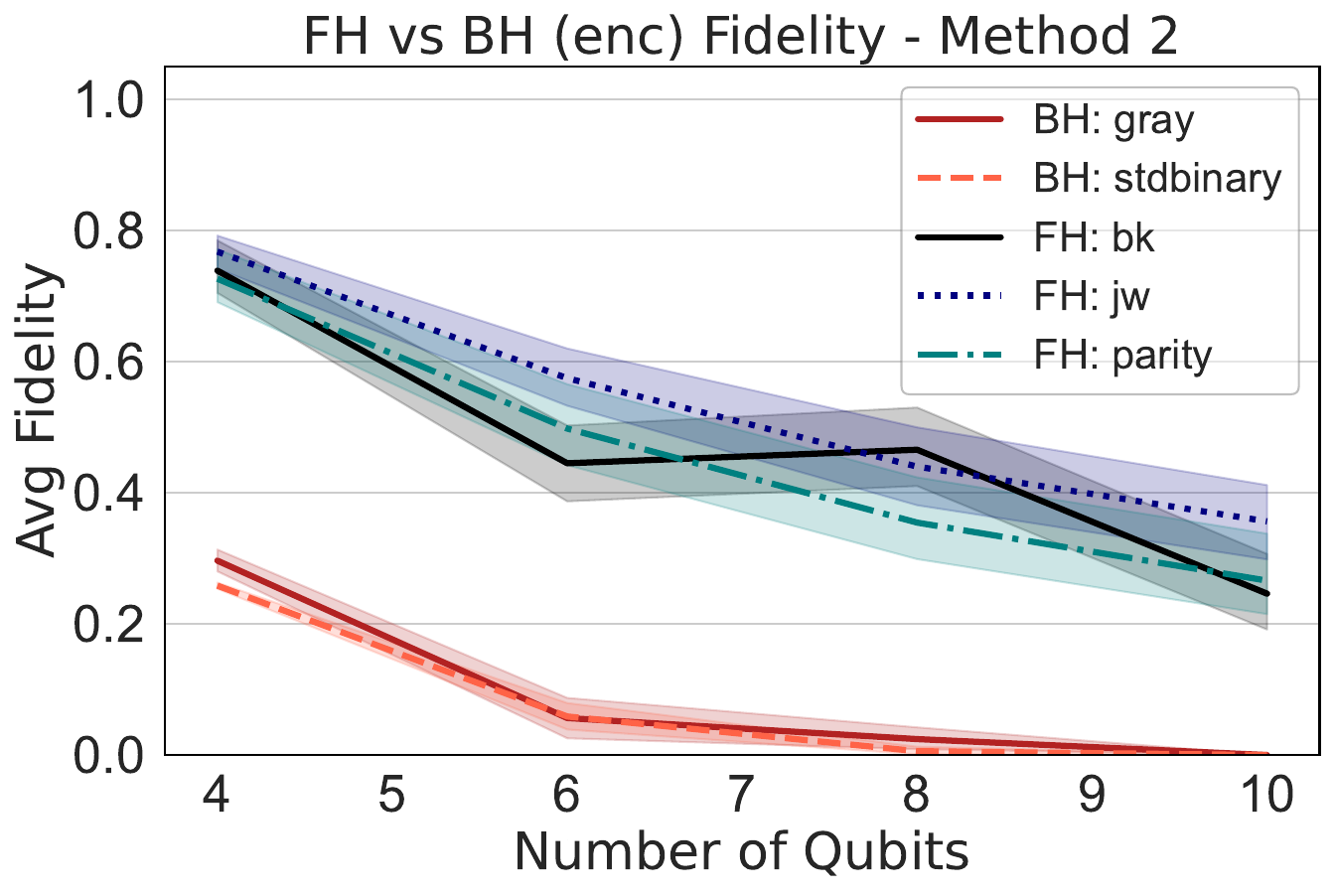}
    \includegraphics[width=0.49\columnwidth]{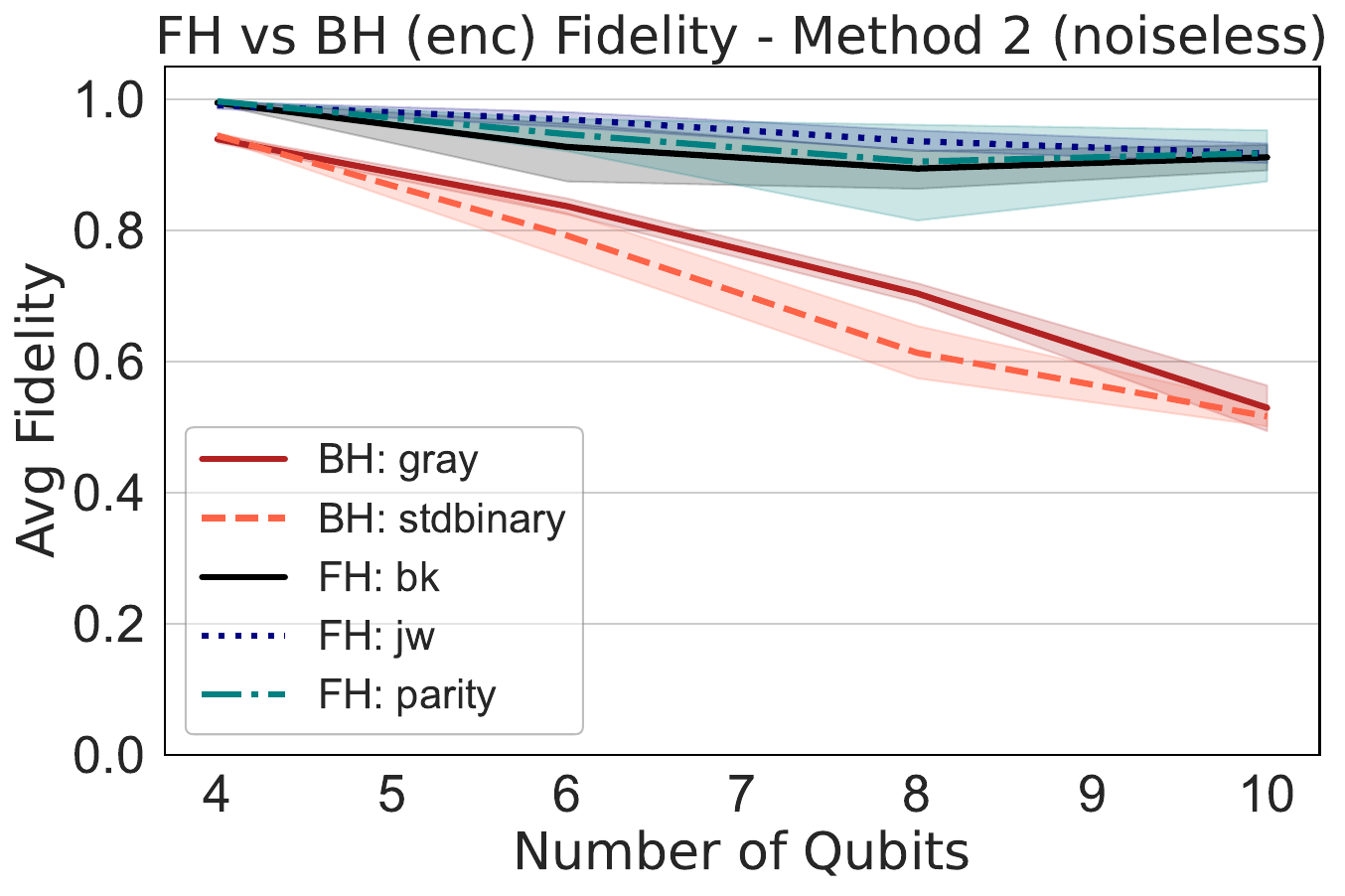}
    \includegraphics[width=0.49\columnwidth]{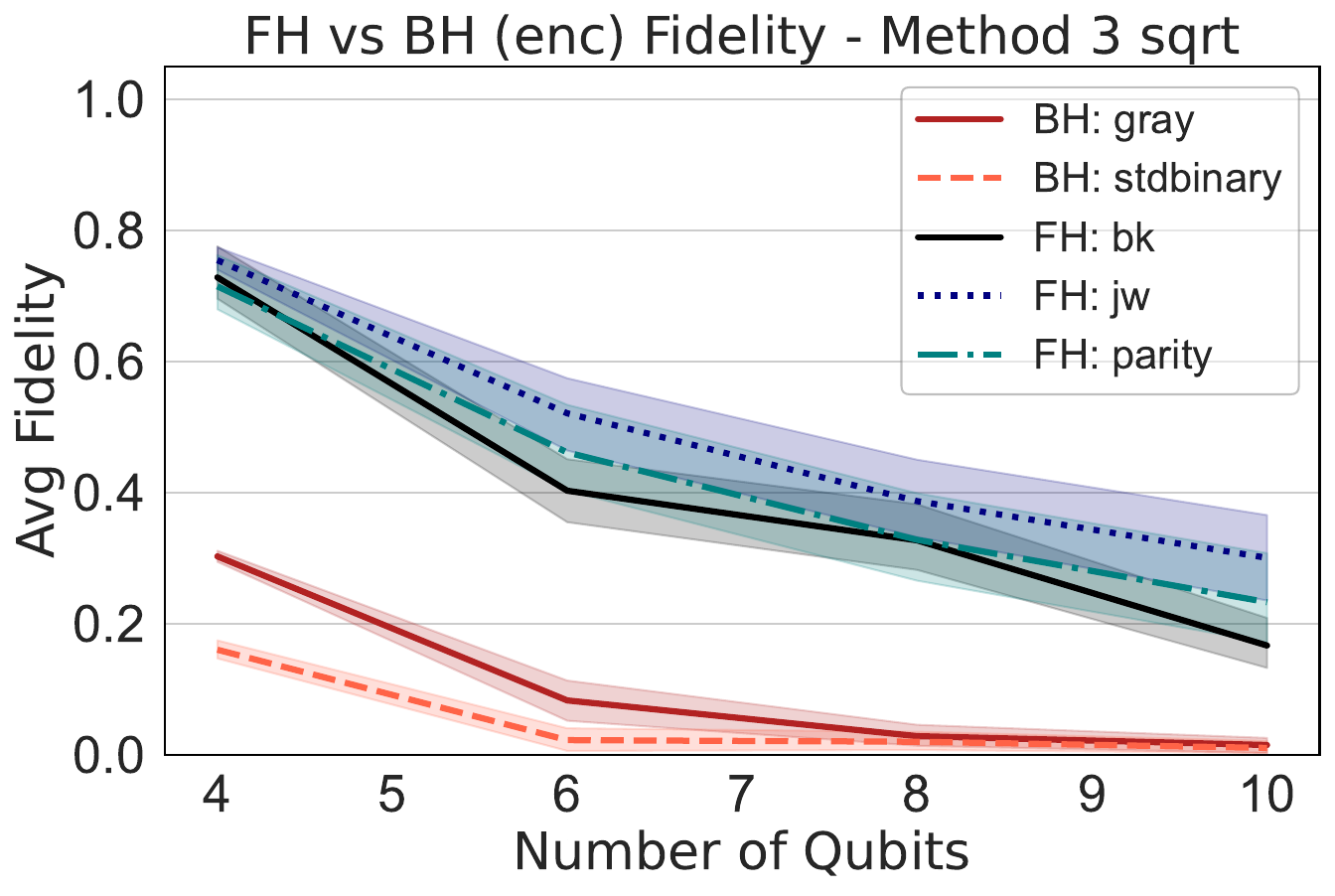}
    \caption{\textbf{Encoding Impact on Fidelity in Hubbard Models.} This figure examines fidelity variations in the Fermi-Hubbard model and the Bose-Hubbard model, each using different encoding strategies. It represents four distinct fidelity calculation methods. The Fermi-Hubbard model uses Bravyi-Kitaev (BK), Jordan-Wigner (JW), and Parity (PAR) encodings, while the Bose-Hubbard model uses Gray code and Standard Binary (StdBinary). These plots compare how each encoding strategy affects fidelity across the models and methods.
    }
    \label{fig:fh_bh_all_methods_all_enc_fidelity}
\end{figure}
% *******************

~\autoref{fig:fh_bh_all_methods_all_pbc_fidelity} illustrates the impact of periodic boundary conditions on the fidelity of both Hubbard models. Configurations with periodic boundary conditions demonstrate lower fidelity than their non-periodic counterparts, a consequence of the increased gate count and resultant greater circuit depth (see~\autoref{fig:new_depth_all_models_all_params} in Appendix~\ref{apdx:subsec:detail_depth_analysis}). We observe that the Fermi-Hubbard model consistently achieves higher fidelity than the Bose-Hubbard model, primarily due to the latter's more complex circuit structure.~\autoref{fig:fh_bh_all_methods_all_enc_fidelity} explores how changes in encoding schemes across the two Hamiltonians affect model fidelity.

Each figure is subdivided to represent the four fidelity calculation methods described in~\autoref{sec:benchmark_ham_sim}. Method 1 fidelities consistently exceed Method 2, as the latter includes both noise degradation and Trotterization error. Method 2 (noiseless) isolates Trotterization error, which is small relative to noise effects. Method 3 sqrt shows slightly lower fidelity, providing a more accurate performance assessment by approximating true process fidelity rather than merely comparing measured and ideal distributions.

For the Fermi-Hubbard model, the choice of encoding mechanism (`bk', `jw', `parity') significantly influences circuit depth. Typically, Bravyi-Kitaev encoding results in greater depth than parity, which exceeds Jordan-Wigner. These depth variations align with fidelity changes observed across encoding schemes in~\autoref{fig:fh_bh_all_methods_all_enc_fidelity}, primarily due to each method's inherent complexity. Increased circuit complexity directly impacts error susceptibility, affecting simulation fidelity.

For the Bose-Hubbard model, encoding choice ('gray', 'stdbinary') also affects circuit depth, with Standard Binary typically yielding greater depth than Gray Code. We exclude 'unary' encoding due to its larger qubit requirements. These depth differences correspond to fidelity variations across encoding schemes (~\autoref{fig:fh_bh_all_methods_all_enc_fidelity}). Deeper circuits, like those from Standard Binary encoding, are more error-prone, resulting in lower fidelity compared to Gray Code's more resilient circuits. This relationship emphasizes how encoding choices impact both operational complexity and the overall performance of quantum simulations.

% ---------------------------------------------------------------------

\subsection{The Max3SAT Problem}
\label{subsec:hamlib_M3S}

% *******************
\begin{figure}[]
    \includegraphics[width=0.49\columnwidth]{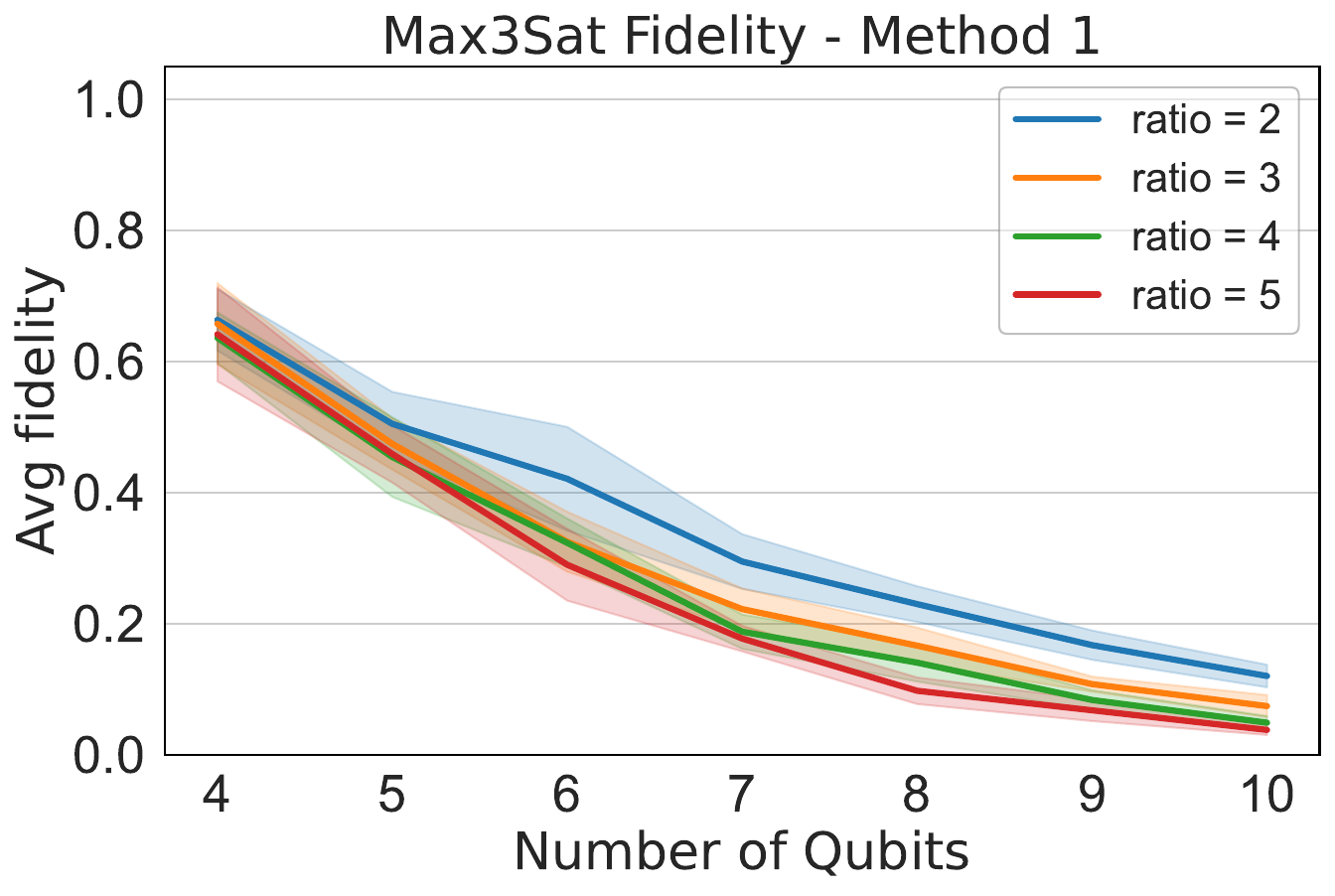}
    \includegraphics[width=0.49\columnwidth]{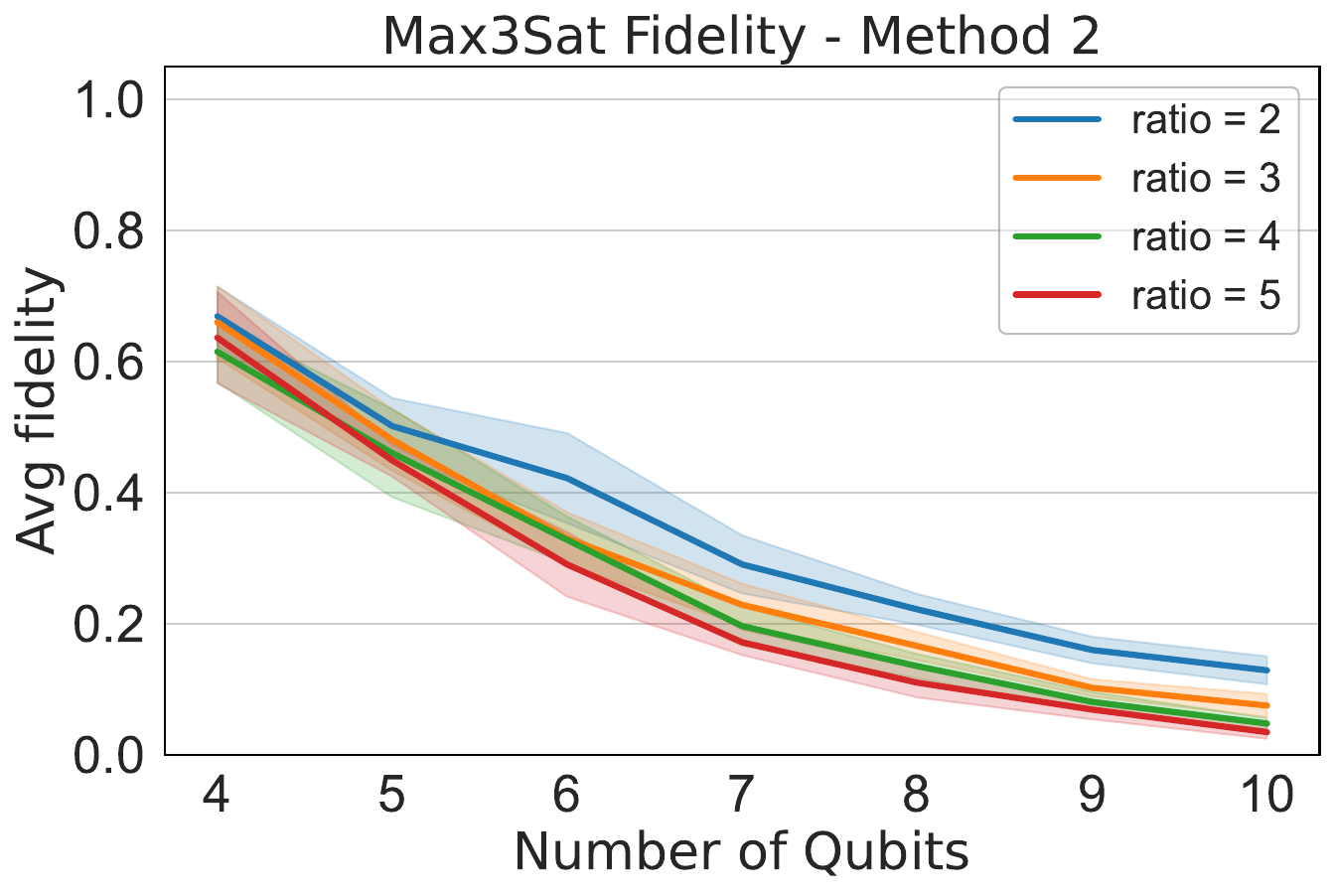}
    \includegraphics[width=0.49\columnwidth]{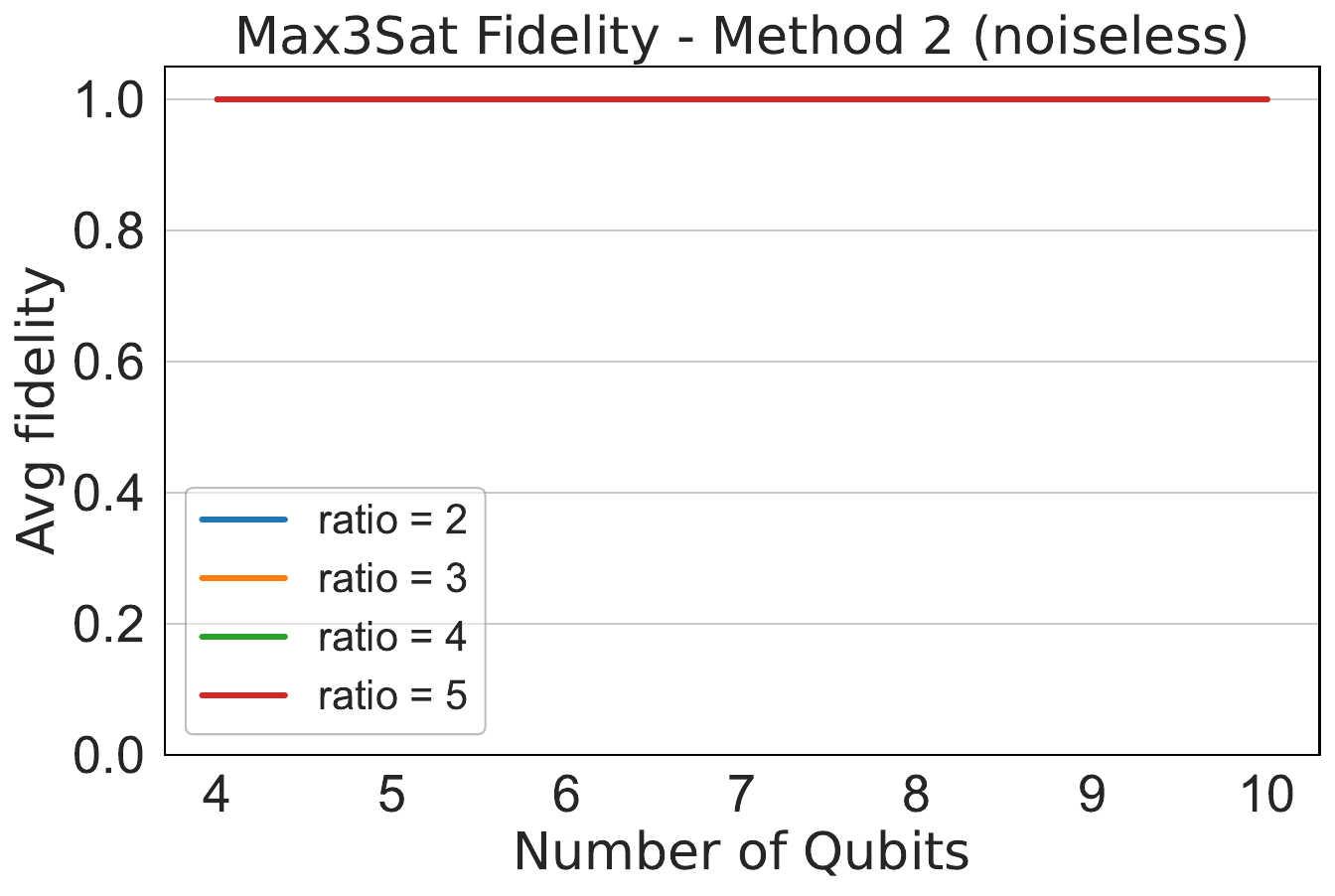}
    \includegraphics[width=0.49\columnwidth]{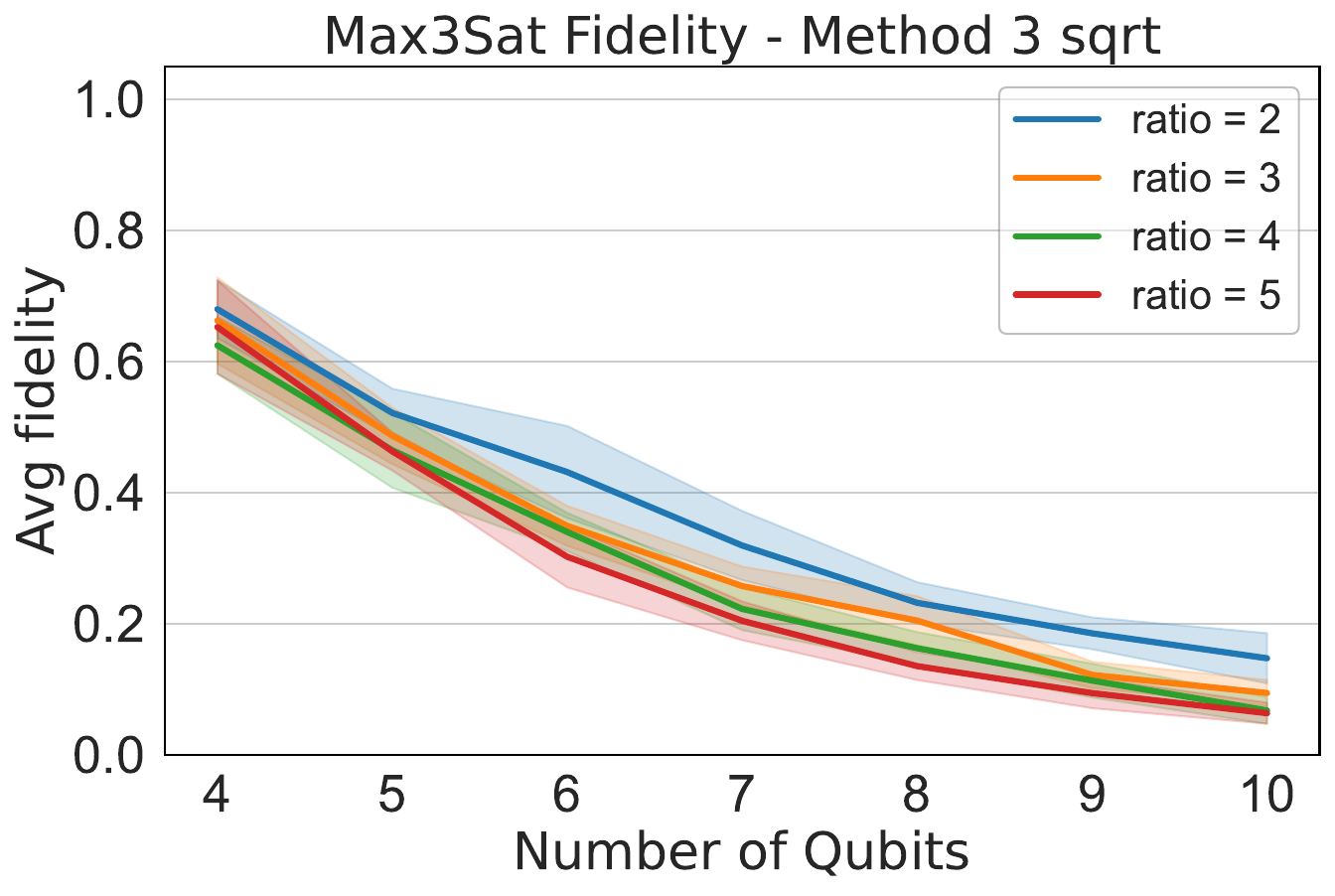}
    \textbf{}\caption{\textbf{Fidelity Trends in Max3SAT Hamiltonians Across Different Clause Ratios.} This figure demonstrates how four distinct methods of computing fidelity vary with the number of qubits and clause ratios in Max3SAT Hamiltonians. The clause ratio of 2 consistently achieves the highest fidelity, which gradually decreases until the clause ratio of 5, where fidelity is the lowest. The fidelity performance of each method remains consistent across all scenarios. Method 2 (noiseless) consistently delivers the highest fidelity, achieving a perfect score of 1 in every scenario because of the lack of Trotter error.
    }
    \label{fig:max3sat_all_methods_fidelity}
    \vspace{-10pt}
\end{figure}
% *******************

Satisfiability problems, particularly 3-SAT, are pivotal in theoretical computer science and industrial optimization. These problems are frequently utilized in complexity theoretic proofs because any NP-Hard problem can be reduced to a 3-SAT problem \cite{arora2009computational}. To represent a 3-SAT problem in quantum computing, one constructs a Hamiltonian by summing terms involving three variables. If no negations are included, the Hamiltonian for a clause \(x_i \lor x_j \lor x_k\) is represented as:
\[
x_i \lor x_j \lor x_k = I - \frac{1}{8} (I + Z_i)(I + Z_j)(I + Z_k),
\]
where \(I\) is the identity matrix and \(Z\) denotes the Pauli-Z operator, reflecting the influence of each variable in the clause. In both classical and quantum contexts, the complexity of satisfiability problems often correlates with the clause ratio \(r = \frac{m}{n}\), where \(m\) is the number of clauses and \(n\) is the number of variables. Research has shown that certain values of \(r\) mark the transition to intractability \cite{achlioptas2005rigorous}. In quantum computational studies, specific clause ratios (e.g., \(r \in \{2,3,4,5\}\)) have been identified that exceed hardness thresholds, as established through both numerical simulations and analytical estimations \cite{akshay2020reachability, zhang2022quantum}. These thresholds serve as benchmarks for testing quantum computational approaches to solving SAT problems.

To delve deeper into the Max3SAT Hamiltonians, we examine~\autoref{fig:max3sat_all_methods_fidelity} that illustrates how four different methods of computing fidelity vary with the number of qubits and clause ratios.~\autoref{fig:new_depth_all_models_all_params} in Appendix~\ref{apdx:subsec:detail_depth_analysis} presents analogous data for circuit depths. The relationship between different methods and fidelity is consistent with this model. Method 2 (noiseless) consistently achieves the highest fidelity, with a perfect fidelity score of 1 in all cases. This exceptional fidelity can be attributed to the classical nature of the Max3SAT problem. Unlike the other quantum Hamiltonians considered here, the Hamiltonian used in Max3SAT simulations consists of commuting terms. This allows for the decomposition of the exponential of the Hamiltonian without introducing any approximation errors.
Meanwhile, Methods 1 and 2 display identical circuit depths, whereas Method 3 shows a circuit depth that is twice as large. Across all scenarios, the clause ratio of 2 consistently yields the highest fidelity, which gradually decreases until the clause ratio of 5, which exhibits the lowest fidelity. This trend is inversely proportional to the circuit depth.

%The deeper circuits are more error-prone due to the increased length of gate sequences and the enhanced exposure to potential quantum noise and operational inaccuracies. This complexity, coupled with higher error rates in deeper circuits, directly contributes to the observed decrease in fidelity, underscoring the delicate balance between problem complexity and circuit architecture in quantum computing efficiency.

%-----------------------------------------------------
% =========================================================

\section{Results and Analysis}
\label{sec:results_analysis}

So far we have discussed the results of using the three fidelity calculation methods to evaluate performance across different Hamiltonians and parameter settings. In this section, we evaluate other practical performance assessments.
First, we compare execution runtime performance across various hardware platforms and simulators. Subsequently, we explore the potential of using the scalable Method 3 and its variants to predict the fidelity outcomes of Method 1, considering that Method 1 is not realistically scalable.

% --------------------------------------------------
\subsection{Observations on Execution Time}
\label{subsec:execution_runtime_performance}
For execution runtime analysis, we conducted simulations of Hamiltonian evolution on three different platforms. Two high-performance quantum simulators from BlueQubit, a CPU cluster (BlueQubit-CPU) and an NVIDIA GPU (BlueQubit-GPU)~\cite{bluequbit2024}, provide a view into the performance of the classically implemented quantum simulation. A state-of-the-art quantum hardware system from IBM Quantum (IBM Fez)~\cite{ibmq2024} yields insights into trends associated with execution on a physical quantum hardware device.
Summary results are presented in this section, while detailed data can be found in Appendix~\ref{apdx:detail_exec_results}.

The QED-C benchmark framework was used to collect key metrics associated with the execution of the simulations. We focused on two distinct time metrics: elapsed time and quantum execution time. Elapsed time encompasses the total duration from the uploading of the circuit to the retrieval of results, including all intermediate processes, such as time spent in the queue, compilation, transpilation, setup for execution, quantum execution, and transfer of data back to the calling program. In contrast, quantum execution time refers specifically to the duration for which the circuit actively runs, free from any extraneous delays. Elapsed time always exceeds quantum execution time due to the additional overheads.

\begin{figure}[t!]
\includegraphics[width=0.85\columnwidth]{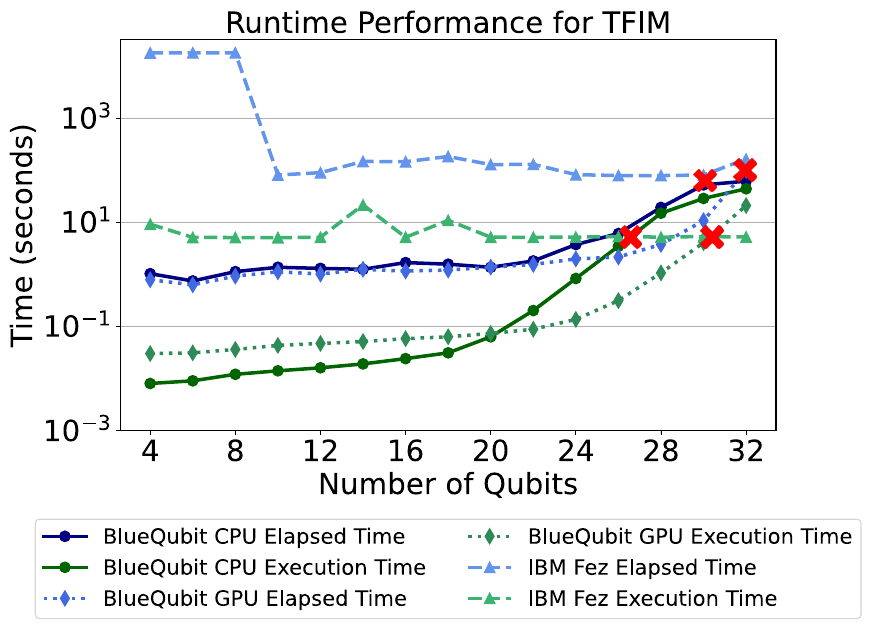}
\caption{
\textbf{Runtime Performance Analysis of TFIM Circuit Across Different Computing Platforms.}
This graph illustrates the runtime performance of the TFIM circuit, comparing elapsed and execution times as a function of circuit width (number of qubits) for BlueQubit-CPU, BlueQubit-GPU, and IBM Fez hardware. The exponential increase in execution and elapsed time for both simulators and the negligible increase on IBM Fez are highlighted. Crossover points (marked with red) indicate where the time on simulators surpasses that on IBM Fez hardware, demonstrating the efficiency of real quantum systems for larger circuits.
}
\label{fig:comparing_time_tfim}
\end{figure}

We executed the TFIM circuit by progressively increasing the number of qubits from 4 to 32.~\autoref{fig:comparing_time_tfim} depicts the relationship between the number of qubits (circuit width) and both elapsed and quantum execution times (in seconds) across different platforms: IBM Fez hardware and noiseless simulations on BlueQubit-CPU and BlueQubit-GPU. On a logarithmic scale, the times taken increase exponentially with the circuit width for the BlueQubit simulations. In contrast, the IBM Fez hardware execution times increase negligibly across varying numbers of qubits. We identify several crossover points, marked with red,  where the classically implemented simulators' times surpass those of the IBM hardware.
Additionally, the elapsed time on all platforms significantly exceeds the quantum execution time, indicating substantial overheads beyond the core quantum processing unit (QPU) computation.

%****************
%****************

From the graph in~\autoref{fig:comparing_time_tfim}, illustrating the runtime performance for the TFIM circuit, several key observations and insights can be drawn, particularly regarding the crossover points and the exponential increase in simulation times.
\begin{enumerate}
    \item \textit{Crossover Points:} These are crucial as they highlight the threshold at which a real quantum system (e.g., IBM Fez) becomes more time-efficient than a high-performance quantum circuit simulator. Notably, the crossover occurs at a circuit width of 24 to 30 qubits for BlueQubit-CPU and BlueQubit-GPU, suggesting that for circuits with fewer than 24 qubits, circuit simulators may be preferable in terms of speed. For more extensive circuits, real quantum hardware provides a clear benefit in terms of run time for shot-based circuit simulation.
    \item \textit{Exponential Increase in Simulation Times:} As the number of qubits increases, the simulation times for both BlueQubit-CPU and BlueQubit-GPU rise exponentially, particularly evident in the elapsed time data. This trend is characteristic of quantum simulations, where the complexity and, thus, the computational load increase exponentially with the number of qubits. The graph shows a sharp upturn in times beyond 20 qubits, underscoring the scaling challenges faced by quantum circuit simulators.
    \item \textit{Stability in IBM Fez Hardware Times:} In contrast to the simulators, the IBM Fez hardware shows only a negligible increase in execution times across different numbers of qubits. This stability is characteristic of many quantum hardware systems where the execution runtime performance does not degrade significantly as the circuit width increases, an essential attribute for practical quantum computing applications~\cite{lubinski2023optimization,wack_clops_2021}.
    \item \textit{Comparison of Simulators:} Both simulators exhibit similar trends, but we note that BlueQubit-GPU has faster execution times than BlueQubit-CPU at larger circuit depths, reflecting the advantage of using GPU resources for simulation purposes. However, both follow the same exponential growth pattern, emphasizing the significant opportunity for quantum hardware that is capable of executing at high fidelity but with sub-exponential cost.
\end{enumerate}

%****************
\begin{figure}[t!]
\includegraphics[width=0.85\columnwidth]{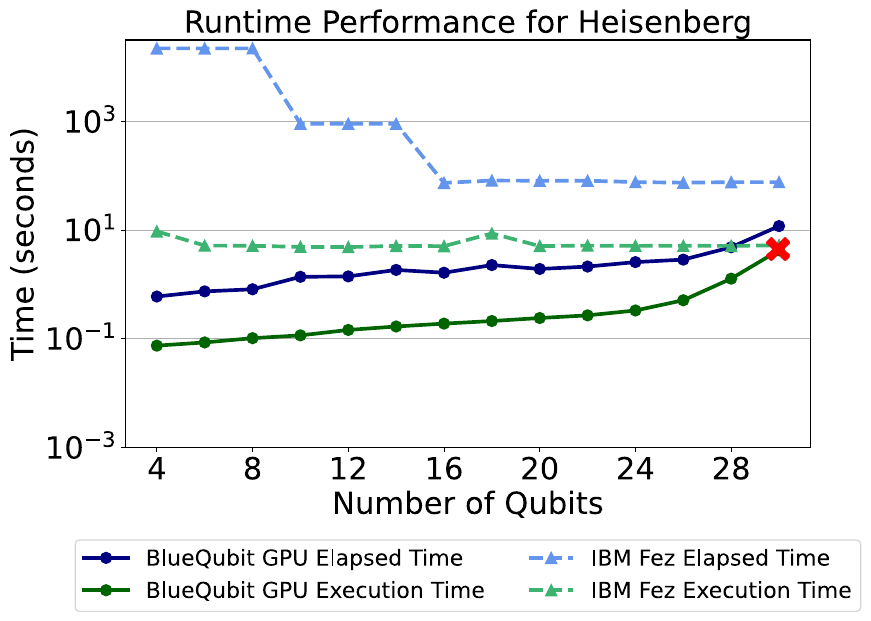}
\caption{
\textbf{Runtime Performance for the Quantum Heisenberg Hamiltonian Model Across Different Computing Platforms.}
This figure illustrates the runtime performance of the Quantum Heisenberg Hamiltonian Model on IBM Fez hardware and BlueQubit GPU, scaling up to 30 qubits. The GPU execution time rises exponentially, while IBM Fez shows minimal variation. A critical crossover point (marked with red) occurs at 30 qubits, where the GPU’s execution time converges with that of IBM Fez, indicating the threshold where the GPU begins to lose its advantage in execution speed. Further scaling to more qubits would likely result in another crossover point, where the elapsed times converge.
}
\label{fig:comparing_time_heis}
\end{figure}
%****************

We conducted a second experiment using the Quantum Heisenberg Hamiltonian model, as illustrated in~\autoref{fig:comparing_time_heis}. This experiment was performed on both the IBM Fez hardware and the BlueQubit GPU, scaling up to $30$ qubits. Similar to our observations with the TFIM circuit, the execution time on the BlueQubit GPU rises exponentially as the number of qubits increases, while the IBM Fez hardware exhibits only negligible variations in execution time. A critical crossover point is marked in red, where the GPU’s execution time meets that of the IBM Fez at $30$ qubits. Notably, had the experiment been extended to include more qubits, the BlueQubit GPU’s elapsed time would likely have also converged with the elapsed time of IBM Fez, potentially resulting in another crossover point.

Despite these similarities, some distinctions between the TFIM and Heisenberg models are evident. In both cases, the BlueQubit GPU shows an exponential rise in execution time as the number of qubits increases, yet this rise is more pronounced for the TFIM model, particularly beyond 24 qubits. In contrast, the Heisenberg model demonstrates a more gradual increase, with a single crossover point occurring at 30 qubits. Nevertheless, both models underscore the stability of the IBM Fez hardware, which consistently shows minimal variations in execution time across varying qubit counts. Furthermore, while the elapsed time consistently exceeds the quantum execution time across all platforms in both models, this gap is more pronounced in the TFIM model. Despite these nuanced differences, the overarching trend of exponential growth in simulation execution time and the stability of IBM Fez remains a common theme between the two models.

%%%%%  These two plots move here for page layout

%****************
\begin{figure*}[t!]
\includegraphics[width=0.68\columnwidth]{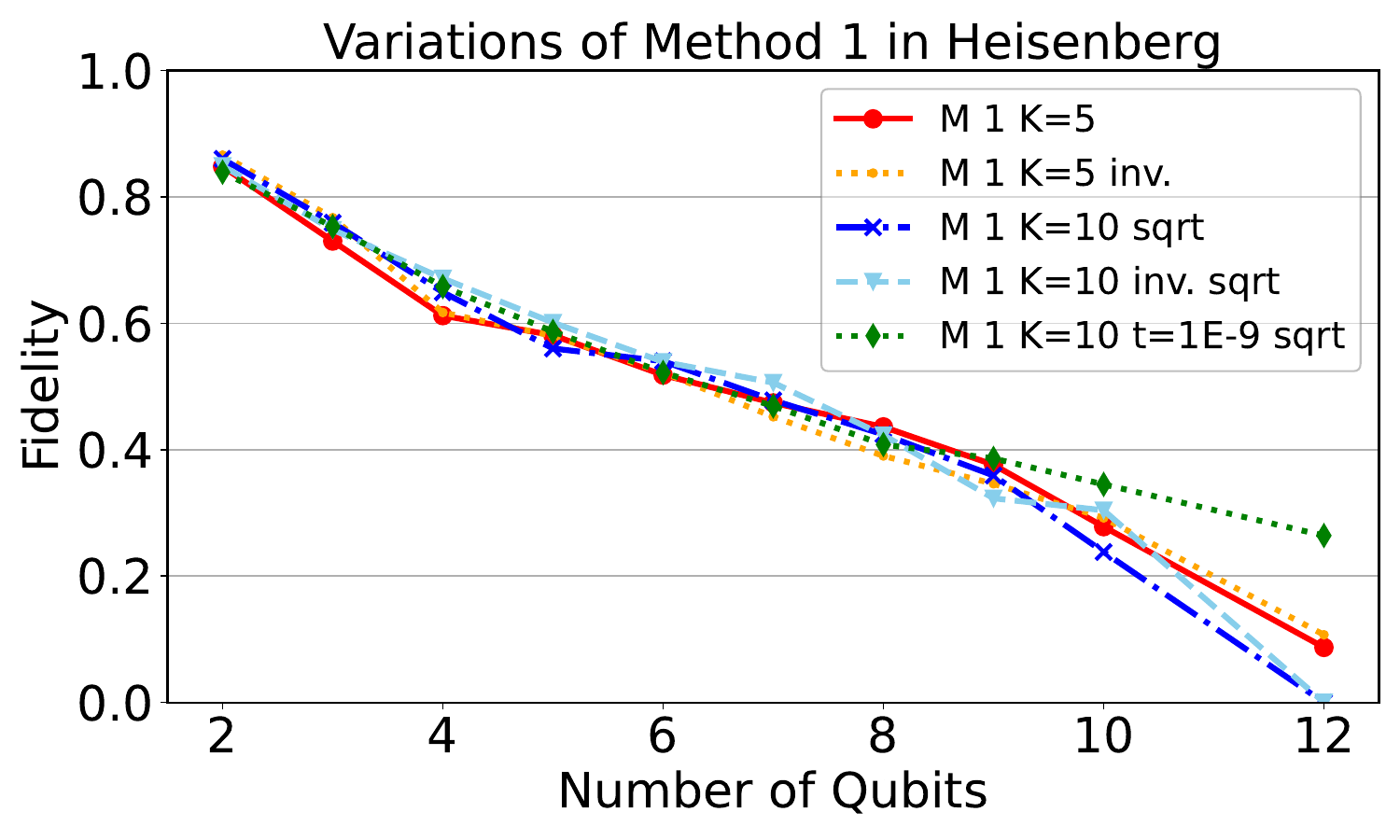}
\includegraphics[width=0.68\columnwidth]{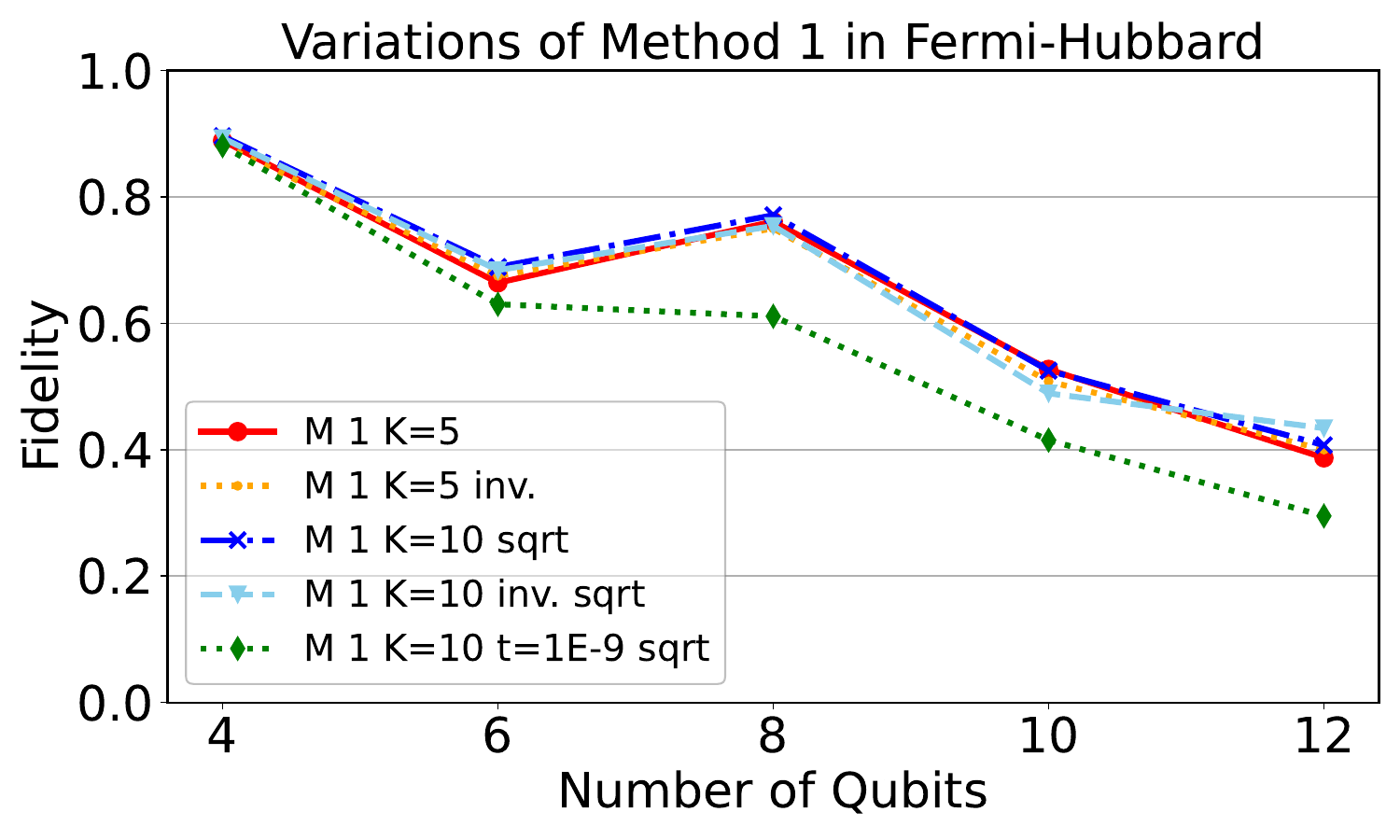}
\includegraphics[width=0.68\columnwidth]{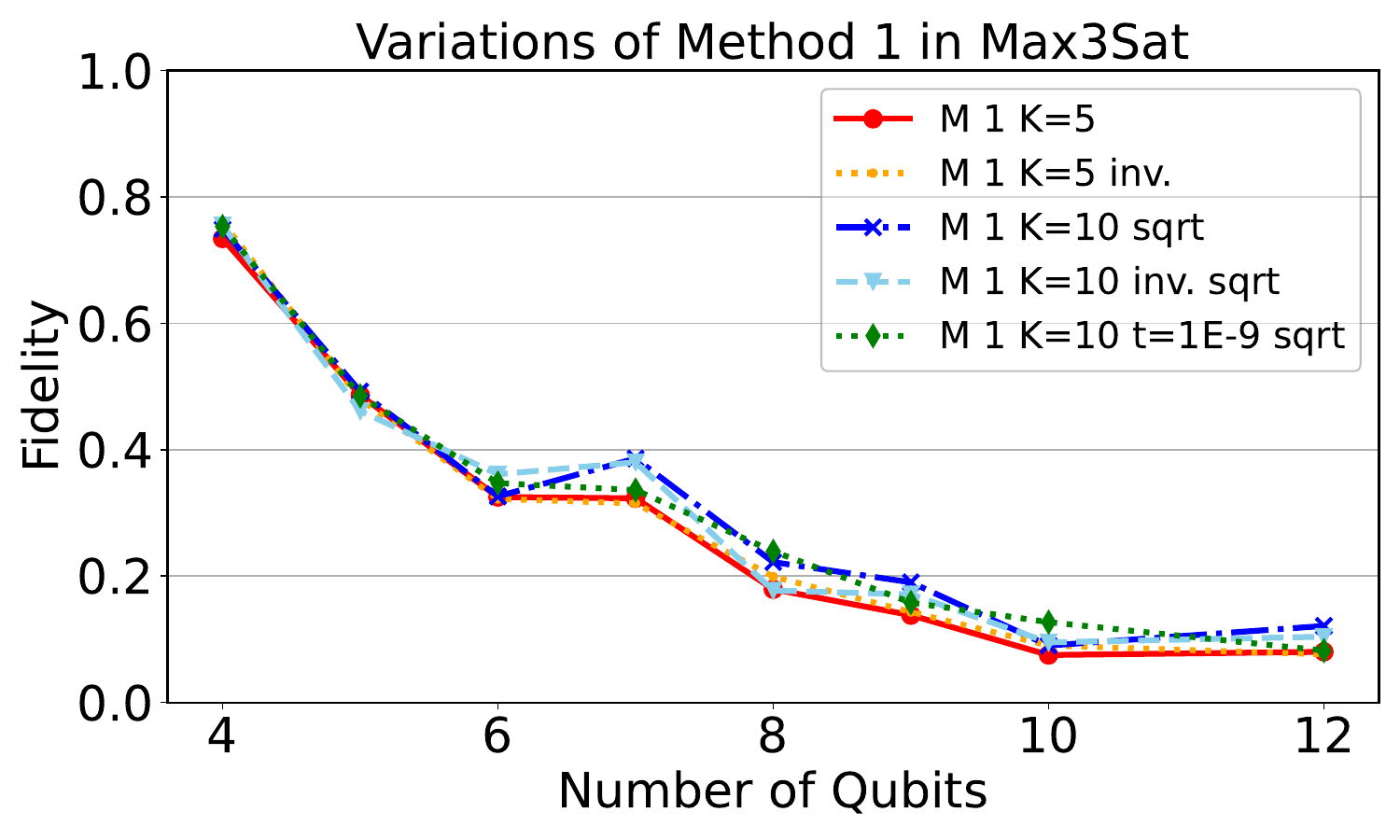}
\caption{
\textbf{Analysis of Method 1 Variants in Quantum Hamiltonian Simulation.}
This figure focuses on examining various configurations of Method 1 to understand its behavior across different settings before comparison with Method 3. This figure highlights the influence of changes in the number of Trotterization steps ($K$) and time parameter ($t$) on the fidelity of quantum simulations. The standard setup, labeled `Method 1 $K = 5$', serves as the baseline, while variations include `Method 1 $K = 5$ inverse' and adjustments where the circuit length is doubled to `Method 1 $K = 10$' and `Method 1 $K = 10$ inverse', with both reporting the square root of the measured fidelity values. These modifications aim to explore the scalability and adaptability of Method 1 by aligning its conditions more closely with those of the mirror method used in Method 3, particularly by matching circuit lengths and minimizing Trotterization time to $1E-9$ to minimize the distribution width. The resulting fidelity measurements from these variations confirm the general effectiveness of the square-root normalization approach and provide insight into the potential for Method 1 to emulate the characteristics of Method 3 under modified conditions.
}
\label{fig:comparing_method_1_variants}
\end{figure*}
%****************

%****************
\begin{figure*}[t!]
\includegraphics[width=0.68\columnwidth]{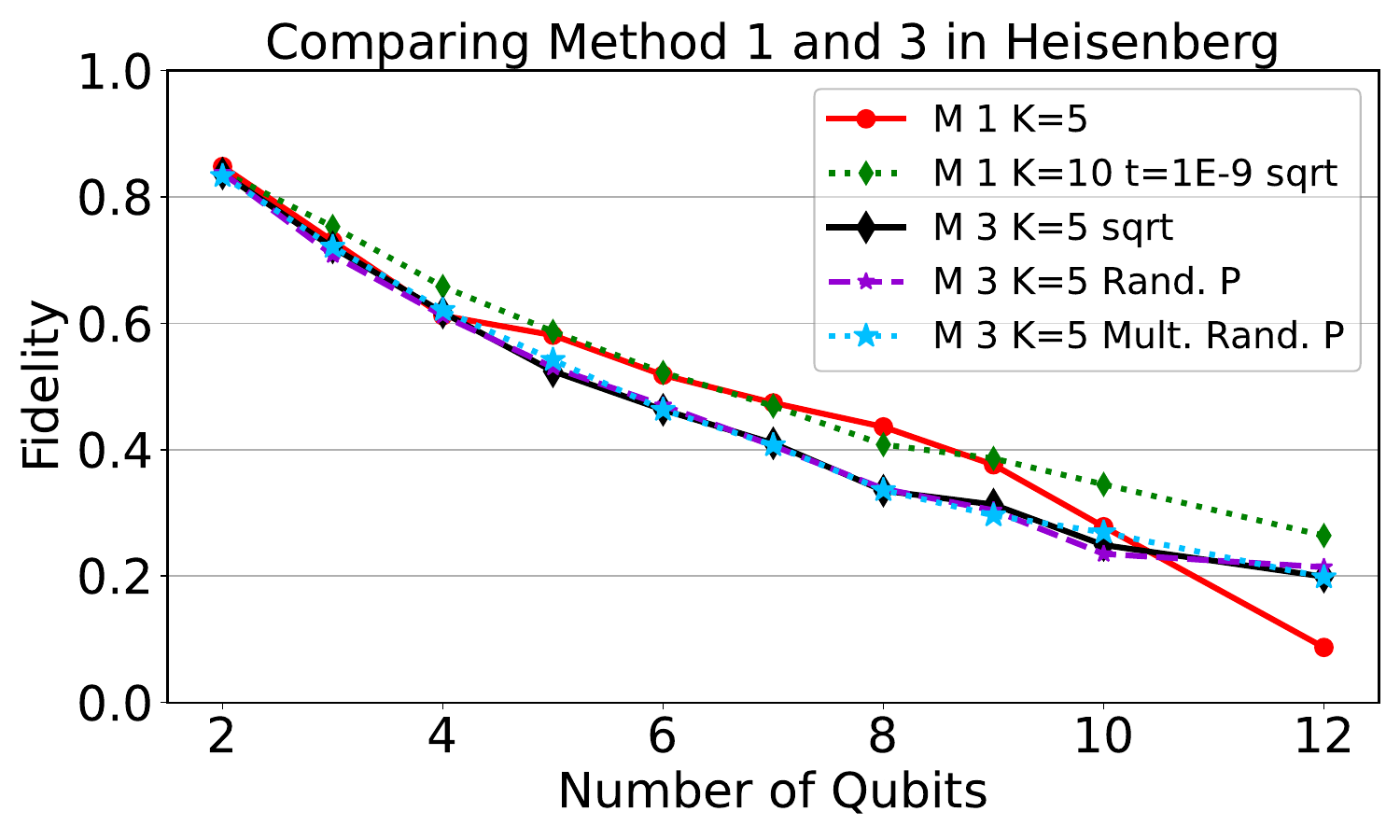}
\includegraphics[width=0.68\columnwidth]{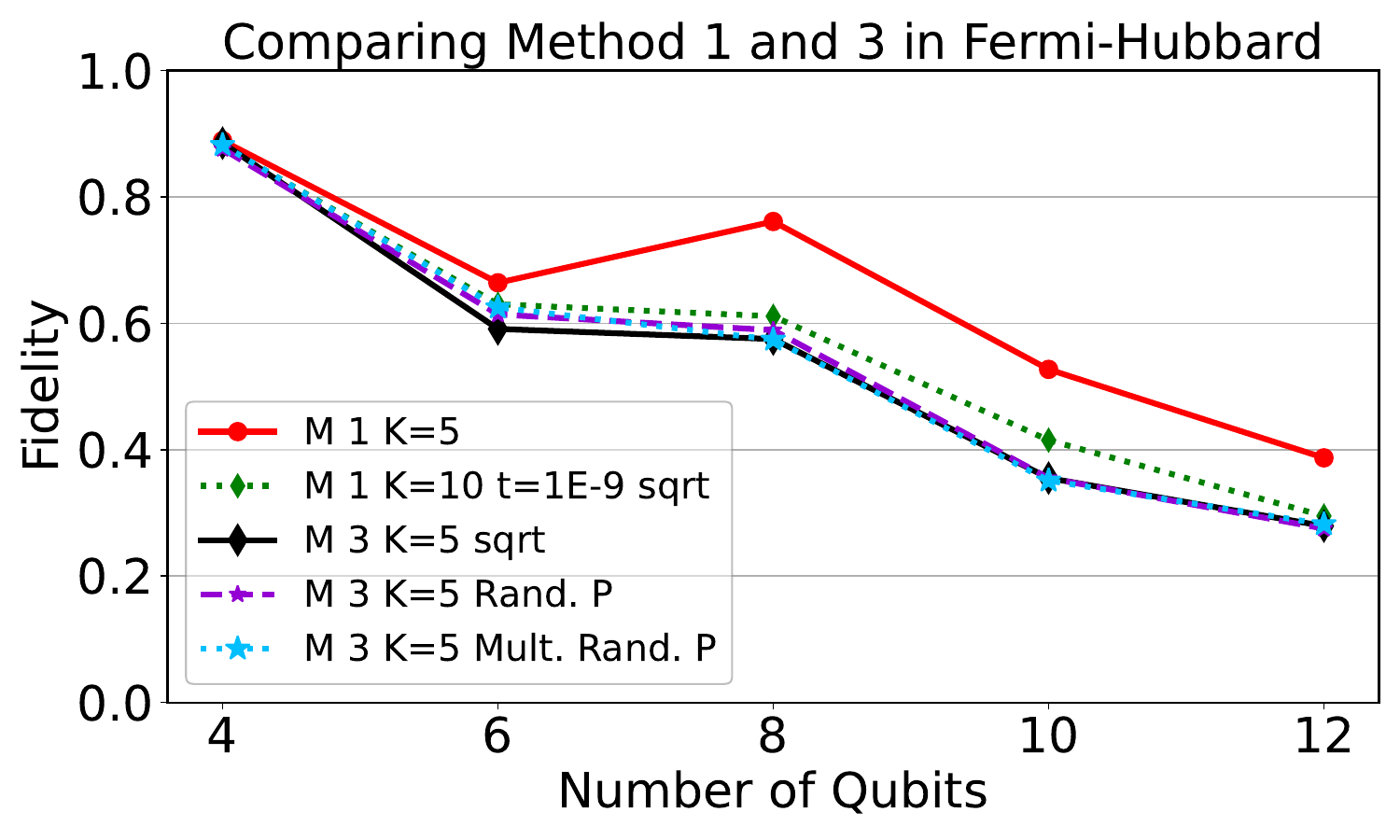}
\includegraphics[width=0.68\columnwidth]{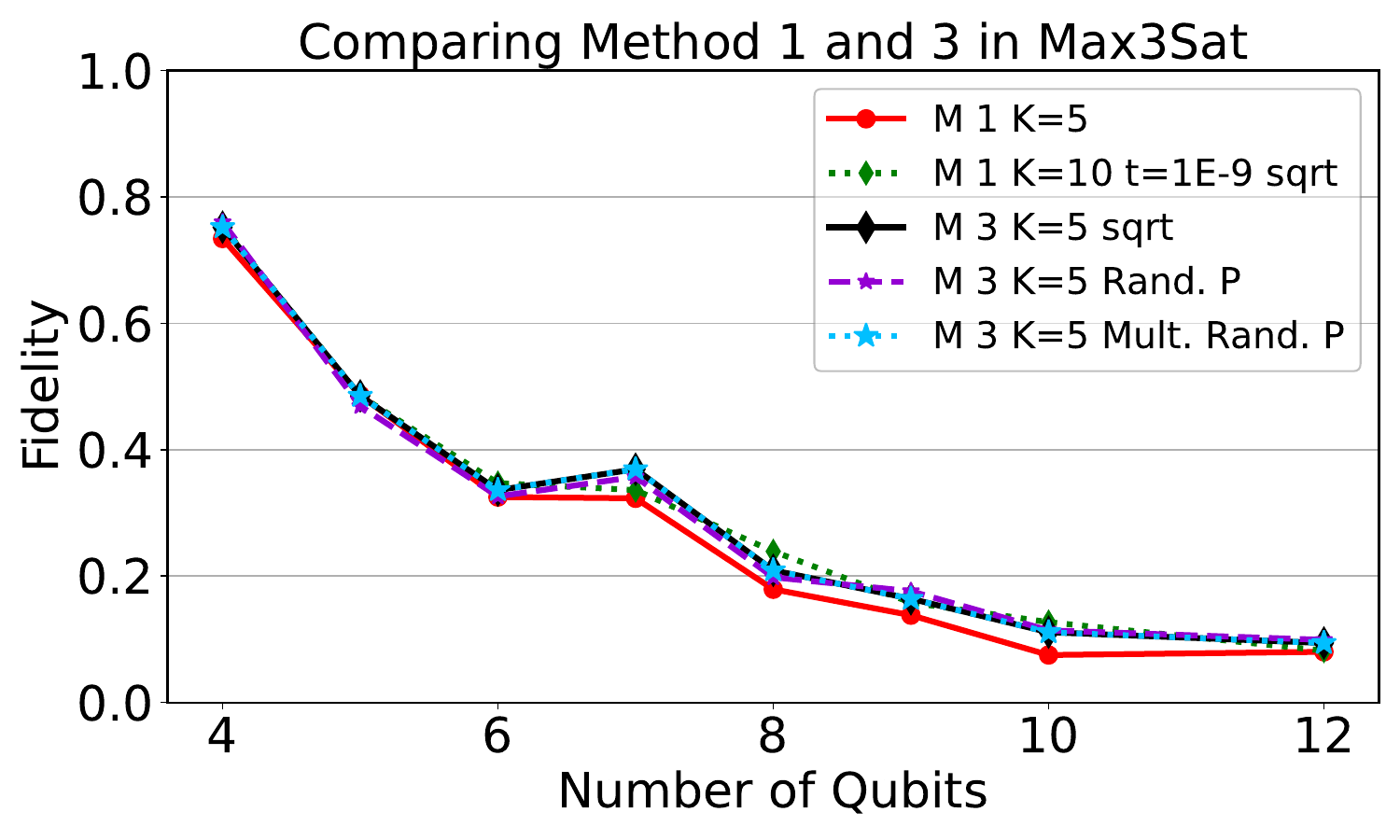}
\caption{
\textbf{Comparative Analysis of Method 1 and Method 3 Fidelity Across Hamiltonian Variations.}
This figure provides a comparative analysis between Method 1 and adaptations of Method 3, focusing on how well the normalized fidelity of Method 3 matches that of Method 1. This analysis highlights `Method 1 $K = 5$'. Notably, the `Method 1 $K = 10$ $t = 1E-9$ sqrt' variant closely aligns with `Method 3 $K = 5$ sqrt', validating our normalization approach and its potential to mimic traditional methods under specific conditions. Additionally, we introduce `Method 3 Random Pauli $K = 5$' and `Method 3 Multiple Random Paulis $K = 5$', with the latter averaging results over 10 different circuits, each with a unique random set of Paulis, to enhance fidelity analysis. This reveals that Method 3's variants provide insights into the scalability and predictive capabilities for fidelity in scenarios beyond the reach of traditional methods.
}
\label{fig:comparing_method_1_and_3}
\end{figure*}
%****************

The data from~\autoref{fig:comparing_time_tfim} and~\autoref{fig:comparing_time_heis} have implications for quantum computing applications, highlighting the critical point at which quantum hardware becomes a necessity over simulators due to performance considerations. This can inform decisions on whether to use simulators or real quantum hardware based strictly on the size of the quantum circuit.

However, the quality of the result obtained from quantum circuit execution is equally, if not more, important for assessing the utility of a quantum computing system. A quantum circuit simulator can produce results with 100\% fidelity at circuit widths over 30 qubits. It is clear from the detailed data shown in Appendix~\ref{apdx:detail_exec_results} and other benchmark studies that physical quantum computers are not yet capable of producing high-quality results for problems of this size. Thus, high-performance GPU-based quantum simulators play an important role in the development of novel quantum algorithm techniques that can compensate for the lower fidelity of quantum hardware while taking advantage of their faster execution times.

It's important to note that these results do not demonstrate a break-even point for hardware in runtime or the performance of a real application. In principle, a classical circuit simulator could be optimized for the specific target application, potentially outperforming quantum hardware in both speed and fidelity and pushing out the break-even point.
Additionally, the fidelity of quantum circuit execution may not directly correlate with the success or failure of an application. Many quantum algorithms rely on expectation values to compute solutions, and the relationship between circuit fidelity and the accuracy of these expectation values can be complex~\cite{Moses_2023}.

%----------------------------------------------------------

\subsection{Scalable Fidelity Prediction}
\label{subsec:scalable_fidelity_prediction}

In~\autoref{sec:benchmark_ham_sim}, we introduced three distinct methods for computing fidelity. We asserted that while two of the fidelity computation methods are not scalable, the third method — once normalized by taking the square root of its fidelity — can be used to predict the fidelity of the base circuit at larger qubit widths. In~\autoref{sec:benchmarking_hamlib}, the three methods were executed across several different Hamiltonians at small numbers of qubits. From the resulting plots, it can be seen that the trend lines of Method 1 and the square root of Method 3 fidelities qualitatively agree, lending considerable support to this proposal.

A careful examination of the data reveals that Method 3 fidelity is slightly lower for several of the Hamiltonians than for Method 1. This is consistent with the fact that Method 3 estimates process fidelity, and Method 1 estimates the difference between classical distributions---as process fidelity is sensitive to more errors. For example, in~\autoref{fig:fh_bh_all_methods_all_pbc_fidelity}, the Fermi-Hubbard fidelities are lower by as much at 15\% for some widths. In this subsection, we describe several tests designed to probe the source of this fidelity difference and to boost confidence in the reliability of the Method 3 predictions. Following this, we explore strategies to enhance the robustness of Method 3, aiming to retain its scalability while improving its resilience to coherent errors in the quantum system.

% \vspace{0.3cm}

~\autoref{fig:comparing_method_1_variants} displays fidelity trends from executing the default Method 1 circuit, labeled `Method 1 $K = 5$' and several variations designed to mimic specific aspects of the Method 3 circuit to isolate the root cause for the difference in fidelity. The first variant, `Method 1 $K = 5$ inv.', replaces each gate in the circuit with its inverse to determine whether the nature of the inverse operation is a factor in the fidelity degradation. The second and third variants, `Method $K = 10$ sqrt' and `Method $K = 10$ inv. sqrt', double the number of steps to $10$, but return the square root of the measured fidelity. This doubling of the circuit depth aligns with the depth of the mirror circuits and tests whether fidelity degrades more than expected with circuit depth.
In a fourth variant, `Method $K = 10$ t = $1E-9$ sqrt', we not only double the circuit depth by setting $K = 10$ but also reduce the Trotterization time to a value of $1E-9$. This makes the Method 1 circuit as similar as possible to the Method 3 circuit, as its circuit depth is nearly identical and the expected distribution after the measurement is sharply peaked, the single-bit string initial state, rather than evolved for time $t$ as in the other variants.

We observe negligible differences in the reported fidelity of the base circuit and each of the first 3 variants. This suggests that neither the inverse gates nor the circuit depth is a factor in the reduced fidelity and confirms that the square root normalization method is effective.
In contrast, the fourth variant, which implements minimal state evolution, shows a significant decrease in fidelity for the Fermi-Hubbard model but one that is small or negligible for the others.
We also note that the relationship between the fidelity trends of the first three variants and the fourth variant across the Hamiltonians tested is similar to the relationship between the fidelities of Method 1 and the square root of Method 3 visible in the plots of~\autoref{fig:fh_bh_all_methods_all_pbc_fidelity}.

% \vspace{0.3cm}

In~\autoref{fig:comparing_method_1_and_3}, we establish this relationship more clearly, indicating where the fidelities of several variations of Method 3 overlap the fidelities of Method 1.
As reference from~\autoref{fig:comparing_method_1_variants}, we show the fidelity trends for the default `Method 1 $K = 5$' and the `Method 1 $K = 10$ $t = 1E-9$ sqrt' variant, which most closely resembles Method 3 in circuit structure. We observe that the normalized `Method 3 $K = 5$ sqrt' trend aligns nearly precisely with that of `Method 1 $K = 10$ $t = 1E-9$ sqrt', except in the Heisenberg case. 

The fidelity of `Method 1 (K = 5)' declines rapidly for circuits wider than 8 qubits in several Hamiltonians, particularly the Heisenberg model shown here and the Bose-Hubbard (BH) model. Hamiltonians which have more matrix elements between computational basis states will typically require a higher number of shots to capture the state distribution~\cite{lubinski2023optimization} accurately after a finite time evolution. Increasing the number of shots beyond our default of 1000 could potentially lead to more precise state estimation and improved fidelity results for these models.

From the data presented in this section and the previous~\autoref{sec:benchmarking_hamlib}, we make several observations. First, given adequate shots, the fidelity trends for the default Method 1 circuit are always slightly greater than or equal to the square root of the Method 3 fidelities. Second, the square root of the un-evolved double-depth Method 1 fidelities can fall anywhere between those of the default Method 1 and the square root of the Method 3 fidelities.

Based on these observations, we propose Method 3 as an effective, scalable benchmark for the lower bound of Method 1 fidelity in Hamiltonian simulation circuits executed on backend target systems. The actual measured Method 1 fidelity of execution often exceeds slightly the measure predicted by Method 3, as several factors influence it. These include the terms in the Hamiltonian, the step size and number of Trotterization steps used, initial state characteristics, the number of shots used, and the type of fidelity computation used. Given our established framework, we recommend further investigation to quantify the impact of these factors, potentially enhancing Method 3's precision and utility.

% \vspace{0.3cm}

This study has shown the square root of the Method 3 fidelity to represent an effective benchmark test for quantum Hamiltonian simulations. However, some questions remain as to when Method 1 and Method 3 will result in significantly different performance metrics. Although results from Method 3 are somewhat comparable to those of `Method 1 $K = 5$', the mirror circuits used in Method 3 Simple will not detect certain types of coherent errors or noise which can be canceled out due to its inverse nature~\cite{proctor2022measuring}. 

To accurately assess and account for the actual behavior of these error modes, we use the two variants of method 3 described in~\autoref{subsec:method_3} (for both of these, we assume the square root). The fidelity trend labeled `Method 3 Random Pauli $K = 5$' shows the result of executing Method 3 a single time with one random set of Pauli gates inserted. The line labeled `Method 3 Multiple Random Paulis $K = 5$' averages the results over $10$ different random Pauli initializations to ensure robustness. The variant that uses a single set of random Paulis under-performs or matches previous results, depending on the Hamiltonian circuit and its inherent error consistency. However, the variant that executes multiple random Pauli initializations aims to retain most system errors, also underperforming or maintaining performance based on the circuit type.

Despite discrepancies, Method 3 generally meets our objectives effectively. Its stringent fidelity assessments and scalability likely provide a more accurate reflection of hardware performance than Method 1, serving as a lower bound in most cases. The Random Pauli and Multiple Random Pauli variants further enhance  Method 3's effectiveness in evaluating quantum hardware systems. This approach offers more accurate and scalable insights into quantum hardware capabilities, particularly where traditional methods may be inadequate. 

Note that~\autoref{fig:comparing_method_1_variants} and~\autoref{fig:comparing_method_1_and_3} focus on three Hamiltonian models: Heisenberg, Fermi-Hubbard, and Max3Sat. Comparisons for the other two Hamiltonians, TFIM and Bose-Hubbard, are shown in~\autoref{fig:other_two_method_1_and_3_analysis}, Appendix~\ref{apdx:subsec:method_analysis_additional}.

% ===================================================

\section{Future Direction}
\label{sec:future}

Further research will focus on broadening the scope of Hamiltonian benchmarking techniques. One significant area involves integrating a more comprehensive array of Hamiltonian models, which would help assess the robustness of current benchmarking methods across various quantum systems. Additionally, refining fidelity normalization techniques is crucial for improving comparative analysis, particularly in large-scale quantum simulations, ensuring more accurate benchmarking results.

Benchmarking Hamiltonians by assessing fidelity and execution time performance holds significant value. 
The authors and their collaborators plan to expand this approach by including problem Hamiltonians at the center of various standardized and relevant applications, as described in HamPerf~\cite{10.1145/3637543.3653431}. 
This strategy would enable the standardized evaluation of end-to-end quantum computational performance across algorithms and hardware. 

Closely linked to this effort is the integration of methods that can reduce runtimes such as techniques to reduce shot counts in expectation value estimation~\cite{sawaya2024nonclifforddiagonalizationmeasurementshot}, and error correction and mitigation strategies to increase accuracy.  These methods will become an essential part of using quantum computers to solve real-life application problems and their evaluation should be integrated into our platform.

As quantum systems grow in size and complexity, examining scalability challenges and solutions will remain a priority, ensuring that benchmarking methods evolve alongside technological advancements. Comparative studies across various quantum computing platforms will also be crucial in determining the suitability of different platforms for specific quantum tasks.

Creating accessible and user-friendly benchmarking tools will facilitate broader participation in quantum benchmarking efforts, enhancing the field's overall growth and development. Collectively, these efforts in quantum benchmarking will help support the development of quantum computing technology.

% ===================================================

\section{Summary and Conclusions}
\label{sec:summary-and-conclusions}

This paper presents a comprehensive framework for benchmarking the performance of gate-model quantum computers for implementing Hamiltonian simulations.
We used fidelity as a key metric to assess the performance and scalability of quantum computing methods. We demonstrate this framework using five Hamiltonian models from the HamLib library: the Fermi and Bose Hubbard models, the transverse field Ising model, the Heisenberg model, and the Max3SAT problem. 
We employed three distinct approaches: comparing noisy and noiseless Hamiltonians, contrasting quantum Hamiltonian simulations with classically computed ones, and implementing the mirror benchmarking method for enhanced scalability. 
Each method contributed complementary insights to our understanding of quantum computational system performance.

Through our analysis, the scalable mirror method emerged as particularly significant. This approach demonstrates the potential for measuring important fidelity metrics for larger quantum systems, where traditional methods often become computationally intractable. As quantum systems continue to grow in size and complexity, such scalable benchmarking techniques will play an increasingly critical role in assessing and improving quantum computational performance.

Our analysis of simulation times on CPU and GPU platforms confirms the well-documented limitations of classical computational resources in simulating large quantum circuits. Using generally available yet high-performance quantum circuit simulation tools, we found that the exponential increase in simulation times restricts their practical use to problems of 30 to 40 qubits, depending on the user's tolerance for extended execution times. In contrast, execution time on physical quantum hardware increased only marginally with circuit size, enabling us to identify crossover points where quantum hardware surpasses any feasible classical simulation of quantum circuit execution.

Nonetheless, high-performance GPU-based quantum simulators can play an important role in developing novel quantum algorithm techniques that compensate for the lower fidelity of quantum hardware while taking advantage of its faster execution times. Innovation in this area could help maximize the utility obtained from the next generation of quantum computers.

In conclusion, the insights gained from this study refine our current understanding of the practical performance of Hamiltonian simulation algorithms and provide a robust foundation for future research.  As the field continues to evolve, the strategies developed in this work will play a crucial role in overcoming the scalability challenges of benchmarking quantum computers.

% Various directions for future research and improvements to the framework are suggested by our work.
% =========================================================

\section*{Code Availability}
\label{sec:data_and_code_availability}

The code for the benchmark suite described in this work is available at 
\href{https://github.com/SRI-International/QC-App-Oriented-Benchmarks}{https://github.com/SRI-International/QC-App-Oriented-Benchmarks}.
Detailed instructions are provided in the repository. 

% =========================================================

\section*{Acknowledgement}

The Quantum Economic Development Consortium (QED-C), a group of commercial organizations, government institutions, and academia, supported by NIST, formed a Technical Advisory Committee (TAC) to study the landscape of standards development in quantum technologies and to identify ways to encourage economic development through standards. In this context, the Standards TAC undertook the creation of the suite of Application-Oriented Performance Benchmarks for Quantum Computing as an open-source project, with contributions from many members of the QED-C involved in Quantum Computing.
We thank the many members of the QED-C for their valuable input in reviewing and enhancing this work.

Timothy Proctor’s contribution to this work was supported by the U.S. Department of Energy, Office of Science, National Quantum Information Science Research Centers, and Quantum Systems Accelerator. Sandia National Laboratories is a multi-program laboratory managed and operated by National Technology and Engineering Solutions of Sandia, LLC., a wholly owned subsidiary of Honeywell International, Inc., for the U.S. Department of Energy's National Nuclear Security Administration under contract DE-NA-0003525. All statements of fact, opinion or conclusions contained herein are those of the authors and should not be construed as representing the official views or policies of the U.S. Department of Energy, or the U.S. Government. Pratik Sathe acknowledges the support of NNSA for the U.S. DoE at LANL under Contract No. DE-AC52-06NA25396, and Laboratory Directed Research and Development (LDRD) for support through 20240032DR. LANL is managed by Triad National Security, LLC, for the National Nuclear Security Administration of the U.S. DOE under contract 89233218CNA000001.

We acknowledge the use of IBM Quantum services for this work. The views expressed are those of the authors and do not reflect the official policy or position of IBM or the IBM Quantum team.
IBM Quantum. https://quantum-computing.ibm.com, 2024.

We acknowledge BlueQubit for providing access to and execution time on their CPU and GPU-based quantum simulation systems. https://app.bluequbit.io, 2024.

We acknowledge Daan Camps for his contribution to code development and editorial contributions to this manuscript.
We thank Nicolas Sawaya, Charlie Baldwin, Mirko Amico, Will Yang, Trevor Keen, Jon Felbinger, Celia Merzbacher, and others for their valuable comments on this manuscript.

%===========================================================
%% Bibliography

% force all figures to be printed before the bibliography
\clearpage

%\refname{}
\renewcommand{\bibsection}{
  \section*{\refname}
}
\renewcommand{\refname}{References}

% use the 'unsrtnat' style so that~\citet works properly
\bibliographystyle{unsrtnat}  
\bibliography{references}

%===========================================================

% --------------------------------------------------
%% Appendices

\clearpage
\appendix
%\clearpage

%%%%%%%%%%%%%%%%%% Insert image here for positioning

%****************
\begin{figure*}[t!]
\includegraphics[width=0.51\columnwidth]{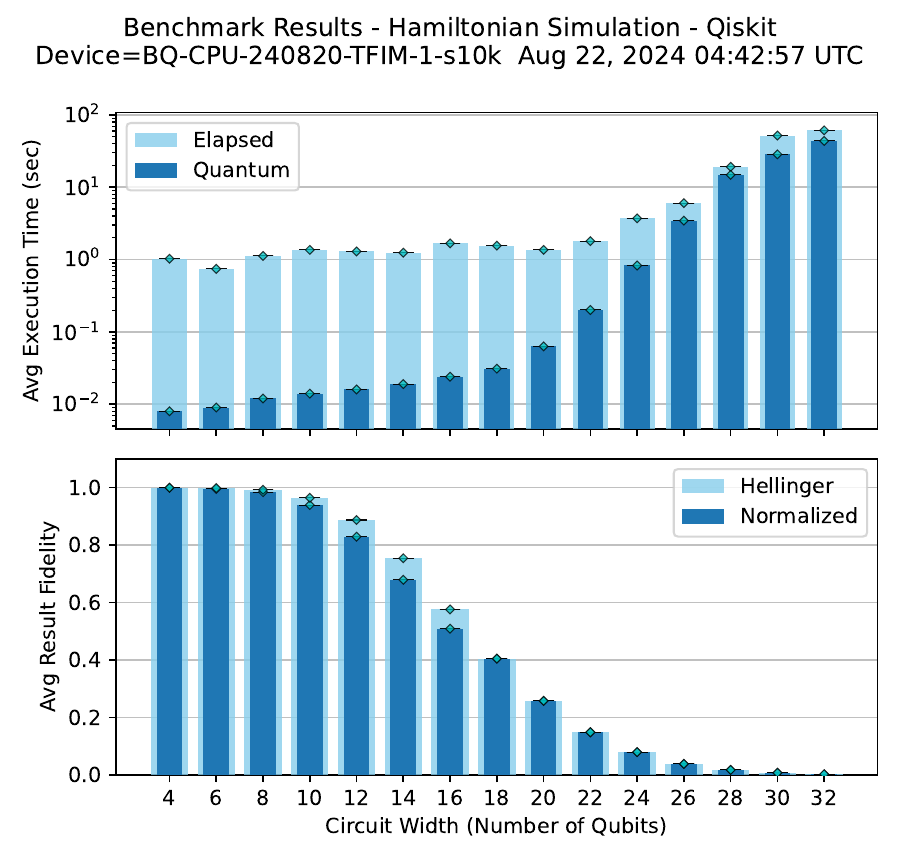}
\includegraphics[width=0.51\columnwidth]{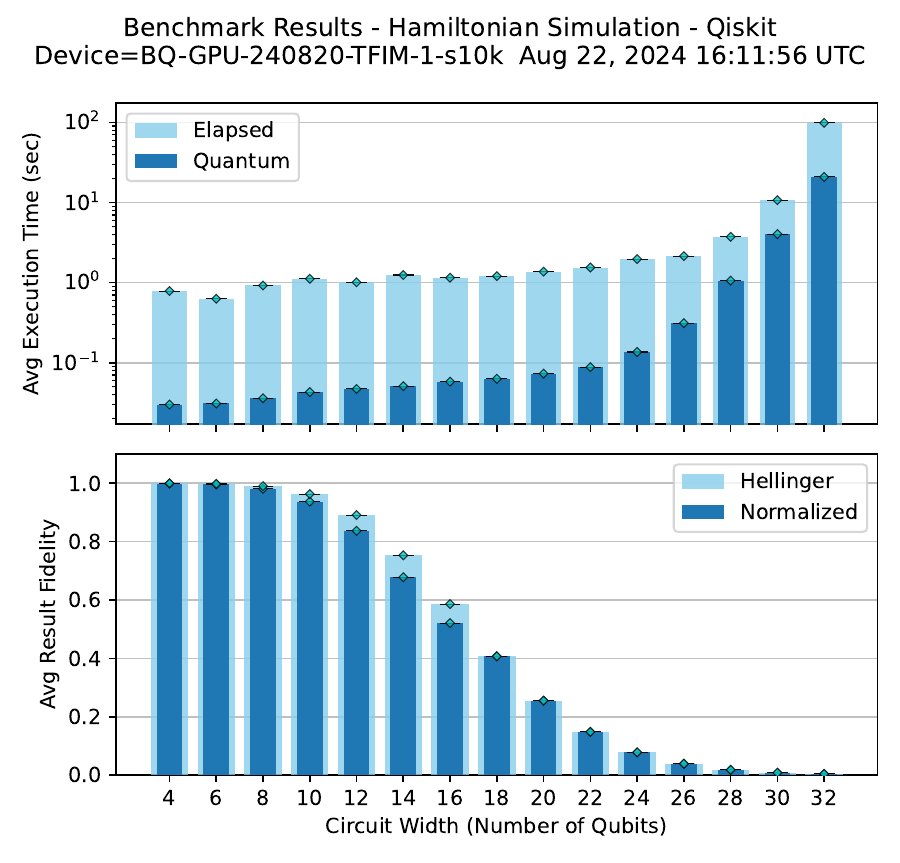}
\includegraphics[width=0.51\columnwidth]{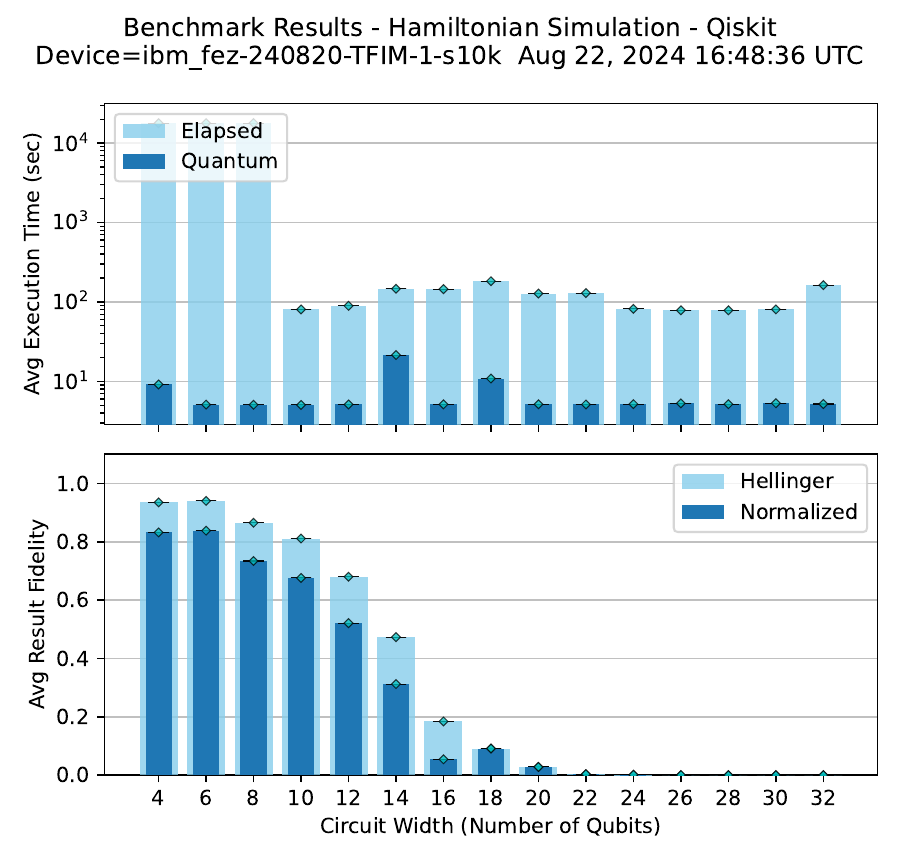}
\includegraphics[width=0.51\columnwidth]{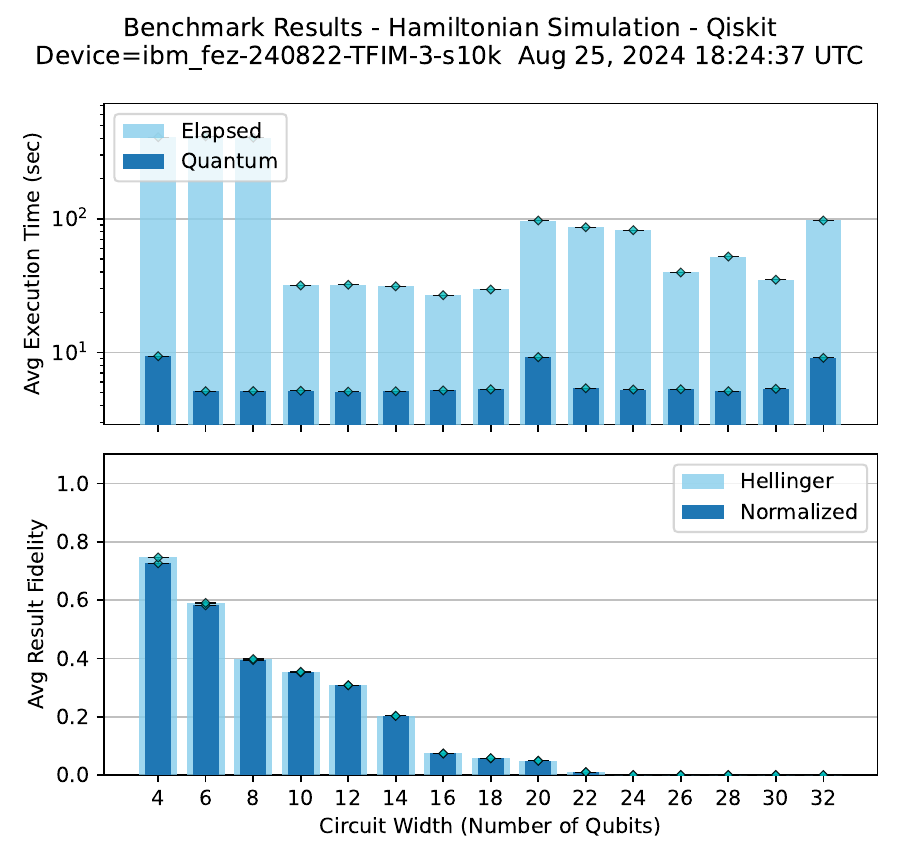}
\caption{
\textbf{HamLib TFIM Performance on Quantum Simulator and Hardware Backend Systems.}
Shown here are the performance metrics for the TFIM Hamiltonian simulation circuits for 4 to 32 qubits, at intervals of 2 qubits, on the BlueQubit-CPU (method 1), BlueQubit-GPU (method 1), and IBM Fez backends (method 1 and method 3), from left to right.
%, an ideal quantum simulator implemented on a high-performance CPU-based classical computer.
The upper subplot from each column shows the average elapsed and execution times for executing the circuit at each qubit width. 
Each lower plot shows the corresponding fidelity, which diminishes rapidly particularly beyond a circuit width of 10 qubits due to insufficient shots, which were chosen to be 10000 for each width.
On the CPU-based simulator, the quantum execution time increases exponentially from~\textasciitilde{0.01} secs to~\textasciitilde{70} secs at 32 qubits, while on the GPU-based system, the time increases from~\textasciitilde{0.02} to~\textasciitilde{10} secs. 
A slowly increasing increment of about 1 to 4 seconds contributes to the total elapsed time, but this becomes less significant at larger qubit numbers.
On the IBM Fez device, the quantum execution time remains nearly constant at~\textasciitilde{2} secs for Method 1 but rises to~\textasciitilde{4} secs, while the total elapsed time varies widely, primarily due to the wait time before execution, which is currently not removed from the elapsed time. 
The result fidelity similarly decreases with qubit width but more rapidly than the noiseless simulators. Note that Method 3 shows lower fidelity than Method 1, as expected.
(\emph{Data collected via cloud service}.)
}
\label{fig:TFIM_hw_1}
\end{figure*}
%****************

%==================================================================

\section{Comparing Quantum Fidelity Metrics}
\label{comparing_fidelity_metrics}

In this manuscript, we use the term `fidelity' to mean the performance metric defined in our original paper on `Application-Oriented Performance Benchmarking for Quantum Computing'~\cite {lubinski2023_10061574}. For analyzing the output quality of our Hamiltonian simulation implementations, we use `Hellinger Fidelity' and a normalized version, the `Normalized Hellinger Fidelity', as a measure of how faithfully a quantum computing backend has executed a quantum algorithm under test. This method is computationally efficient in the few-qubit setting, reasonably accurate, and well-accepted in the field. However, it is not the only possible well-motivated metric. 
In this section, we will describe some of the other options and compare them with the metrics we use. 

There are various, inequivalent ways to quantify how well a noisy quantum computer implements some circuit $C$ (such as one of our Hamiltonian simulation circuits)~\cite{hashim2024practicalintroductionbenchmarkingcharacterization}.
While `fidelity' is often used as a catch-all term for quality, it is also a technical term that is (i) not the unique answer to how well a noisy quantum computer executes a circuit, and (ii) has multiple specific technical definitions. There are primarily three types of fidelities, those which compare measurement outcomes, those which compare quantum states, and those which compare quantum processes.
Let us review each of these below.

\vspace{0.3cm}

\paragraph{Comparing Measurement Distributions:}
Method 1 involves repeatedly measuring all the qubits at the end of a circuit $C$, with the qubits set to the state $\ket{1010\dotsc}$ at the beginning of each run.
Each measurement yields a bit string, and enough measurements are done to obtain an estimate of the outcome distribution $P$ up to the desired precision.
We also obtain $P_{\text{ideal}}$, the probability distribution that a perfect quantum computer would generate.
Comparing the experimentally obtained distribution $P$ with the ideal distribution $P_{\text{ideal}}$ enables us to assess the quality of the implementation.
This is quantified by the classical or Hellinger fidelity $F(P, P_{\text{ideal}})$.
Here, we ignore the normalization rescaling done on the classical fidelity since it is not crucial for our discussion.

While we use Hellinger fidelity throughout our analysis, let us note that other, mutually inequivalent, general-purpose metrics could instead be used to quantify the difference between two probability distributions.
One quantifier is the total-variation distance or TVD, which has properties different from Hellinger fidelity.
Various metrics that are specialized to the application of interest can also be designed.
For example, if obtaining the expectation value of a particular Pauli operator is involved in the algorithm of interest, then that observable can be used to define a metric that compares $P$ with $P_{\text{ideal}}$.
The primary benefit of using classical statistical measures such as Hellinger fidelity and TVD is that they require knowledge only of the probability distributions over the computational basis states.
These can be estimated using significantly fewer circuit measurements than the other measures described below, especially for circuits with many qubits.
 
\vspace{0.3cm}

\paragraph{Comparing Quantum States:}
The quality of a noisy quantum computer's implementation of a circuit $C$ can also be inferred by comparing its output state $\rho$, to the (pure) state $\psi$ that a perfect quantum computer would generate after implementing the same circuit $C$.
Again, there are many possibilities for how to compare $\rho$ to $\psi$. 
Here, we state two important options:

\begin{enumerate}

\item \textit{Quantum state fidelity} between $\rho$ and $\psi$, typically denoted $F(\rho, \psi)$.We note that in general, $F(\rho, \psi)\neq F(P, P_{\text{ideal}})$, since they quantify the differences between categorically different kinds of objects, and they generally have different numerical values.

\item \textit{Trace distance} between $\rho$ and $\psi$.
\end{enumerate}
The trace distance is a measure of how easy it is to tell two quantum states apart, while state fidelity tells us how similar or alike two quantum states are.

Quantum state fidelity directly compares the quantum states themselves, not the probability distributions. It is a fundamentally quantum-mechanical concept that (for pure states) captures the overlap between two quantum states. Computing quantum state fidelity is often more computationally intensive than the classical statistical measures.

\vspace{0.3cm}

\paragraph{Comparing Quantum Processes:}
A quantum circuit defines a map or a rule for transforming the initial state of the qubits to the corresponding final state depending on the details of the circuit.
Technically, such a map is a so-called super-operator that maps density matrices to density matrices.
When a noiseless, ideal quantum computer implements a circuit $C$, it essentially implements a map, say $\Sigma$, that depends only on the unitary operator corresponding to the circuit.
However, when a noisy quantum computer implements a circuit, it implements some super-operator $\Lambda$, that differs from the ideal implementation $\Sigma$.
This difference can be attributed to the types and magnitudes of the various noise sources present in the hardware implementation.
Hence, the quality of a noisy quantum computer's implementation of $C$ can also be assessed by comparing $\Lambda$ with $U$.
Two of the most common measures for this are:
\begin{enumerate}
\item \textit{Process fidelity}, also typically denoted $F(\Lambda, U)$. Despite the similarity in the notation, process fidelity is not equivalent to quantum state fidelity or classical fidelity.

\item \textit{Diamond distance}, which is a kind of ‘worst case’ metric. 
\end{enumerate}
Process fidelity measures how well a quantum operation performs for typical input states, while the diamond distance conveys the worst possible error that could occur when using that operation.

Process fidelity thus examines the quantum circuit and evaluates how faithfully it implements the desired unitary transformation. 
While all the fidelity measures provide insight into the performance of a quantum system, process fidelity focuses on the quality of the quantum evolution implemented by a circuit, and it offers a perspective that is complementary to the others. However, process fidelity is more complex and can be more computationally intensive to calculate than the other measures.
Typically, measuring a circuit $C$'s process fidelity requires measuring the final quantum state in multiple bases.

There are some simple circumstances under which all the most widely used metrics will give similar results (e.g., simple rescalings of each other). 
In particular, with global depolarizing noise, there are simple relationships between classical fidelity, process fidelity, and state fidelity (and diamond distance).

Mirror circuits are designed for estimating \emph{process fidelity}. Specifically, the ``mirror circuit fidelity estimation'' method of Ref.~\cite{proctor2022establishing} is proven (under certain assumptions) to estimate a circuit $C$’s process fidelity accurately. Other simpler mirror circuit procedures (e.g., the `simple mirror circuits') are a less robust version of this procedure and can be regarded as a less precise method for estimating process fidelity.

In summary, Method 1 Hellinger fidelity is a useful method for measuring classical fidelity. In many (but not all) realistic circumstances, process fidelity will be smaller than classical fidelity – as seen in some of the results presented in this work.  
We use the Hellinger computation by default in our benchmark suite, primarily due to its lower cost of computation relative to state fidelity or process fidelity.
We have also introduced several implementations of Method 3, the mirror method, which is intended to be scalable and an estimator of process fidelity. 
However, it is important to be aware of the differences between the methods and the trade-offs.%==================================================================

%****************
\begin{figure*}[t!]
\includegraphics[width=0.69\columnwidth]{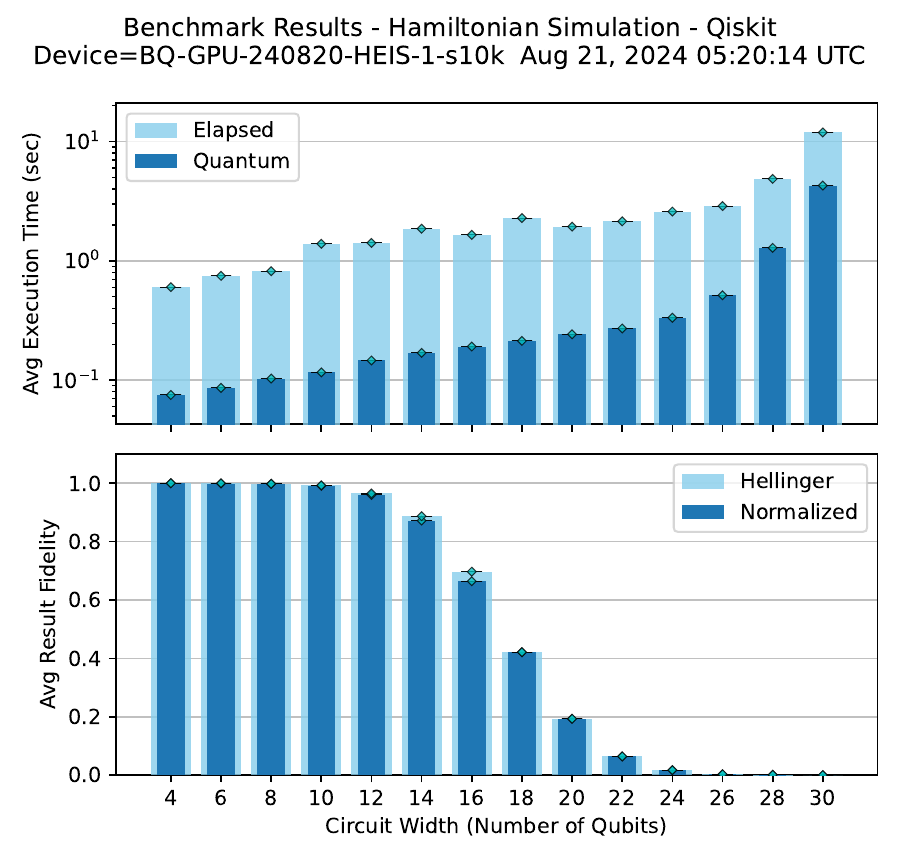}
\includegraphics[width=0.69\columnwidth]{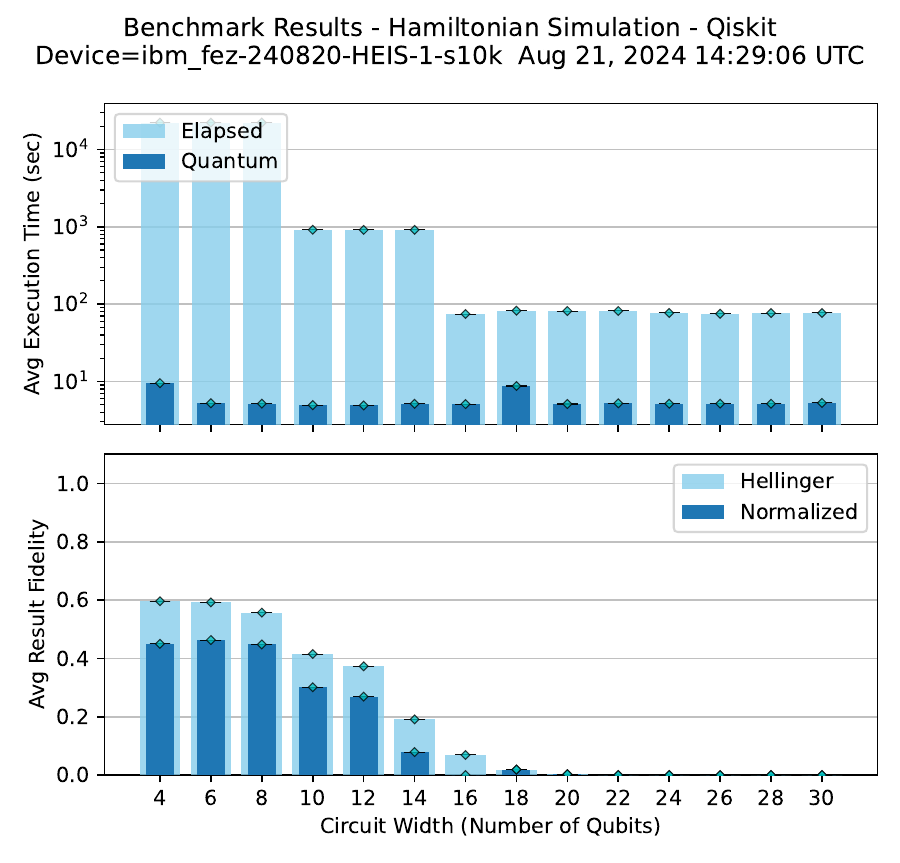}
\caption{
\textbf{HamLib Heisenberg Performance on Quantum Simulator and Hardware Backend Systems.}
Here, we show the execution of the Heisenberg Hamiltonian simulation circuit from 4 to 30 qubits, skipping 2 qubits each time, on the BlueQubit-GPU and IBM Fez backends for method 1.
%, an ideal quantum simulator implemented on a high-performance GPU-based accelerator.
On the GPU-based simulator, the quantum execution time increases from~\textasciitilde{0.05} secs to~\textasciitilde{4} secs at 30 qubits, with a growing increment of about one second to~\textasciitilde{5} seconds contributing to the total elapsed time, due to the transfer of measurement data. On the IBM Fez device, quantum execution time remains nearly constant at~\textasciitilde{2} seconds, while the total elapsed time is at least 90 seconds at any qubit width. 
The second plot displays the resulting fidelity, which diminishes as circuit width increases beyond 10 qubits due to insufficient shots (10000). Unlike the TFIM result, the result fidelity of the Heisenberg circuits is significantly lower than that of the ideal quantum simulator.
(\emph{Data collected via cloud service}.)
}
\label{fig:HEIS_hw_1}
\end{figure*}
%****************

\section{Detailed Execution Results}
\label{apdx:detail_exec_results}

In this section, we will provide some supporting details regarding the execution of our benchmarks on backend simulators and hardware systems.
In particular, we present the raw data used to create the plots in~\autoref{fig:comparing_time_tfim} and~\autoref{fig:comparing_time_heis} in the main text. 
All of the other figures in this manuscript were generated from datasets collected similarly, but using the Aer Simulator with its default noise model.

In ~\autoref{fig:TFIM_hw_1}, we show plots of the raw data collected from executing the HamLib TFIM Hamiltonian simulation circuits on a CPU-based quantum simulator (BlueQubit-CPU), a GPU-based quantum simulator (BlueQubit-GPU), and the IBM Fez backend hardware device (ibm\_fez). 
We executed circuits of widths 4 to 32 qubits at intervals of 2, and used 10,000 shots for greater resolution instead of the 1,000 shots used by default in the benchmark suite.
No error mitigation techniques were used, as the cost of error mitigation for larger numbers of qubits can be high on quantum hardware devices.

On both the simulators, in line with expectations, the execution times increased exponentially with qubit width, ranging from~\textasciitilde{0.01} secs to~\textasciitilde{12} secs for the GPU and~\textasciitilde{70} secs for the CPU.
On the IBM QPU, the execution times remained nearly constant, averaging~\textasciitilde{2-3} secs for all circuit widths. 
Total elapsed times varied due to other system factors, such as network data transfer times and wait times.

The result fidelity for all the system sizes and implementations diminished as the circuit width increased beyond 10 qubits due to insufficient shots (10,000). However, the result fidelity decreased more rapidly with qubit width on the IBM hardware device due to the inherent noise in the execution of the circuit.
For example, at 14 qubits, the normalized fidelity is at $\approx 0.66$ for both the simulators, while it is $\approx 0.30$ on the IBM hardware device. The fidelity measured by Method 3 was found to decrease more rapidly than that of Method 1 on the IBM hardware device. 
The shapes of these fidelity curves are consistent with those in~\autoref{fig:heis_tfim_all_methods_all_pbc_fidelity}, although the latter shows higher fidelities due to the lower noise characteristics of the simulator used for data collection.

% \vspace{0.3cm}

We see similar trends for execution time and fidelity in~\autoref{fig:HEIS_hw_1}, where the results from the execution of the Heisenberg Hamiltonian simulation are shown for BlueQubit-GPU and IBM Fez.
This time, we executed the benchmark circuits for sizes ranging from 4 to 30 qubits, at intervals of 2 qubits each time, and using 10,000 shots as with TFIM.
Quantum execution time on the GPU-based simulator increases from 0.05 secs to~\textasciitilde{4} secs at 30 qubits, with a growing increment of about one-second to~\textasciitilde{5} seconds contributing to the total elapsed time, due to the transfer of measurement data. On the IBM Fez device, quantum execution time remains nearly constant at~\textasciitilde{2} seconds, while the total elapsed time is at least 90 seconds at any qubit width. 

On both systems, the result fidelity diminishes as circuit width increases beyond 10 qubits due to insufficient shots (10000). The result fidelity of the Heisenberg circuits is significantly lower on the quantum hardware device than we saw for the TFIM circuits. The increased depth of the Heisenberg circuits means that noise plays a larger role in the degradation of fidelity than for the TFIM circuits.

%==================================================================

\section{Extended Analysis of Quantum Hamiltonians}
% \label{apdx:detail_depth_analysis}
\label{apdx:extended_analysis_hamiltonian}

In this section, we provide a more detailed exploration of the quantum Hamiltonian models beyond the primary analysis presented in the main text. Specifically, we delve into the circuit depth characteristics and extend our method comparisons to all Hamiltonians by varying their parameters. The following subsections present the circuit depth analysis across various models and offer supplementary comparisons of Method 1 and Method 3 for Hamiltonians not covered in the main text.

%----------------------------------------------------------

\subsection{Parametric Circuit Depth Analysis}
\label{apdx:subsec:detail_depth_analysis}

%****************
\begin{figure*}[t!]
\includegraphics[width=0.51\columnwidth]{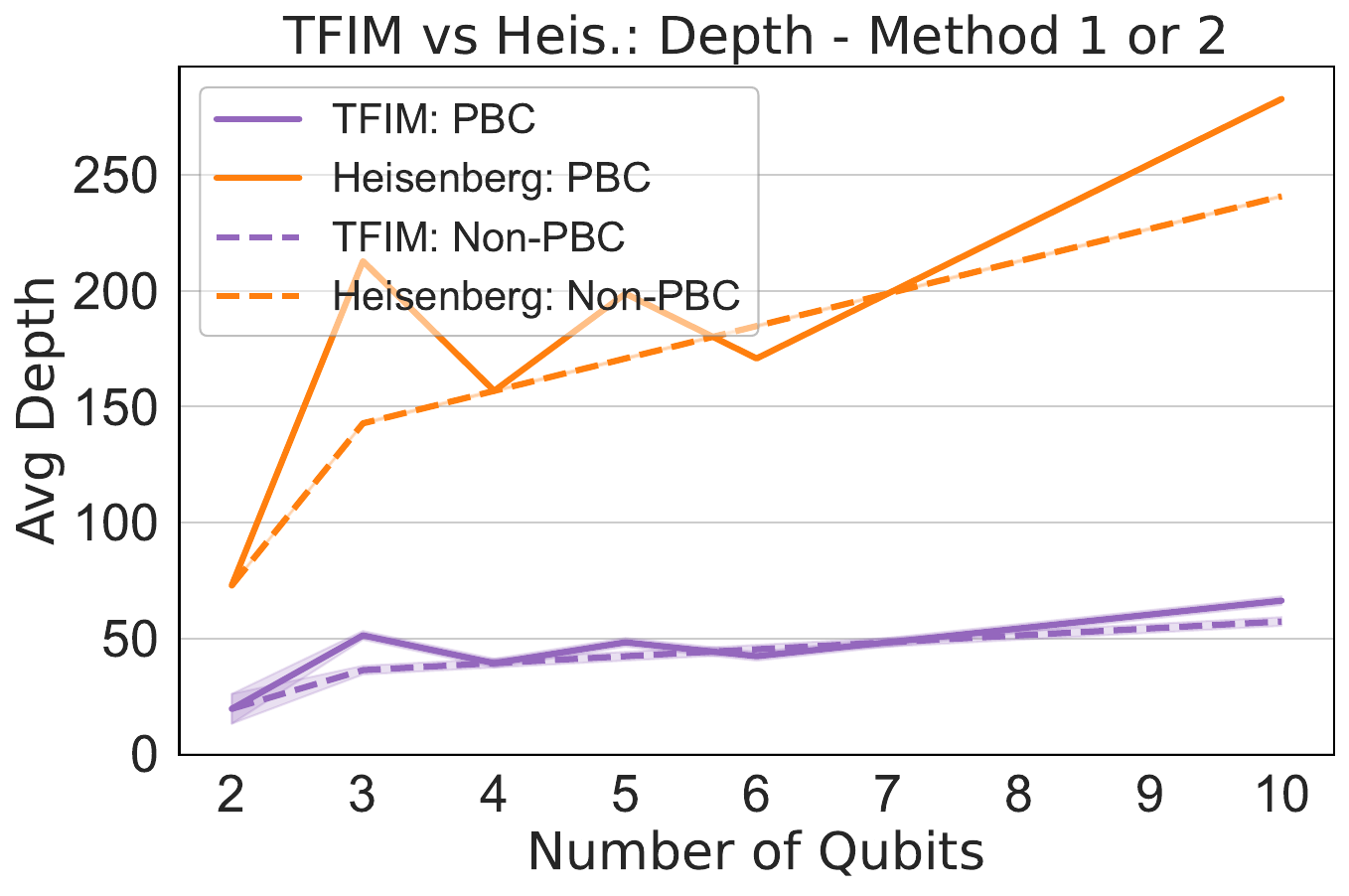}
\includegraphics[width=0.51\columnwidth]{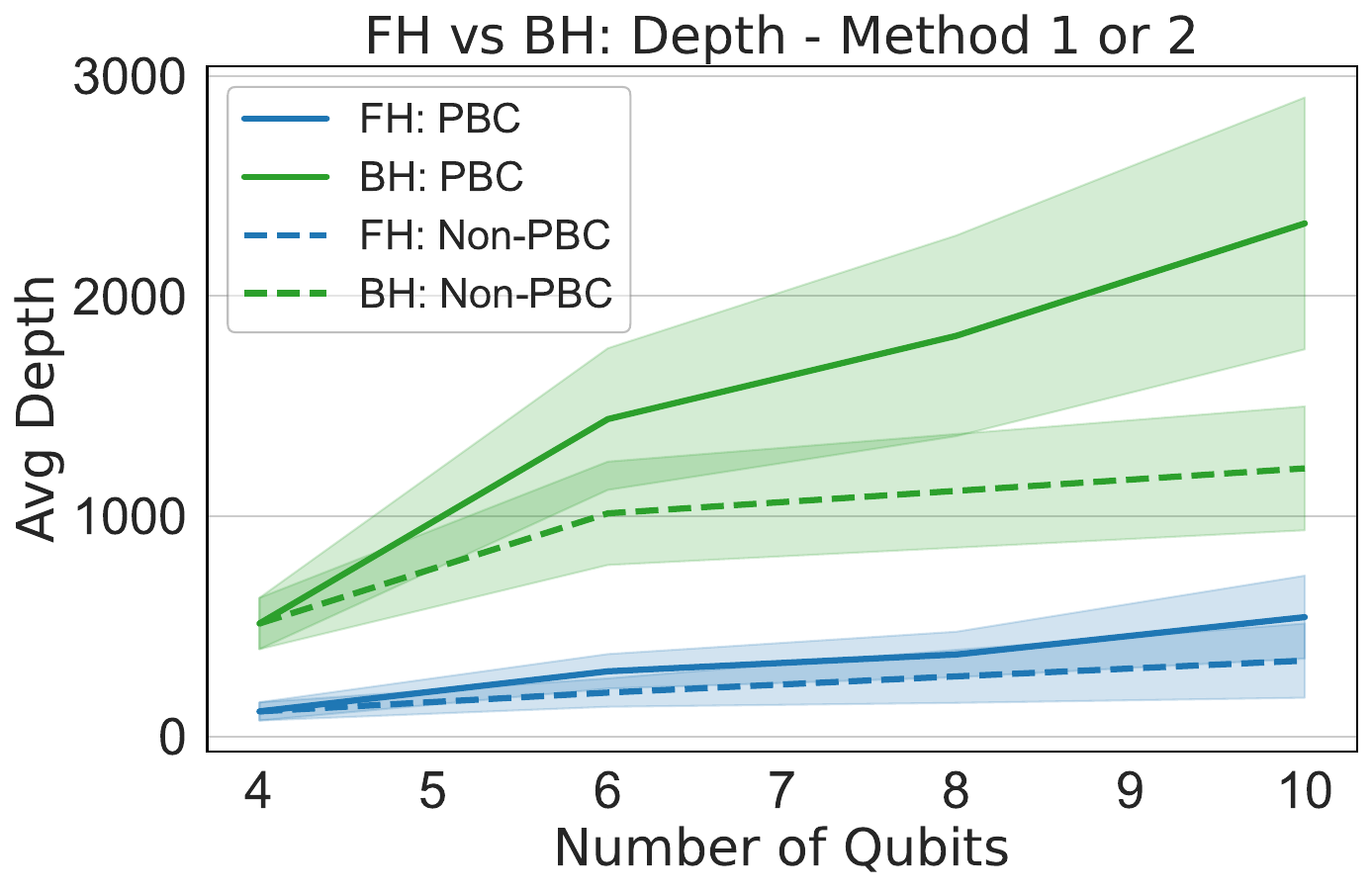}
\includegraphics[width=0.51\columnwidth]{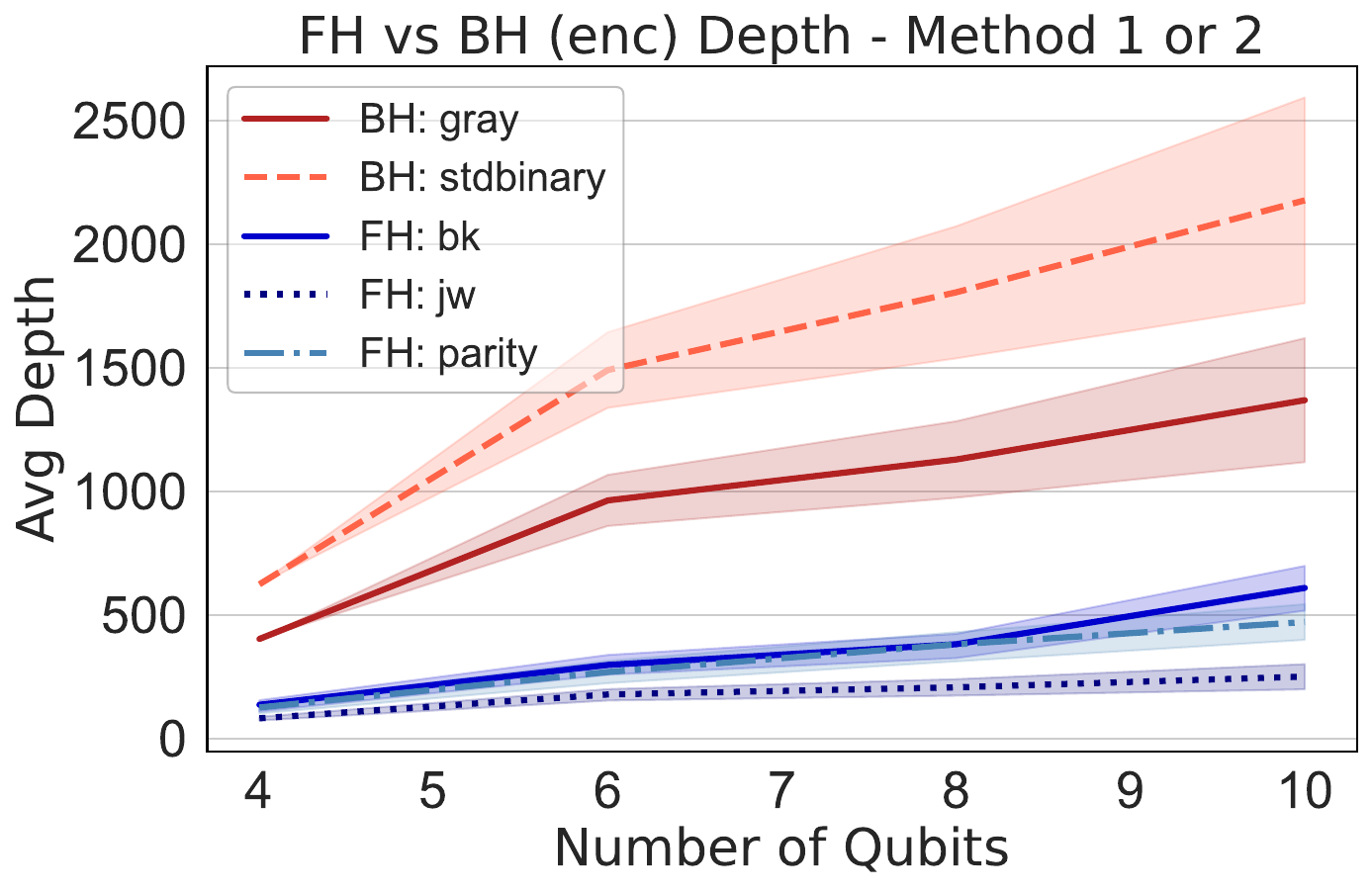}
\includegraphics[width=0.51\columnwidth]{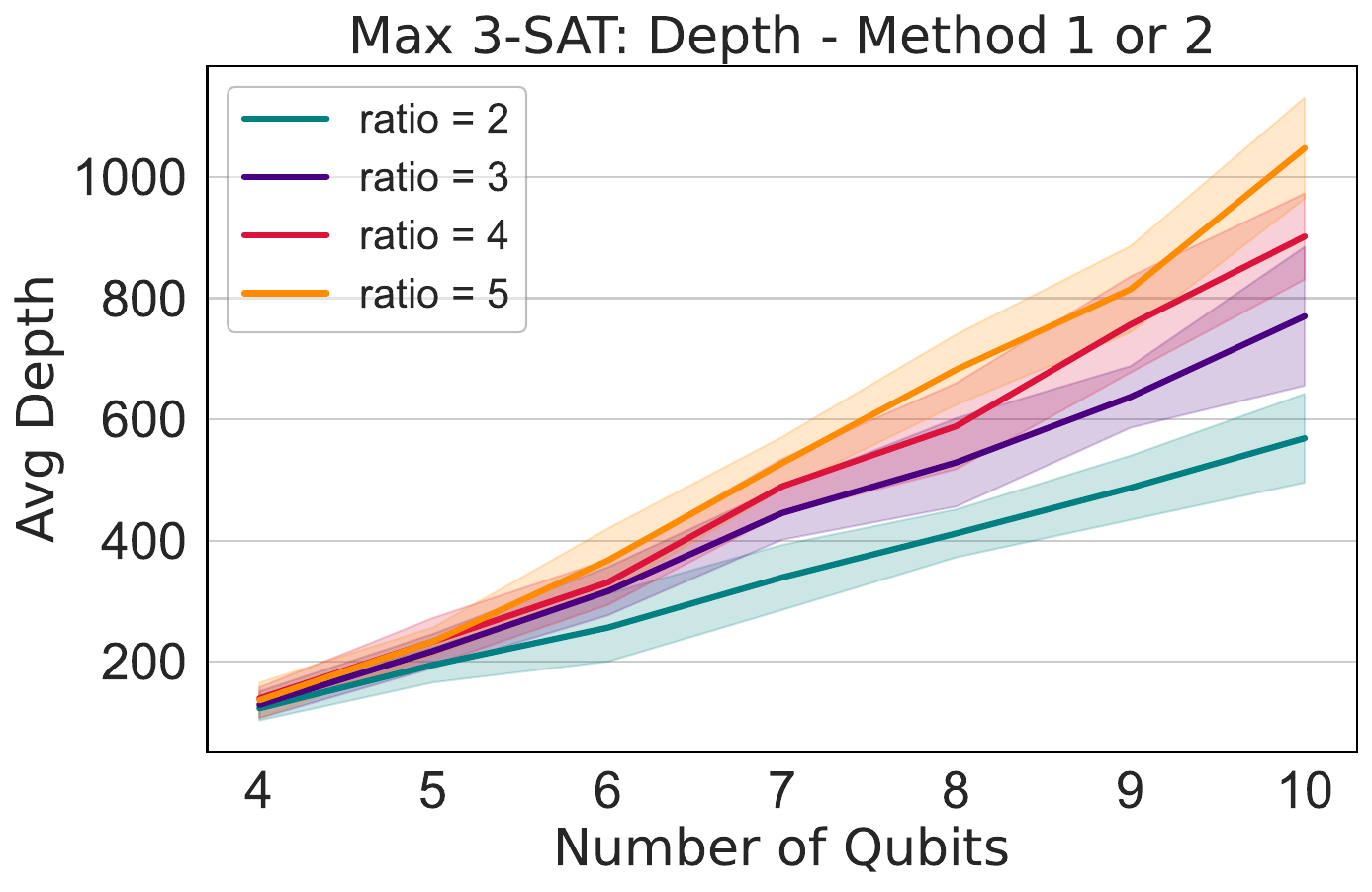}
\caption{
\textbf{Circuit Depth Comparisons Across Different Hamiltonian Models Based on Various Parameters.}
This figure illustrates the circuit depth comparisons for five Hamiltonian models—TFIM, Heisenberg, Fermi-Hubbard (FH), Bose-Hubbard (BH), and Max3SAT—using Method 1 or 2. The first plot compares TFIM and Heisenberg, showing that TFIM consistently has a lower depth, with periodic boundary conditions (PBC) increasing the depth for both models. The second plot compares FH and BH, indicating that BH requires more depth, with PBC contributing to higher depths for both. The third plot focuses on different encoding schemes for FH, where Bravyi-Kitaev (BK) encoding results in the highest depth, followed by Parity and Jordan-Wigner (JW). The fourth plot shows the BH model, where standard binary encoding leads to greater depth than Gray code. The final plot examines the Max3SAT model, revealing that circuit depth increases with the clause ratio, with a ratio of 5 resulting in the highest depth.
}
\label{fig:new_depth_all_models_all_params}
\end{figure*}
%****************

In this section, we compare the circuit depths of the five Hamiltonian models based on their specific parameters, as illustrated in~\autoref{fig:new_depth_all_models_all_params}. Our analysis focuses on Methods 1 and 2, as they exhibit similar circuit depths. Although Method 3 employs mirror circuits, resulting in double the circuit depth, it follows the same trends. Therefore, we do not explicitly show the circuit depths for Method 3 to avoid redundancy. The Transverse-Field Ising Model (TFIM) has a lower circuit depth compared to the Heisenberg model, as seen in the first plot of~\autoref{fig:new_depth_all_models_all_params}. This is due to the simpler interactions in TFIM, leading to less complex circuits. In both TFIM and Heisenberg models, periodic boundary conditions (PBC) result in higher depths than non-periodic conditions because PBC requires additional gates to account for the wrapping connections between the first and last qubits \cite{dutta2010quantum, biamonte2008realizable, white2004real}.

The Bose-Hubbard (BH) model has a higher circuit depth than the Fermi-Hubbard (FH) model, as shown in the second plot of~\autoref{fig:new_depth_all_models_all_params}. This difference arises because BH involves more complex interactions, particularly with particle hopping and on-site interactions. Similar to TFIM and Heisenberg, PBC increases the circuit depth for both BH and FH models due to the additional gates needed for periodic interactions. Generally, Hamiltonians that capture more detailed interactions between particles in the system will lead to circuits with higher depth. \cite{wecker2014gate, byrnes2006simulating,ortiz2001quantum}.

For the Fermi-Hubbard model, as illustrated in the third plot of~\autoref{fig:new_depth_all_models_all_params}, the choice of encoding mechanism—Bravyi-Kitaev (BK), Jordan-Wigner (JW), and Parity—significantly influences the circuit depth. The Bravyi-Kitaev encoding typically results in the deepest circuits since it employs a non-local mapping between fermionic operators in the Hamiltonian and qubits \cite{bravyi2002fermionic, seeley2012bravyi}. Parity encoding produces circuits with intermediate depth, while Jordan-Wigner encoding, known for its straightforward mapping of fermionic operators to qubit operators, results in the shallowest circuits \cite{jordan1993paulische}. Additionally, encodings that better manage fermionic anti-symmetry, like BK and Parity, might require extra SWAP gates, further increasing circuit depth when implemented on hardware \cite{whitfield2011simulation, tranter2018comparison}.

Similarly, for the Bose-Hubbard model, the choice of encoding mechanism—Standard Binary and Gray Code—affects the circuit depth, as shown in the third plot of~\autoref{fig:new_depth_all_models_all_params}. Standard Binary encoding generally leads to greater circuit depth compared to Gray Code due to the more complex gate operations required by the binary encoding scheme. Standard Binary encodes quantum states as binary numbers, requiring multiple bit flips between transitions, leading to more quantum gates and thus deeper circuits \cite{qian2015entanglement}. Conversely, Gray Code, where only one-bit changes between consecutive states, minimizes the number of necessary bit flips and reduces the quantum gate count, resulting in shallower circuits \cite{di2021improving}. Although unary encoding is another option for bosonic or vibrational simulations, it is not used here because it requires a significantly larger number of qubits to simulate the entire Hilbert space, which would further increase the circuit depth.

For the Max3SAT model, as illustrated in the last plot of~\autoref{fig:new_depth_all_models_all_params}, the circuit depth increases with the clause ratio. This trend can be understood through the lens of circuit complexity and error rates. A higher clause ratio implies more terms in the Hamiltonian requiring deeper circuits with more quantum gates \cite{rodriguez2024implementing}. 

%This simplicity translates to reduced opportunities for errors, as shallower circuits are less susceptible to decoherence, operational errors, and noise, thus maintaining higher fidelity. In contrast, as the clause ratio increases, the problem becomes more constrained and complex, necessitating deeper circuits . Additionally, the optimization landscape becomes more challenging with higher clause ratios featuring more local minima, complicating finding the global minimum.

In summary, our analysis of circuit depths across various Hamiltonian models highlights the impact of model-specific interactions, boundary conditions, and encoding mechanisms. The results underscore the significance of carefully selecting encoding schemes and understanding model intricacies when designing quantum circuits, particularly for complex models like Bose-Hubbard and Fermi-Hubbard. These findings provide valuable insights into optimizing Hamiltonian simulation implementations, particularly in balancing circuit depth with computational efficiency and fidelity. 
% Future work could explore further optimizations in encoding strategies and boundary condition management to reduce circuit depth while maintaining or improving simulation accuracy, especially as quantum hardware continues to advance and support larger, more complex circuits.

%----------------------------------------------------------

\subsection{Method Analysis for Additional Hamiltonians}
\label{apdx:subsec:method_analysis_additional}

%****************
\begin{figure}[t!]
\includegraphics[width=0.49\columnwidth]{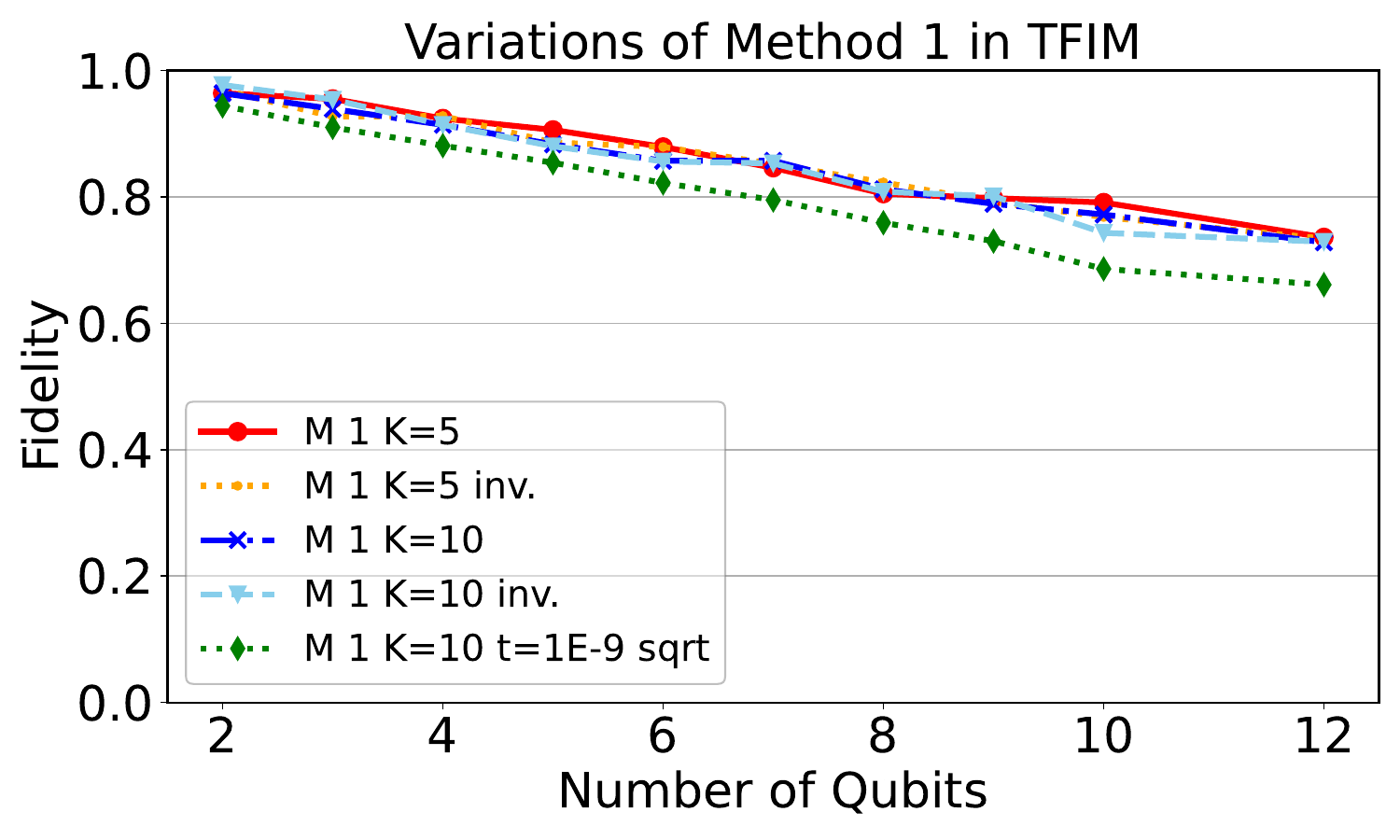}
\includegraphics[width=0.49\columnwidth]{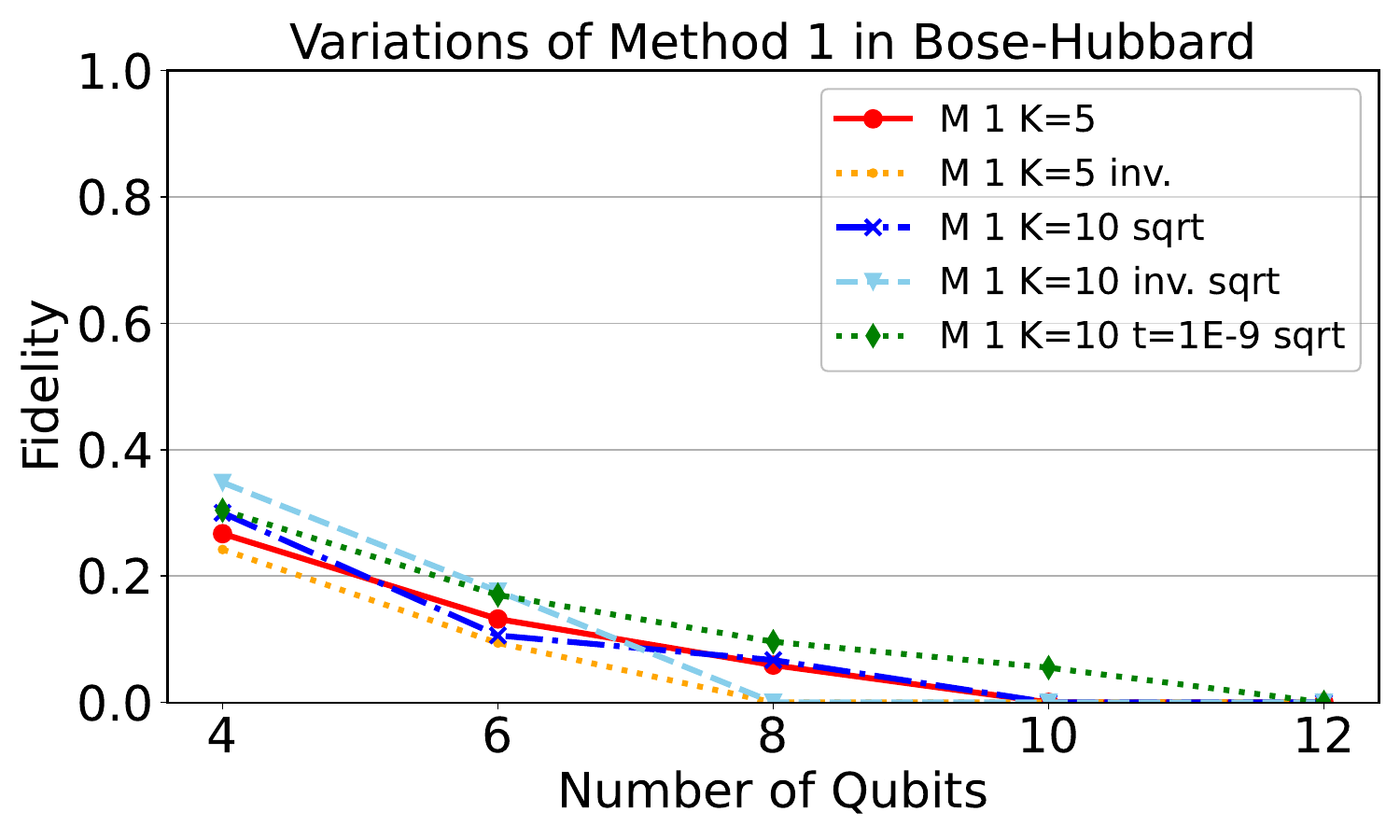}
\includegraphics[width=0.49\columnwidth]{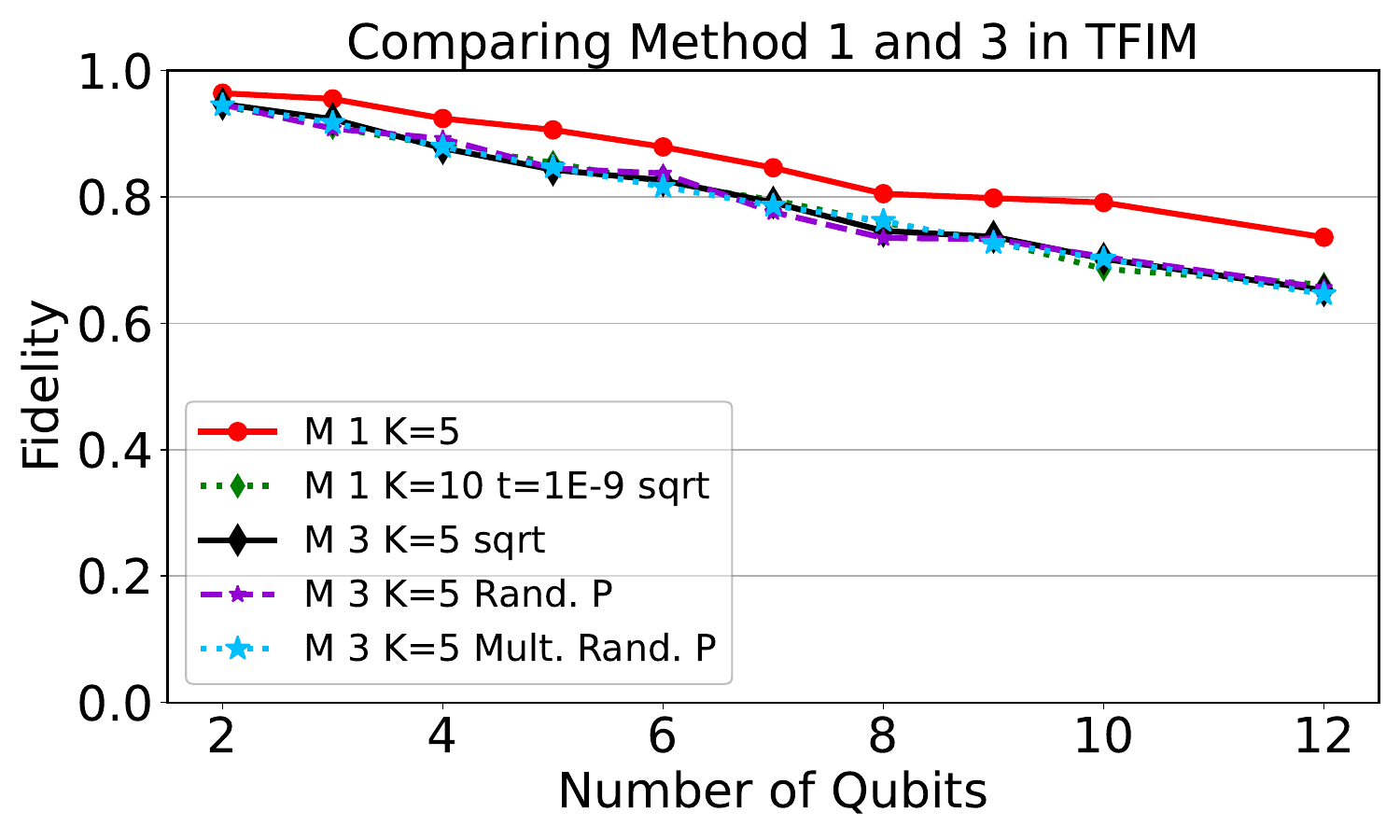}
\includegraphics[width=0.49\columnwidth]{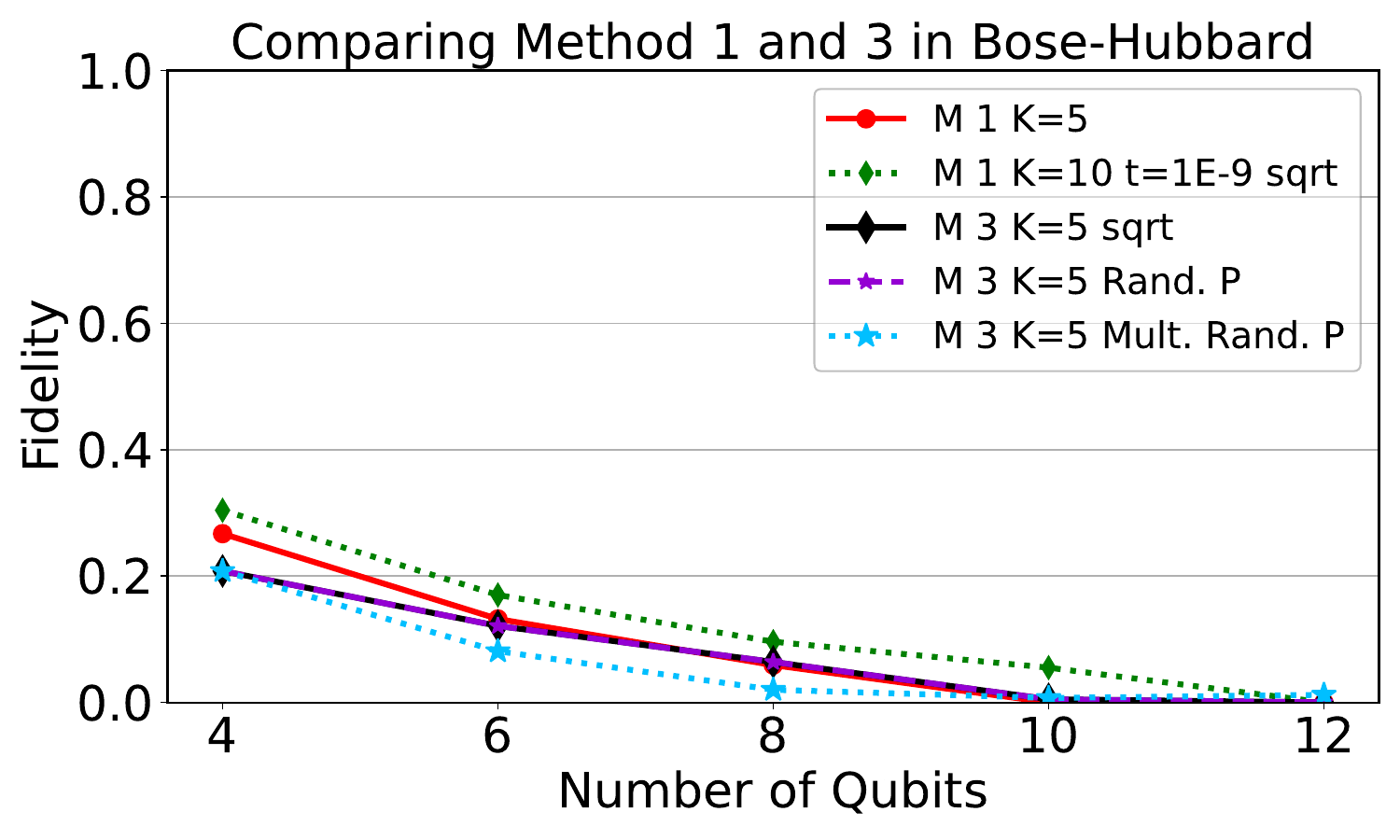}
\caption{
\textbf{Variations and Comparative Analysis of Method 1 and Method 3 Fidelity Across TFIM and Bose-Hubbard Hamiltonians.}
Fidelity comparison across the TFIM and Bose-Hubbard Hamiltonians. The top row shows the fidelity variations of different Method 1 configurations, with the TFIM model on the left and the Bose-Hubbard model on the right. The bottom row compares the fidelity trends between `Method 1' and the normalized `Method 3' for both Hamiltonians. In the TFIM model, the fidelity remains relatively stable across `Method 1' variations, with the normalized `Method 3' aligning closely with `Method 1 $K=10$ t=1E-9 sqrt'. In contrast, the Bose-Hubbard model shows a sharper fidelity decline, with Method 3's random Pauli variants failing to improve fidelity.
}
\label{fig:other_two_method_1_and_3_analysis}
\end{figure}
%****************

In~\autoref{subsec:scalable_fidelity_prediction}, we discuss the variants of Method 1 and provide a comparison between Method 1 and Method 3 across three Hamiltonian models: Heisenberg, Fermi-Hubbard, and Max3Sat. In this subsection, we extend our comparative analysis of Method 1 and Method 3 fidelities to two additional Hamiltonians: the TFIM and the Bose-Hubbard model. The fidelity trends for both Hamiltonians are presented in~\autoref{fig:other_two_method_1_and_3_analysis}.

For the TFIM Hamiltonian, as shown in the top-left panel of~\autoref{fig:other_two_method_1_and_3_analysis}, the variations in Method 1 circuits exhibit a consistent decline in fidelity as the number of qubits increases. The `Method 1 $K = 5$' variant remains the most stable, whereas the `Method 1 $K = 10$ t = $1E-9$ sqrt' variant shows a noticeable reduction in fidelity, aligning closely with the normalized `Method 3 $K = 5$ sqrt' variant, as depicted in the bottom-left panel. This close alignment further corroborates the effectiveness of the square root normalization method in Method 3, particularly for Hamiltonians like TFIM where circuit depth plays a significant role.

The Bose-Hubbard model, illustrated in the top-right and bottom-right panels of~\autoref{fig:other_two_method_1_and_3_analysis}, presents a more complex fidelity landscape. Here, the fidelity drop is more pronounced across all variants of Method 1 as the qubit count increases. The `Method 1 $K = 10$ t = $1E-9$ sqrt' variant shows a significant drop, which is reflected similarly in the normalized `Method 3 $K = 5$ sqrt' fidelity. However, unlike in the TFIM analysis, the random Pauli variants (`Method 3 Random Pauli $K = 5$' and `Method 3 Multiple Random Paulis $K = 5$') do not improve the fidelity trends, suggesting that Bose-Hubbard circuits are more susceptible to certain types of coherent errors that are not mitigated by these variants.

Overall, these results reinforce our earlier observations from the Heisenberg, Fermi-Hubbard, and Max3Sat Hamiltonians: the square root of the Method 3 fidelity serves as a reliable benchmark for the lower bound of Method 1 fidelity, even in more complex Hamiltonians like Bose-Hubbard. Nonetheless, the differences observed, particularly in the random Pauli variants, indicate that further refinement of Method 3 may be necessary to fully capture the fidelity nuances across different quantum systems.

\end{document}